\documentclass[10pt,aps,prd,twocolumn,notitlepage,superscriptaddress]{revtex4-1}
\usepackage{amsthm,amsfonts,verbatim,color,times}
\usepackage{amsmath,amssymb}
\usepackage{graphicx}
\usepackage{bm}
\usepackage{epsfig,slashed}
\usepackage{amssymb}
\usepackage{amsfonts}
\usepackage{tikz}
\usepackage{xparse}
\usepackage{ifthen}
\usepackage{lipsum}
\usepackage[colorlinks=true,citecolor=blue,
linkcolor=blue,urlcolor=blue]{hyperref}
\usepackage[caption=false]{subfig}
\allowdisplaybreaks

\begin{document}

\newcommand{\tr}{\mathop{\mathrm{tr}}}
\newcommand{\bsigma}{\boldsymbol{\sigma}}
\newcommand{\bphi}{\boldsymbol{\phi}}
\newcommand{\re}{\mathop{\mathrm{Re}}}
\newcommand{\im}{\mathop{\mathrm{Im}}}
\renewcommand{\b}[1]{{\boldsymbol{#1}}}
\newcommand{\diag}{\mathrm{diag}}
\newcommand{\sign}{\mathrm{sign}}
\newcommand{\sgn}{\mathop{\mathrm{sgn}}}

\newcommand{\halfs}{\mbox{\small{$\frac{1}{2}$}}} 
\newcommand{\Nf}{N_{\!f}}
\newcommand{\partialslash}{\partial \! \! \! /}
\newcommand{\xslash}{x \! \! \! /}
\newcommand{\yslash}{y \! \! \! /}

\newcommand{\cl}{\mathrm{cl}}
\newcommand{\mb}{\bm}
\newcommand{\ua}{\uparrow}
\newcommand{\da}{\downarrow}
\newcommand{\ra}{\rightarrow}
\newcommand{\la}{\leftarrow}
\newcommand{\mc}{\mathcal}
\newcommand{\bs}{\boldsymbol}
\newcommand{\lra}{\leftrightarrow}
\newcommand{\nn}{\nonumber}
\newcommand{\half}{{\textstyle{\frac{1}{2}}}}
\newcommand{\mf}{\mathfrak}
\newcommand{\MF}{\text{MF}}
\newcommand{\IR}{\text{IR}}
\newcommand{\UV}{\text{UV}}
\newcommand{\sech}{\mathrm{sech}}

\newcommand*{\vcenteredhbox}[1]{\begingroup
\setbox0=\hbox{#1}\parbox{\wd0}{\box0}\endgroup}
\newcommand{\picscalefactor}{0.5}


\title{Critical properties of the N\'eel--algebraic-spin-liquid transition}

\author{Nikolai Zerf}
\affiliation{Institut f\"ur Physik, Humboldt-Universit\"at zu Berlin, Newtonstra{\ss}e 15, D-12489 Berlin, Germany}

\author{Rufus Boyack}
\affiliation{Department of Physics, University of Alberta, Edmonton, Alberta T6G 2E1, Canada}
\affiliation{Theoretical Physics Institute, University of Alberta, Edmonton, Alberta T6G 2E1, Canada}

\author{Peter Marquard}
\affiliation{Deutsches Elektronen Synchrotron (DESY), Platanenallee 6, D-15738 Zeuthen, Germany}

\author{John A. Gracey}
\affiliation{Theoretical Physics Division, Department of Mathematical Sciences, University of Liverpool, P.O. Box 147, Liverpool, L69 3BX, United Kingdom}

\author{Joseph Maciejko}
\affiliation{Department of Physics, University of Alberta, Edmonton, Alberta T6G 2E1, Canada}
\affiliation{Theoretical Physics Institute, University of Alberta, Edmonton, Alberta T6G 2E1, Canada}
\affiliation{Canadian Institute for Advanced Research, Toronto, Ontario M5G 1Z8, Canada}

\date\today

\begin{abstract}
The algebraic spin liquid is a long-sought-after phase of matter characterized by the absence of quasiparticle excitations, a low-energy description in terms of emergent Dirac fermions and gauge fields interacting according to (2+1)D quantum electrodynamics (QED$_3$), and power-law correlations with universal exponents. The prototypical algebraic spin liquid is the Affleck-Marston $\pi$-flux phase, originally proposed as a possible ground state of the spin-1/2 Heisenberg model on the 2D square lattice. While the latter model is now known to order antiferromagnetically at zero temperature, recent sign-problem-free quantum Monte Carlo simulations of spin-1/2 fermions coupled to a compact $U(1)$ gauge field on the square lattice have shown that quantum fluctuations can destroy N\'eel order and drive a direct quantum phase transition to the $\pi$-flux phase. We show this transition is in the universality class of the chiral Heisenberg QED$_{3}$-Gross-Neveu-Yukawa model with a single $SU(2)$ doublet of four-component Dirac fermions (i.e., $N_f=1$), pointing out important differences with the corresponding putative transition on the kagom\'e lattice. Using an $\epsilon$ expansion below four spacetime dimensions to four-loop order, and a large-$N_f$ expansion up to second order, we show the transition is continuous and compute various thermodynamic and susceptibility critical exponents at this transition, setting the stage for future numerical determinations of these quantities. As a byproduct of our analysis, we also obtain charge-density-wave and valence-bond-solid susceptibility exponents at the semimetal-N\'eel transition for interacting fermions on the honeycomb lattice.
\end{abstract}

\maketitle

\section{Introduction}
\label{sec:intro}

The discovery of fractionalized phases of correlated quantum matter which fall outside the standard broken-symmetry classification pioneered by Landau is a prime goal of modern condensed matter physics. The inability to describe such phases in terms of conventional local order parameters suggests that continuous transitions among them, or between them and conventional phases, are themselves exotic and may fall outside the traditional Landau-Ginzburg-Wilson (LGW) paradigm according to which critical properties are solely determined by the long-wavelength, low-energy dynamics of critical order parameter fluctuations.

The paradigmatic example of fractionalized phase of matter is the algebraic spin liquid (ASL) or Dirac spin liquid, originally proposed by Affleck and Marston~\cite{affleck1988,marston1989} as a possible ground state of the spin-1/2 Heisenberg model on the 2D square lattice. This state can be envisaged as the result of quantum disordering a magnetically ordered ground state, e.g., the antiferromagnetic (AF) N\'eel state, in such a way that the bosonic spin-1 magnons of the ordered state fractionalize into a pair of electrically neutral spin-1/2 fermions governed by a Dirac dispersion and interacting with an emergent gauge field. However, it was shown early on~\cite{anderson1952,huse1988,reger1988,huse1988b} that due to the bipartite nature of the square lattice, the nearest-neighbor spin-1/2 Heisenberg model on this lattice develops N\'eel AF order at zero temperature, and the search for quantum spin liquid ground states rapidly switched its focus to frustrated magnetism~\cite{balents2010}, with much emphasis on spin models defined on non-bipartite lattices or with anisotropic interactions.

Surprisingly, a recent sign-problem-free quantum Monte Carlo (QMC) study~\cite{Meng2019} demonstrated the realization of an ASL-like phase on the 2D square lattice. The model simulated is a model of spinful fermions coupled to a lattice $U(1)$ gauge field (see Sec.~\ref{sec:square}); a gauge constraint (Gauss' law) enforced dynamically at low temperatures ensures there are no gauge-invariant local fermionic excitations in the low-energy Hilbert space, thus the ground state of such a Hamiltonian corresponds to that of a bosonic spin model. A magnetic flux of $\pi$ per plaquette is spontaneously generated at the mean-field level, similarly to the Affleck-Marston $\pi$-flux phase~\cite{FluxPhasesASL}. A parameter $J$ in the model controls the strength of gauge fluctuations such that for $J$ below a finite critical value there is no symmetry breaking, and AF and valence-bond-solid (VBS) correlations exhibit a power-law decay with distance, as expected for the ASL~\cite{kim1999,rantner2002,hermele2005,*hermele2007}. The low-energy effective theory of the model in this regime---(2+1)D quantum electrodynamics (QED$_3$) with $N_f=1$ flavor of spinful Dirac fermions, where $N_f$ is defined precisely here as the number of spin $SU(2)$ doublets of four-component Dirac spinors---is identical to that of the ASL~\cite{FluxPhasesASL}, and for simplicity we will refer to the phase observed numerically as the ASL. At low energies the ASL is believed~\footnote{To formally derive the low-energy effective field theory of the ASL one should successively integrate out short-wavelength degrees of freedom from a microscopic spin model, or a lattice gauge theory such as Eq.~(\ref{Hlattice}), to arrive at a coarse-grained continuum description. However, rigorously implementing such a sequence of exact renormalization group transformations in an interacting many-body system is practically impossible. 
Generally accepted arguments for the  QED$_{3}$ description starting from a microscopic spin model are presented in Ref.~\cite{hermele2005}. Furthermore, numerical studies using variational QMC~\cite{iqbal2011,iqbal2013,iqbal2014} and the density-matrix renormalization group (DMRG)~\cite{he2017,zhu2018} support the description of the ASL phase in terms of Dirac fermions coupled to a $U(1)$ gauge field} to be described by a strongly coupled (2+1)D conformal field theory whose universal properties, notably the critical exponents controlling the power-law correlations mentioned above, can be systematically computed using field-theoretic approaches such as the $\epsilon$-expansion~\cite{dipietro2016,dipietro2017,dipietro2018,zerf2018} and the large-$N_f$ expansion~\cite{gracey1993,gracey1994,rantner2002,hermele2005,*hermele2007,chester2016}. As alluded to in the abstract, the ASL phase is characterized by an absence of quasiparticle excitations. As pointed out in the classic papers on the ASL~\cite{wen1996,rantner2001,wen2002,hermele2005,*hermele2007}, the standard low-energy description of the ASL is in terms of Dirac fermions coupled to a fluctuating gauge field, i.e., QED$_{3}$. At high energies in this theory the coupling is weak, the physical excitations of the system are in one-to-one correspondence with free Dirac fermions, and a quasiparticle description holds. By contrast, at low energies---where the QED$_{3}$ description of the ASL is believed to hold---the coupling is strong and the quasiparticle description in terms of free Dirac fermions breaks down. Mathematically, spectral functions of physical (i.e., gauge-invariant) observables do not exhibit quasiparticle peaks at low energies, but rather are characterized by power laws with universal exponents.

Remarkably, Ref.~\cite{Meng2019} reported not only the first unbiased numerical observation of the ASL, but also that of a direct quantum phase transition from the ASL to the AF N\'eel state at a critical value of the parameter $J$ introduced above. (The existence of this transition is nontrivial because no intermediate phases appear, by contrast with the case with twice as many fermion flavors where the ASL-like phase first transitions to a VBS phase, then to an AF phase.) Numerical evidence further points to a continuous transition, as opposed to a first-order transition, implying the existence of universal critical phenomena. On general grounds one expects the low-energy theory of the transition to be of the QED$_3$-Gross-Neveu-Yukawa (GNY) type~\cite{lu2017,Meng2019,song2018}, as for the recently studied transitions between the ASL and a gapped chiral spin liquid~\cite{janssen2017,ihrig2018,zerf2018,boyack2019} or a gapped $\mathbb{Z}_2$ spin liquid~\cite{boyack2018}, but with the spontaneous breaking of an $O(3)$ symmetry as appropriate for N\'eel order~\cite{ghaemi2006}, as opposed to the Ising and XY symmetries appropriate for those two transitions, respectively. In theories of this type, gapless $U(1)$ gauge fluctuations as well as gapless Dirac fermions couple strongly to bosonic order parameter fluctuations and can give rise to novel universality classes of critical behavior distinct from the standard (Wilson-Fisher) Ising, XY, and Heisenberg universality classes, provided {\it bona fide} critical fixed points of the renormalization group (RG) exist in those theories. While a stable critical point was found for the chiral Ising QED$_3$-GNY model both in the $\epsilon$-expansion~\cite{janssen2017,ihrig2018,zerf2018} for all $N_f$ and in the large-$N_f$ expansion~\cite{gracey1992,gracey1993c,*gracey2018b,gracey1993b,gracey2018,boyack2019}, in the chiral XY QED$_3$-GNY model~\cite{boyack2018} a fixed point found at one-loop order in the $\epsilon$ expansion was shown to disappear below a certain critical value of $N_f$, in analogy to the phenomenon of fluctuation-induced first-order transitions~\cite{halperin1974}, while the large-$N_f$ expansion predicts a stable critical point. Whether the chiral Heisenberg QED$_3$-GNY model supports a critical point that remains stable at all $N_f$, and if so, what its universal critical properties are, are thus important open questions in light of the numerical discoveries reported in Ref.~\cite{Meng2019}. In Ref.~\cite{ghaemi2006}, the chiral Heisenberg QED$_3$-GNY model was proposed as the critical theory for a transition between the algebraic spin liquid, in its staggered flux phase realization on the square lattice, and a N\'eel antiferromagnet. The authors studied this model in the $\epsilon$ expansion below four dimensions at one-loop order, for the specific case of $N_f=1$, and found a critical fixed point describing a continuous phase transition. However, as we will show below, their RG beta functions and critical exponents disagree with ours. Our combined $\epsilon$-expansion and large-$N_f$ analyses allows us to perform several nontrivial consistency checks, detailed below, which lead us to believe our results are correct.

In this work, we revisit the problem of the N\'eel--algebraic-spin-liquid transition, motivated by the numerical results of Ref.~\cite{Meng2019}, and present a detailed study of the critical properties of the chiral Heisenberg QED$_3$-GNY model for a generic number $N_f$ of flavors of four-component Dirac fermions carrying an additional $SU(2)$ spin index. After deriving the specific form of the low-energy Lagrangian from microscopic considerations (Sec.~\ref{sec:Model}), we employ two complementary approaches to study the critical regime: an $\epsilon$ expansion in $d=4-\epsilon$ spacetime dimensions carried out to four-loop order for arbitrary $N_f$ (Sec.~\ref{sec:epsilon}), as well as a large-$N_f$ expansion carried out up to second order in fixed but arbitrary $d$ dimensions (Sec.~\ref{sec:LargeN}). The use of arbitrary $d$ in the latter approach allows us to establish consistency between the two methods, and gives us confidence in the correctness of our analysis. We do, however, pay attention to contributions specific to $d=3$ (Aslamazov-Larkin diagrams) which arise for certain quantities in the large-$N_f$ expansion. Through the $\epsilon$ expansion we establish the existence of a stable critical fixed point for all $N_f$, including the $N_f=1$ case relevant for the N\'eel-ASL transition on the 2D square lattice observed in the QMC study~\cite{Meng2019}. Gauge-invariant universal critical properties including the order parameter anomalous dimension $\eta_\phi$, the (inverse) correlation length exponent $1/\nu$, and the stability critical exponent $\omega$, but also the universal exponents $\Delta_\text{CDW}$, $\Delta_\text{VBS}$, and $\Delta_\text{QAH}$ characterizing the power-law decay of charge-density-wave (CDW), VBS, and quantum anomalous Hall (QAH) correlations at the transition, are computed to $\mathcal{O}(\epsilon^4)$ and up to $\mathcal{O}(1/N_f^2)$. While the CDW, VBS, and QAH exponents have been computed in the ASL phase before~\cite{hermele2005,*hermele2007}, the N\'eel-ASL critical point and its arbitrary-$N_f$ generalization correspond to a different (2+1)D conformal field theory than that describing the ASL phase, and is thus characterized by different---but equally universal---values for these exponents. We then use Pad\'e and Pad\'e-Borel resummation techniques to obtain numerical estimates of the critical exponents (Sec.~\ref{sec:Pade}). As a byproduct of our calculations, we also obtain CDW and VBS exponents at the semimetal-AF insulator quantum phase transition seen in the Hubbard model on the honeycomb lattice~\cite{sorella1992,herbut2006,honerkamp2008,herbut2009,sorella2012,ulybyshev2013,
assaad2013,janssen2014,parisen2015,otsuka2016,buividovich2018} (see Sec.~\ref{sec:O3GNY4Lbilinears} and \ref{sec:O3GNYlargeNf}). These correspond to the scaling dimensions of certain fermion bilinear operators in the chiral Heisenberg GNY model, which to our knowledge have not been computed before despite recent studies of this model~\cite{zerf2017,gracey2018c}.

\section{The N\'eel--algebraic-spin-liquid transition}
\label{sec:Model}

In this section we derive the continuum quantum field theory describing the N\'eel-ASL transition on the square lattice (Sec.~\ref{sec:square}) as observed in the QMC study of Ref.~\cite{Meng2019}, and contrast it with the field theory description of similar magnetic ordering transitions on the kagom\'e lattice (Sec.~\ref{sec:kagome}).

\subsection{Square lattice}
\label{sec:square}

The starting point is the compact $U(1)$ lattice gauge theory model considered in Ref.~\cite{Meng2019}, here restricted to two fermion flavors:
\begin{align}\label{Hlattice}
H&=\frac{J}{4}\sum_{r,\mu}L_{r,r+\hat{\mu}}^2-t\sum_{r,\mu,\sigma}\left(c_{r\sigma}^\dag e^{i\theta_{r,r+\hat{\mu}}}c_{r+\hat{\mu},\sigma}^{\phantom{\dagger}}+\text{H.c.}\right)\nn\\
&\phantom{=}+K\sum_\square\cos(\Delta\times\boldsymbol\theta).
\end{align}
Here $c_{r\sigma}^{(\dag)}$ is the annihilation (creation) operator for a fermion with spin $\sigma=\uparrow,\downarrow$ on site $r$ of a 2D square lattice, $\theta_{r,r+\hat{\mu}}\in[0,2\pi)$ is an angular variable living on the nearest-neighbor links of the square lattice, with $\mu=x,y$ denoting its two orthogonal directions, and $L_{r,r+\hat{\mu}}$ is the operator conjugate to $\theta_{r,r+\hat{\mu}}$, i.e., $[\theta_{r,r+\hat{\mu}},L_{r',r'+\hat{\nu}}]=i\delta_{rr'}\delta_{\mu\nu}$. Both $\theta_{r,r+\hat{\mu}}$ and $L_{r,r+\hat{\mu}}$ can be interpreted as vector fields $\theta_{r,\mu}$ and $L_{r,\mu}$, and $\Delta\times\boldsymbol{\theta}=\Delta_x\theta_{r,y}-\Delta_y\theta_{r,x}$ denotes the lattice curl of $\theta_{r,\mu}$, corresponding to its circulation around an elementary plaquette $\square$, where $\Delta_\mu\theta_{r,\nu}=\theta_{r+\hat{\mu},\nu}-\theta_{r,\nu}$ is the lattice derivative. The first term, proportional to $J$, is an electric-field term that controls the strength of the gauge fluctuations, while the last term, proportional to $K>0$, is a magnetic-field term that favors the maximal (i.e., $\pi$) amount of magnetic flux per plaquette. The fermion chemical potential is set to zero, corresponding to half filling.

When $J=0$, $\theta_{r,\mu}$ is a classical variable and the fermions are noninteracting. The ground state at half-filling is guaranteed by Lieb's theorem~\cite{lieb1994} to have $\pi$ flux per plaquette. When $J$ is nonzero but small, the gauge field acquires dynamics but the QMC results~\cite{Meng2019} indicate the $\pi$-flux phase remains stable up to some critical value, given by $J_c=1.43(46)$ for $t=K=1$. In the presence of gauge fluctuations the $\pi$-flux phase is an ASL described by the conformally invariant, infrared-stable fixed point of QED$_3$ with two flavors of four-component fermions. For $J>J_c$ the model develops long-range $(\pi,\pi)$ AF order, consistent with the known N\'eel ground state of the $SU(2)$-symmetric Heisenberg model, to which model (\ref{Hlattice}) reduces in the $J\rightarrow\infty$ limit. The smooth behavior at $J=J_c$ of both the AF correlation ratio and the flux energy per plaquette observed in the numerical results~\cite{Meng2019} suggests the N\'eel-ASL transition is continuous, and should be described by a critical fixed point of a continuum quantum field theory. To derive the precise form of this quantum field theory, which is expected to be of the chiral Heisenberg QED$_3$-GNY type, we simply need to express the N\'eel order parameter in terms of the low-energy Dirac fermions in the $\pi$-flux phase. This in turn will dictate the correct form of the Yukawa coupling between Dirac fermions and the $O(3)$ vector field describing order parameter fluctuations. As will be seen in detail below and in Sec.~\ref{sec:kagome}, owing to the precise relation between the low-energy Dirac fermion fields and the original microscopic lattice fermions, which differs on the square and kagom\'e lattices, the (2+1)D N\'eel-ASL transition on the square lattice can be accessed by dimensional continuation from the four-dimensional Lorentz-invariant chiral Heisenberg QED$_3$-GNY model, while conceptually similar magnetic ordering transitions on the kagom\'e lattice cannot. This has important implications for $\epsilon$-expansion studies of such transitions as will be discussed below. Similar difficulties in dimensional continuation for critical points described by the chiral Ising QED$_{3}$-GNY model were discussed in Ref.~\cite{boyack2019}, but in a purely field-theoretical context without reference to a microscopic lattice. Related issues were discussed recently for pure QED$_{3}$ in the high-energy literature~\cite{dipietro2016,dipietro2017}, and for the pure chiral XY GNY model in Ref.~\cite{wamer2018}.

\begin{figure}[t]
\includegraphics[width=0.75\columnwidth]{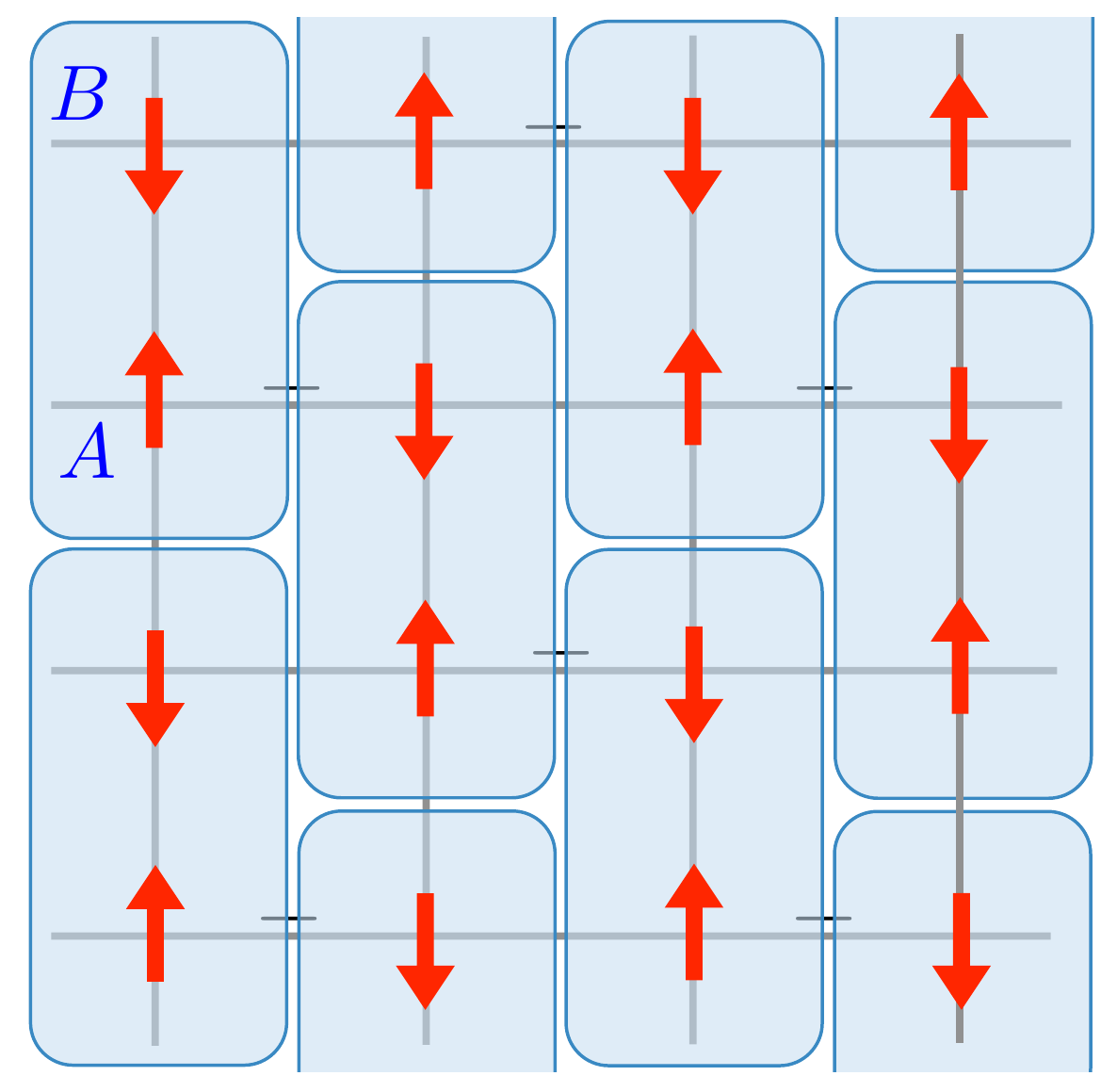}
\caption{The $\pi$-flux phase on the square lattice: the minus signs indicate a particular choice of gauge to realize $\pi$ flux per plaquette, and $A$ and $B$ denote the two sites in the enlarged unit cell (blue rectangles). N\'eel order (red arrows) does not lead to a further enlargement of this gauge-dependent choice of unit cell.}
\label{fig:square}
\end{figure}

We begin by studying the $\pi$-flux phase in the absence of gauge fluctuations, described by the fermionic part of the $J=0$ limit of the model (\ref{Hlattice}). Choosing a gauge such that the hopping amplitude on alternating links of the square lattice acquires an additional phase factor of $-1$ (Fig.~\ref{fig:square}), the Hamiltonian can be written as
\begin{align}\label{Hpiflux}
H_0=\sum_\sigma\int_\text{1BZ}\frac{d^2k}{(2\pi)^2}c_{\b{k}\sigma}^\dag h(\b{k})c_{\b{k}\sigma}^{\phantom{\dagger}},
\end{align}
where $c_{\b{k}\sigma}=(c_{\b{k},A,\sigma},c_{\b{k},B,\sigma})$ is a two-component spinor, $A,B$ denote the two sublattices of the square lattice, and
\begin{align}
h(\b{k})=-t\left(
\begin{array}{cc}
0 & f(\b{k}) \\
f^*(\b{k}) & 0
\end{array}
\right),
\end{align}
where
\begin{align}
f(\b{k})=1-e^{i(k_x-k_y)}+e^{-i(k_x+k_y)}+e^{-2ik_y}.
\end{align}
The integral is over wavevectors $\b{k}$ (measured in inverse units of the lattice constant) in the first Brillouin zone, which is reduced due to the doubling of the unit cell stemming from the choice of gauge and forms a square of side length $\pi\sqrt{2}$ rotated by 45$^\circ$ with respect to the original Brillouin zone. The spectrum is given by $E_\pm(\b{k})=\pm 2t\sqrt{\sin^2 k_x+\cos^2 k_y}$ and exhibits two inequivalent Dirac points at $\pm\b{Q}$ with $\b{Q}\equiv\left(0,\frac{\pi}{2}\right)$. At half filling the low-energy excitations are particle-hole excitations of the Dirac fermions, and we can linearize the Hamiltonian near the Dirac points,
\begin{align}\label{LinearizedH}
h(\pm\b{Q}+\b{p})=v_F(\pm\mu_1 p_x+\mu_2 p_y)+\mathcal{O}(\b{p}^2),
\end{align}
where $|\b{p}|$ is much less than the dimensions of the Brillouin zone, we denote Pauli matrices acting on the Dirac components by $\mu_1,\mu_2,\mu_3$, and $v_F\equiv 2t$. The low-energy Hamiltonian is thus given by
\begin{align}\label{HDirac}
H_0\approx v_F\sum_{\alpha,\sigma}\int\frac{d^2p}{(2\pi)^2}\psi_{\alpha\sigma}^\dag(\b{p})\left(\mu_1 p_x+\mu_2 p_y\right)\psi_{\alpha\sigma}^{\phantom{\dagger}}(\b{p}),
\end{align}
where we define the slow (Dirac) fields
\begin{align}\label{SlowFieldsSquare}
\psi_{+\sigma}(\b{p})=\left(\begin{array}{c}
c_{\b{Q}+\b{p},A,\sigma} \\
c_{\b{Q}+\b{p},B,\sigma}
\end{array}\right),\hspace{3mm}
\psi_{-\sigma}(\b{p})=\left(\begin{array}{c}
c_{-\b{Q}+\b{p},B,\sigma} \\
-c_{-\b{Q}+\b{p},A,\sigma}
\end{array}\right).
\end{align}
Note that defining $\psi_{-\sigma}(\b{p})$ as $i\mu_2$ acting on the two-component spinor $c_{-\b{Q}+\b{p},\sigma}$ is necessary to absorb the extra minus sign in front of $\mu_1$ in the Hamiltonian matrix (\ref{LinearizedH}) linearized near $-\b{Q}$. Introducing the two-component Dirac adjoint $\overline{\psi}_{\alpha\sigma}\equiv\psi^\dag_{\alpha\sigma}\gamma_0$ with $\gamma_0=\mu_3$, and Fourier transforming $H_0$ back to (continuous) real space, we find that the imaginary-time Lagrangian density $\mathcal{L}_0=\sum_{\alpha,\sigma}\psi_{\alpha\sigma}^\dag\partial_0\psi_{\alpha\sigma}+H_0$ can be written as
\begin{align}\label{L0square}
\mathcal{L}_0=\sum_{\alpha,\sigma}\overline{\psi}_{\alpha\sigma}\gamma_\mu\partial_\mu\psi_{\alpha\sigma},
\end{align}
where $(\gamma_0,\gamma_1,\gamma_2)=(\mu_3,\mu_2,-\mu_1)$ are $2\times 2$ Euclidean gamma matrices, and we have rescaled the spatial coordinates to set the Dirac velocity $v_F$ to unity.

We now turn to N\'eel order. Referring to Fig.~\ref{fig:square}, the N\'eel order parameter $\b{N}$ is an $SU(2)$ triplet of fermion bilinears given by
\begin{align}
\b{N}\equiv\frac{1}{\mathcal{N}}\sum_{\b{R}}\left(\b{S}_{\b{R}A}-\b{S}_{\b{R}B}\right),
\end{align}
where $\mathcal{N}$ is the total number of sites, $\b{R}=n_1\b{a}_1+n_2\b{a}_2$ with $n_1,n_2\in\mathbb{Z}$ denotes sites of the underlying Bravais lattice with primitive lattice vectors $\b{a}_1\equiv\hat{\b{x}}-\hat{\b{y}}$ and $\b{a}_2\equiv\hat{\b{x}}+\hat{\b{y}}$, measured in units of the lattice constant, and $\b{S}_{\b{R}A(B)}=\frac{1}{2}\sum_{\sigma\sigma'}c_{\b{R}A(B)\sigma}^\dag\bsigma_{\sigma\sigma'}c_{\b{R}A(B)\sigma'}^{\phantom{\dagger}}$ is the local spin operator on sublattice $A(B)$, where $\bsigma=(\sigma_1,\sigma_2,\sigma_3)$ denotes Pauli matrices acting on the (physical) spin components. Fourier transforming and keeping only the low-energy Dirac degrees of freedom near $\pm\b{Q}$, we find
\begin{align}
\b{N}\sim\sum_{\sigma\sigma'}\int\frac{d^2p}{(2\pi)^2}\Psi_\sigma^\dag(\b{p})
\left(\begin{array}{cc}
\gamma_0 & 0 \\
0 & -\gamma_0
\end{array}\right)
\bsigma_{\sigma\sigma'}\Psi_{\sigma'}(\b{p}),
\end{align}
where we have defined the four-component Dirac spinors
\begin{align}
\Psi_\sigma=\left(\begin{array}{c}
\psi_{+\sigma} \\
\psi_{-\sigma}
\end{array}\right),\hspace{5mm}\sigma=\uparrow,\downarrow.
\end{align}
This motivates the following choice of (reducible) four-dimensional representation of the $SO(3)$ Clifford algebra,
\begin{align}\label{gammaSO3}
\Gamma_\mu=\left(\begin{array}{cc}
\gamma_\mu & 0 \\
0 & -\gamma_\mu
\end{array}\right),\hspace{5mm}\mu=0,1,2,
\end{align}
and the corresponding definition of four-component Dirac adjoint $\overline{\Psi}_\sigma\equiv\Psi_\sigma^\dag\Gamma_0$, in terms of which the N\'eel order parameter is written as
\begin{align}
\b{N}\sim\int d^2x\,\overline{\Psi}\bsigma\Psi,
\end{align}
with trace over spin indices understood.

N\'eel order is not explicit in the microscopic Hamiltonian~(\ref{Hlattice}) but arises from the spontaneous breaking of its spin rotation symmetry. Having identified the form of the N\'eel order parameter bilinear in terms of the low-energy Dirac fields, we can write down an effective LGW Lagrangian valid near the N\'eel-ASL transition in physical 2+1 dimensions,
\begin{align}\label{L2+1square}
\mathcal{L}_{2+1}=\sum_{\alpha,\sigma}\overline{\psi}_{\alpha\sigma}\gamma_\mu D_\mu\psi_{\alpha\sigma}+\frac{1}{4}F_{\mu\nu}^2+g\boldsymbol{\phi}\cdot\overline{\Psi}\frac{\bsigma}{2}\Psi+\mathcal{L}_{\boldsymbol{\phi}},
\end{align}
where $\boldsymbol{\phi}$ is a Lorentz-scalar $O(3)$ vector field that has the interpretation of dynamical order parameter field, and $\mathcal{L}_{\boldsymbol{\phi}}$ is the Lagrangian controlling its dynamics, i.e., the $O(3)$ vector model. In the language of this effective field theory, the ASL phase corresponds to a phase in which $\langle\b{\phi}\rangle=0$ and the $O(3)$ symmetry is unbroken, while in the N\'eel phase $\langle\b{\phi}\rangle\neq 0$ and the $O(3)$ symmetry is spontaneously broken to an $O(2)$ subgroup of spin rotations about the direction of $\langle\b{\phi}\rangle$. We have accounted for the general situation with gauge fluctuations ($J\neq 0$) by replacing $\partial_\mu$ in Eq.~(\ref{L0square}) by the gauge-covariant derivative $D_\mu=\partial_\mu-ieA_\mu$, with $e$ the gauge coupling and $A_\mu$ the continuum limit of the $U(1)$ lattice gauge field $\theta_{r,r+\hat{\mu}}$. We have also added a Maxwell term for the gauge field with $F_{\mu\nu}=\partial_\mu A_\nu-\partial_\nu A_\mu$ the field-strength tensor, which originates microscopically from the terms proportional to $J$ and $K$ in the lattice Hamiltonian (\ref{Hlattice}). The Lagrangian (\ref{L2+1square}) is equivalent to that studied in Ref.~\cite{ghaemi2006}, although the fermionic mean-field state considered in the latter reference is the staggered flux state rather than the $\pi$-flux state, and thus the projective symmetry group characterizing the corresponding spin liquid state is different (see Appendix~\ref{app:PSG}). 

As in previous work on related theories~\cite{dipietro2016,dipietro2017,dipietro2018,janssen2017,ihrig2018,zerf2018,boyack2018}, we will be interested in studying the critical properties of Eq.~(\ref{L2+1square}) in the standard $\epsilon$ expansion below four (spacetime) dimensions, and thus require a four-dimensional Lorentz-invariant Lagrangian which dimensionally continues to Eq.~(\ref{L2+1square}). In $d=4$ dimensions the reducible four-dimensional representation of the $SO(3)$ Clifford algebra becomes an irreducible representation of the $SO(4)$ Clifford algebra when supplemented with a fourth gamma matrix $\Gamma_3$, which for the choice of $SO(3)$ gamma matrices in Eq.~(\ref{gammaSO3}) is given by
\begin{align}\label{Gamma3}
\Gamma_3=\left(\begin{array}{cc}
0 & -i \\
i & 0
\end{array}\right).
\end{align}
One can then check explicitly that the four-dimensional Lagrangian
\begin{align}\label{L3+1square}
\mathcal{L}_{3+1}=\sum_{\sigma}\overline{\Psi}_\sigma\Gamma_\mu D_\mu\Psi_\sigma+\frac{1}{4}F_{\mu\nu}^2+g\boldsymbol{\phi}\cdot\overline{\Psi}\frac{\bsigma}{2}\Psi+\mathcal{L}_{\boldsymbol{\phi}},
\end{align}
is invariant under {\it four-dimensional} Lorentz transformations, and reduces to the physical Lagrangian (\ref{L2+1square}) when all terms involving $x_3$ and $A_3$ are ignored. Thus the critical properties of the N\'eel-ASL transition on the square lattice can be consistently studied in an 
$\epsilon$ expansion of the chiral Heisenberg QED-GNY theory (\ref{L3+1square}) with a single $SU(2)$ doublet $(\Psi_\uparrow,\Psi_\downarrow)$ of four-component Dirac fermions.

Finally, we note that while our field-theoretic study is motivated by the numerical study of a specific model, Hamiltonian~(\ref{Hlattice}), it is more broadly applicable as it relies only on the universal low-energy description of the ASL phase in terms of QED$_{3}$, and the broken symmetries in the N\'eel phase. In other words, if a N\'eel-ASL transition is found in any other microscopic spin model on the square lattice, its universal critical exponents would be given by those we have calculated, regardless of the particular microscopic model. We choose to focus on the model~(\ref{Hlattice}) as it is the only microscopic model in which a direct, continuous transition from the ASL to the N\'eel state has been observed in unbiased, controlled QMC simulations~\cite{Meng2019}.

\subsection{Kagom\'e lattice}
\label{sec:kagome}

We now briefly review the slave-fermion description of the ASL on the kagom\'e lattice~\cite{hastings2000,ran2007,hermele2008} and its possible magnetic ordering instabilities. Similar to the $\pi$-flux phase on the square lattice, ignoring gauge fluctuations the ASL on the kagom\'e lattice can be understood as the ground state of noninteracting fermions with nearest-neighbor hopping on the kagom\'e lattice and $\pi$ flux through each hexagonal plaquette, but zero flux through each triangular plaquette. The simplest gauge choice doubles the unit cell in (say) the $\b{a}_1$ direction, where $\b{a}_1=(2,0)$ and $\b{a}_2=(1,\sqrt{3})$ are the primitive vectors of the underlying triangular Bravais lattice, with lengths measured in units of the side length of a kagom\'e triangle. Correspondingly the original (hexagonal) first Brillouin zone is halved (and becomes rectangular); at half filling the Fermi energy intersects two inequivalent Dirac points at $\pm\b{Q}=\frac{\pi}{2\sqrt{3}}(0,\pm 1)$. Effective Hamiltonians similar to Eq.~(\ref{LinearizedH}) can be obtained in first-order degenerate perturbation theory,
\begin{align}
h(\pm\b{Q}+\b{p})=v_F(\mu_1 p_x+\mu_2 p_y)+\mathcal{O}(\b{p}^2),
\end{align}
with $v_F=\sqrt{2}t$. Slow Dirac fields can be defined in analogy to Eq.~(\ref{SlowFieldsSquare}), albeit with a more complicated dependence on the microscopic lattice fields (the doubled unit cell has six sites), and the imaginary-time Lagrangian density is again given by Eq.~(\ref{L0square}) with the same set of $2\times 2$ gamma matrices, $(\gamma_0,\gamma_1,\gamma_2)=(\mu_3,\mu_2,-\mu_1)$~\cite{hermele2008}. As before, gauge fluctuations are incorporated by promoting $\partial_\mu$ to the gauge-covariant derivative $D_\mu$ in the Dirac Lagrangian and adding a Maxwell term for the emergent continuum gauge field $A_\mu$.

We now turn to magnetic ordering instabilities, i.e., instabilities accompanied by spontaneous breaking of the spin $SU(2)$ symmetry (which excludes VBS states) of an underlying microscopic spin model on the kagom\'e lattice, such as the spin-1/2 Heisenberg antiferromagnet~\cite{hastings2000,ran2007,hermele2008,iqbal2011,iqbal2013,iqbal2014,he2017,zhu2018}. We further focus on patterns of ordering that can be achieved through the condensation of Dirac fermion bilinears which gaps out the Dirac fermions (once the fermions are gapped, monopole proliferation typically ensues, which further gaps out the dynamical photon and leads to confinement~\cite{song2018,song2018b}---see Sec.~\ref{sec:monopole}). This leaves two possibilities~\cite{hermele2008}: a three-fold degenerate collinear stripe AF state with ordering wavevectors corresponding to the three symmetry-related $M$ points of the physical (gauge-invariant) hexagonal Brillouin zone, and a zero-wavevector noncollinear state with nonzero vector spin chirality~\cite{lu2017}.

We first consider the stripe AF state (i.e., Fig.~4 in Ref.~\cite{hermele2008}), which corresponds to the condensation of either of the three degenerate spin-triplet, time-reversal-odd bilinears $\Psi^\dag\mu_3\tau_i\bsigma\Psi$, $i=1,2,3$, where $\tau_1,\tau_2,\tau_3$ are Pauli matrices acting on the nodal/valley index $\alpha=\pm$ distinguishing the two Dirac nodes~\cite{hermele2008}. Introducing a corresponding set of Lorentz-scalar $O(3)$ vector fields $\boldsymbol{\phi}_1,\boldsymbol{\phi}_2,\boldsymbol{\phi}_3$ describing the fluctuating order parameters, the form of the LGW Lagrangian for the stripe AF-ASL transition is dictated by the transformation properties of the stripe AF fermion bilinears under space group operations, which have been worked out in detail in Ref.~\cite{hermele2008}. One obtains
\begin{align}\label{L2+1stripe}
\mathcal{L}_{2+1}&=\sum_{\alpha,\sigma}\overline{\psi}_{\alpha\sigma}\gamma_\mu D_\mu\psi_{\alpha\sigma}+\frac{1}{4}F_{\mu\nu}^2+g\sum_i\boldsymbol{\phi}_i\cdot\Psi^\dag\gamma_0\tau_i\bsigma\Psi\nonumber\\
&\phantom{=}+\mathcal{L}_{\boldsymbol{\phi}_i},
\end{align}
where the order parameter Lagrangian $\mathcal{L}_{\boldsymbol{\phi}_i}$ is given by
\begin{align}\label{Lphi1phi2phi3}
\mathcal{L}_{\boldsymbol{\phi}_i}&=\sum_i\left[\frac{1}{2}(\partial_\mu\boldsymbol{\phi}_i)^2+\frac{r}{2}\boldsymbol{\phi}_i^2+u(\boldsymbol{\phi}_i^2)^2\right]\nonumber\\
&\phantom{=}+\sum_{i<j}\left[u'\boldsymbol{\phi}_i^2\boldsymbol{\phi}_j^2
+u''(\boldsymbol{\phi}_i\cdot\boldsymbol{\phi}_j)^2\right]+\mathcal{O}\left((\boldsymbol{\phi}_i^2)^3\right),
\end{align}
where $r,u,u',u'',\ldots$ are phenomenological coupling constants. The $\mathbb{Z}_3$ anisotropy introduced by the quartic terms in Eq.~(\ref{Lphi1phi2phi3}) is a consequence of the discrete $C_6$ symmetry of the kagom\'e lattice which rotates the three $M$ points into each other.

In order to study the critical properties of the stripe AF-ASL transition in the $4-\epsilon$ expansion one must lift the (2+1)D theory (\ref{L2+1stripe}) to a four-dimensional Lorentz-invariant theory with four-component Dirac spinors. The main issue, as compared to the theory (\ref{L2+1square}) for the N\'eel-ASL transition on the square lattice, is that the Yukawa coupling involves the nodal matrices $\tau_i$. These matrices generate an $SU(2)_\text{nodal}$ subgroup of the enlarged $SU(4)$ flavor symmetry of the pure ASL state, i.e., the internal symmetry of the first term in Eq.~(\ref{L2+1stripe}),  but in the four-dimensional theory elements of this subgroup act as {\it Lorentz group} elements on four-component Dirac spinors. In other words, the Yukawa coupling breaks the $SU(2)_\text{nodal}$ symmetry in the (2+1)D theory, which means it breaks Lorentz symmetry in the four-dimensional theory. For instance, suppose we choose the $4\times 4$ gamma matrix representation given in Eq.~(\ref{gammaSO3}) and Eq.~(\ref{Gamma3}): then the three bilinears $\Psi^\dag\gamma_0\tau_i\bsigma\Psi$, $i=1,2,3$ are proportional to $i\overline{\Psi}\Gamma_3\bsigma\Psi$, $-i\overline{\Psi}\Gamma_5\bsigma\Psi$, and $\overline{\Psi}\bsigma\Psi$, respectively, only the last two of which are invariant under proper orthochronous Lorentz transformations (recall that $\Gamma_5\equiv\Gamma_0\Gamma_1\Gamma_2\Gamma_3$ is a Lorentz pseudoscalar). Different choices of $4\times 4$ gamma matrices that preserve the equality between the first term in Eq.~(\ref{L2+1stripe}) and its four-component counterpart (i.e., the first term of Eq.~(\ref{L3+1square})) in the three-dimensional limit simply permute the assignments of $\Psi^\dag\gamma_0\tau_i\bsigma\Psi$ to those same three bilinears. 

A slightly more general argument of the above conclusion is as follows. There are three independent stripe AF ground states related by point-group rotations, described in the field theory by the condensation of three fermion bilinears. For such bilinears to open a gap in the Dirac fermion spectrum, they must be mass terms, i.e., they must be given by linearly independent matrices which anticommute with the massless Dirac Hamiltonian. In 3+1 dimensions, there are only two such Lorentz-invariant mass terms: the normal mass $\overline{\Psi}\Psi$ and the axial mass $i\overline{\Psi}\Gamma_5\Psi$. (Since AF order breaks spin rotation symmetry, the appropriate mass terms are actually $\overline{\Psi}\bsigma\Psi$ and $i\overline{\Psi}\Gamma_5\bsigma\Psi$, but spin is an internal degree of freedom which does not play a role in this argument.) Therefore, one of the three mass terms describing stripe AF order must necessarily violate Lorentz invariance.

We finally turn to the $\b{q}=0$ noncollinear state with vector spin chirality order (i.e., Fig.~7 in Ref.~\cite{hermele2008}), which corresponds to the condensation of a unique spin-triplet bilinear $\Psi^\dag\mu_3\bsigma\Psi$ even under time reversal~\cite{hermele2008}. The LGW theory for the transition from the ASL to this state was given in Ref.~\cite{lu2017},
\begin{align}\label{LGWnoncollinear}
\mathcal{L}_{2+1}=\sum_{\alpha,\sigma}\overline{\psi}_{\alpha\sigma}\gamma_\mu D_\mu\psi_{\alpha\sigma}+\frac{1}{4}F_{\mu\nu}^2+g\boldsymbol{\phi}\cdot\Psi^\dag\gamma_0\bsigma\Psi+\mathcal{L}_{\boldsymbol{\phi}},
\end{align}
and involves a single $O(3)$ vector $\boldsymbol{\phi}$. In four-component notation, the bilinear appearing in the Yukawa coupling is proportional to $i\overline{\Psi}\Gamma_3\Gamma_5\bsigma\Psi$ for standard choices of $4\times 4$ gamma matrices that preserve the form of the pure ASL Lagrangian; in such cases the theory (\ref{LGWnoncollinear}) thus breaks Lorentz invariance when lifted to four dimensions.

To demonstrate generally that the lift from $d=2+1$ to $d=3+1$ for the $\b{q}=0$ noncollinear order requires breaking Lorentz invariance  is more subtle, as there is a unique ground state (modulo global $SU(2)$ spin rotations), which implies only one mass term is required. In general, one must consider the fate of the theory's symmetries upon dimensional continuation. In $d=2+1$, the ASL has an emergent $SU(4)\supset SU(2)_\text{nodal}\times SU(2)_\text{spin}$ ``flavor'' symmetry which rotates the four two-component Dirac spinors $(\psi_{R\uparrow},\psi_{R\downarrow},\psi_{L\uparrow},\psi_{L\downarrow})$ into one another. (Here we denote the nodal indices by $R,L$ instead of $+,-$ used previously, for reasons that will become clear shortly.) The spin subgroup $SU(2)_\text{spin}$ acts only on the spin indices $\uparrow,\downarrow$ while the nodal subgroup $SU(2)_\text{nodal}$ acts only on the nodal indices $R,L$. In QED$_4$, the two-component Dirac fermions become two-component Weyl fermions: $\psi_{R\sigma}$ forms an $SU(2)$ doublet of two-component right-handed Weyl fermions, while $\psi_{L\sigma}$ forms an $SU(2)$ doublet of left-handed Weyl fermions. (The two Weyl components must have opposite chirality, otherwise the $d=3+1$ theory would have a gauge anomaly, which would make it inconsistent.) Among other symmetries, massless QED$_4$ has a $U(1)_{R-L}$ chiral symmetry, under which the Weyl spinors transform as $\psi_{R\sigma}\rightarrow e^{i\alpha}\psi_{R\sigma}$ and $\psi_{L\sigma}\rightarrow e^{-i\alpha}\psi_{L\sigma}$. It is well known that this chiral symmetry protects the masslessness of the four-component Dirac fermion $\Psi_\sigma=(\psi_{R\sigma},\psi_{L\sigma})$. (In a condensed matter setting, this is simply the topological protection of Weyl semimetals.) Conversely, mass terms---such as the only two Lorentz-invariant mass terms, $\overline{\Psi}\Psi$ and $i\overline{\Psi}\Gamma_5\Psi$---couple the right-handed and left-handed Weyl components and break this chiral symmetry. Now, the $U(1)_{R-L}$ transformation can be written as
\begin{align}
\left(\begin{array}{c}
\psi_{R\sigma} \\
\psi_{L\sigma}
\end{array}\right)
\rightarrow\left(\begin{array}{cc}
e^{i\alpha} & 0 \\
0 & e^{-i\alpha}
\end{array}\right)
\left(\begin{array}{c}
\psi_{R\sigma} \\
\psi_{L\sigma}
\end{array}\right),
\end{align}
which is an $SU(2)_\text{nodal}$ rotation in the (2+1)-dimensional interpretation. However, the bilinear $\Psi^\dag\mu_3\bsigma\Psi$ corresponding to the $\b{q}=0$ noncollinear order is proportional to the identity in nodal space (i.e., it does not contain nodal Pauli matrices $\tau_i$); by contrast with N\'eel order on the square lattice, it produces the same mass (amplitude and sign) at both Dirac nodes. In other words, this bilinear preserves the $SU(2)_\text{nodal}\supset U(1)_{R-L}$ symmetry, and thus cannot correspond to a Lorentz-invariant mass term in 3+1 dimensions.\\

\section{$\epsilon$-expansion}
\label{sec:epsilon}

We have shown above that the LGW theories for transitions from the ASL to either the stripe AF or the $\b{q}=0$ noncollinear state on the kagom\'e lattice break Lorentz invariance when lifted from three to four spacetime dimensions. This creates difficulties when applying the standard $\epsilon$ expansion as a variety of Lorentz-breaking terms not present in the original Lagrangian will be generated under renormalization (for a recent discussion of similar issues in the pure GNY model, see Ref.~\cite{wamer2018}). By contrast, the LGW theory (\ref{L2+1square}) for the N\'eel-ASL transition on the square lattice can be extended to a four-dimensional Lorentz-invariant theory, Eq.~(\ref{L3+1square}). This allows for an RG analysis in the usual $\epsilon$ expansion, to which we now turn.

\subsection{Renormalization group}
\label{sec:RG}

The starting point is the Lagrangian of the four-dimensional chiral Heisenberg QED-GNY model,
\begin{align}\label{O3QEDGNY}
\mathcal{L}&=\sum_{i=1}^{N_f}\overline{\Psi}_i\slashed{D}\Psi_i+\frac{1}{4}F_{\mu\nu}^2+\frac{1}{2\xi}(\partial_\mu A_\mu)^2\nn\\
&\phantom{=}+\frac{1}{2}(\partial_\mu\boldsymbol{\phi})^2+\frac{1}{2}m^2\bphi^2+\lambda^2(\bphi^2)^2+g\bphi\cdot\sum_{i=1}^{N_f}\overline{\Psi}_i\frac{\bsigma}{2}\Psi_i,
\end{align}
where $\slashed{D}\equiv\Gamma_\mu(\partial_\mu-ieA_\mu)$ is the gauge-covariant derivative and $\Gamma_\mu$ are $4\times 4$  Euclidean gamma matrices representing the $SO(4)$ Clifford algebra $\{\Gamma_\mu,\Gamma_\nu\}=2\delta_{\mu\nu}$. We denote by $N_f$ the number of flavors of $SU(2)$ doublets of four-component Dirac fermions, i.e.,
\begin{align}
\Psi_i\equiv\left(\begin{array}{c}
\Psi_{i\uparrow} \\
\Psi_{i\downarrow}
\end{array}\right),
\end{align}
where $\Psi_{i\sigma}$, $\sigma=\uparrow,\downarrow$ are four-component Dirac fermions. With this convention the Lagrangian (\ref{L3+1square}) for the N\'eel-ASL transition on the square lattice corresponds to $N_f=1$. The triplet of real fields $\bphi=(\phi_1,\phi_2,\phi_3)$ transforms in the vector representation of $SU(2)$, and $\xi$ is a gauge-fixing parameter ($\xi=0$ is the Landau gauge). By contrast, the chiral Ising QED-GNY model studied in Ref.~\cite{janssen2017,ihrig2018,zerf2018} possesses a single real scalar field coupled to a fermion mass term, instead of a spin bilinear, and the chiral XY QED-GNY model studied in Ref.~\cite{boyack2018} possesses a complex scalar field coupled to a superconducting mass term.

We study the critical (i.e., at $m^2=0$) properties of the model (\ref{O3QEDGNY}) in $d=4-\epsilon$ spacetime dimensions with field-theoretic RG using the modified minimal subtraction ($\overline{\text{MS}}$) prescription. We consider a bare Lagrangian,
\begin{align}
\mathcal{L}_0&=\sum_{i=1}^{N_f}\overline{\Psi}_i^0\slashed{D}^0\Psi_i^0+\frac{1}{4}\left(F_{\mu\nu}^0\right)^2+\frac{1}{2\xi_0}\left(\partial_\mu A_\mu^0\right)^2\nn\\
&\phantom{=}+\frac{1}{2}\left(\partial_\mu\boldsymbol{\phi}_0\right)^2+\frac{1}{2}m^2_0\bphi^2_0+\lambda^2_0\left(\bphi^2_0\right)^2\nn\\
&\phantom{=}+g_0\bphi_0\cdot\sum_{i=1}^{N_f}\overline{\Psi}_i^0\frac{\bsigma}{2}\Psi_i^0,
\end{align}
in terms of bare fields $\Psi_i^0$, $\bphi_0$, $A_\mu^0$ and coupling constants $e_0$, $\xi_0$, $m_0$, $\lambda_0$, $g_0$, with $\slashed{D}^0=\Gamma_\mu(\partial_\mu-ie_0A_\mu^0)$, and a renormalized Lagrangian
\begin{align}
\mathcal{L}_R&=\sum_{i=1}^{N_f}Z_\Psi\overline{\Psi}_i\slashed{D}\Psi_i+\frac{1}{4}Z_AF_{\mu\nu}^2+\frac{1}{2\xi}(\partial_\mu A_\mu)^2\nn\\
&\phantom{=}+\frac{1}{2}Z_\phi(\partial_\mu\boldsymbol{\phi})^2+\frac{1}{2}Z_{\phi^2}m^2\mu^2\bphi^2+Z_{\lambda^2}\lambda^2\mu^\epsilon(\bphi^2)^2\nn\\
&\phantom{=}+Z_gg\mu^{\epsilon/2}\bphi\cdot\sum_{i=1}^{N_f}\overline{\Psi}_i\frac{\bsigma}{2}\Psi_i,
\end{align}
where $\slashed{D}=\Gamma_\mu(\partial_\mu-ie\mu^{\epsilon/2}A_\mu)$. We define renormalized fields $\Psi_i=Z_\Psi^{-1/2}\Psi_i^0$, $A_\mu=Z_A^{-1/2}A_\mu^0$, $\bphi=Z_\phi^{-1/2}\bphi_0$ and dimensionless renormalized coupling constants,
\begin{align}
e^2&=e_0^2\mu^{-\epsilon}Z_A,\\
g^2&=g_0^2\mu^{-\epsilon}Z_\Psi^2 Z_\phi Z_g^{-2},\\
\lambda^2&=\lambda_0^2\mu^{-\epsilon}Z_\phi^2 Z_{\lambda^2}^{-1},\\
m^2&=m_0^2\mu^{-2}Z_\phi Z_{\phi^2}^{-1},\\
\xi&=\xi_0 Z_A^{-1},
\end{align}
where $\mu$ is an arbitrary renormalization scale. The renormalization constants $Z_X$, $X=\Psi,A,\phi,\phi^2,\lambda^2,g$ are calculated up to four-loop order using an automated setup, technical details of which can be found in previous publications~\cite{zerf2016,mihaila2017,zerf2017,zerf2018}. The number of diagrams that arise during the perturbative calculation of the renormalization constants
is exactly the same as in the chiral-Ising GNY-QED$_{3}$ case~\cite{zerf2018}, because one needs only to ``dress" the chiral-Ising Lorentz amplitudes with $SU(2)$ weight-factors.

\subsection{Beta functions}

The flow of the running couplings $e^2$, $g^2$, $\lambda^2$ is governed by the following RG equations,
\begin{align}
\beta_{e^2}&=(-\epsilon+\gamma_A)e^2,\label{be2}\\
\beta_{g^2}&=(-\epsilon+2\gamma_\Psi+\gamma_\phi-2\gamma_g)g^2,\\
\beta_{\lambda^2}&=(-\epsilon+2\gamma_\phi-\gamma_{\lambda^2})\lambda^2,\label{bl2}
\end{align}
where the beta functions are defined as
\begin{align}
\beta_{\alpha^2}=\mu\frac{d\alpha^2}{d\mu},
\end{align}
for $\alpha=e,g,\lambda$, and we absorb a factor of $1/(4\pi)^2$ in the definition of the couplings, $\alpha^2/(4\pi)^2\rightarrow\alpha^2$. Likewise, the anomalous dimensions $\gamma_X$ are defined as
\begin{align}\label{gammaX}
\gamma_X=\mu\frac{d\ln Z_X}{d\mu}.
\end{align}
Note that Eq.~(\ref{be2})-(\ref{bl2}) have the same form as for the chiral Ising QED-GNY model~\cite{zerf2018}, but the explicit form of the renormalization constants, and thus the anomalous dimensions and beta functions, are different for the two theories. As in Ref.~\cite{zerf2018} we express the beta functions as a sum over contributions at fixed loop order,
\begin{align}\label{betaalpha2}
\beta_{\alpha^2}=-\epsilon\alpha^2+\beta_{\alpha^2}^{\text{(1L)}}+\beta_{\alpha^2}^{\text{(2L)}}
+\beta_{\alpha^2}^{\text{(3L)}}+\beta_{\alpha^2}^{\text{(4L)}},
\end{align}
for $\alpha=e,g,\lambda$. We give here the beta functions up to and including three-loop order; contributions at four-loop order are lengthy and made available in electronic format~\cite{SuppMat}. The beta function for the gauge coupling $e^2$ is given by
\begin{align}
\beta_{e^2}^\text{(1L)}&=\frac{16N_f}{3}e^4,\label{be21L}\\
\beta_{e^2}^\text{(2L)}&=16N_fe^6-6N_fe^4g^2,\\
\beta_{e^2}^\text{(3L)}&=-9N_fe^6g^2+\frac{3N_f}{2}(7N_f+3)e^4g^4\nn\\
&\phantom{=}-\frac{8N_f}{9}(44N_f+9)e^8.\label{be23L}
\end{align}
The beta function for the Yukawa coupling $g^2$ is given by
\begin{align}
\beta_{g^2}^\text{(1L)}&=-12e^2g^2+\frac{1}{2}(4N_f+1)g^4,\\
\beta_{g^2}^\text{(2L)}&=-\frac{1}{32}(96N_f-47)g^6+2(5N_f+2)e^2g^4\nn\\
&\phantom{=}+\frac{2}{3}(40N_f-9)e^4g^2-40g^4\lambda^2+160g^2\lambda^4,\\
\beta_{g^2}^\text{(3L)}&=\frac{2}{27}[8N_f(140N_f+621-648\zeta_3)-3483]e^6g^2\nn\\
&\phantom{=}+\frac{1}{512}[4N_f(608N_f+139-144\zeta_3)-2731\nn\\
&\phantom{=}+1008\zeta_3]g^8-10(60N_f-89)g^4\lambda^4\nn\\
&\phantom{=}-\frac{1}{4}[N_f(29+48\zeta_3)+61-42\zeta_3]e^2g^6+40e^2g^4\lambda^2\nn\\
&\phantom{=}-\frac{1}{8}[4N_f(64N_f-17+432\zeta_3)-301-432\zeta_3]e^4g^4\nn\\
&\phantom{=}+75(2N_f+1)g^6\lambda^2-3520g^2\lambda^6.
\end{align}
Finally, the beta function for the four-scalar coupling $\lambda^2$ is given by
\begin{align}
\beta_{\lambda^2}^\text{(1L)}&=88\lambda^4+4N_fg^2\lambda^2-\frac{N_f}{4}g^4,\label{bl21L}\\
\beta_{\lambda^2}^\text{(2L)}&=-4416\lambda^6-176N_fg^2\lambda^4-\frac{3N_f}{2}g^4\lambda^2-N_fe^2g^4\nn\\
&\phantom{=}+20N_fe^2g^2\lambda^2+\frac{N_f}{2}g^6,\\
\beta_{\lambda^2}^\text{(3L)}&=-\frac{N_f}{512}(1208N_f-365+384\zeta_3)g^8+8272N_fg^2\lambda^6\nn\\
&\phantom{=}+\frac{N_f}{64}(4720N_f-6339-912\zeta_3)g^6\lambda^2\nn\\
&\phantom{=}-\frac{N_f}{2}(528N_f-3067-1272\zeta_3)g^4\lambda^4\nn\\
&\phantom{=}+\frac{N_f}{8}(232N_f+131-96\zeta_3)e^4g^4\nn\\
&\phantom{=}-N_f(64N_f+119-144\zeta_3)e^4g^2\lambda^2\nn\\
&\phantom{=}-\frac{N_f}{16}(17-96\zeta_3)e^2g^6+\frac{N_f}{4}(841-1200\zeta_3)e^2g^4\lambda^2\nn\\
&\phantom{=}-132N_f(17-16\zeta_3)e^2g^2\lambda^4+64(6023+3552\zeta_3)\lambda^8.\label{bl23L}
\end{align}
In those expressions $\zeta_z$ denotes the Riemann zeta function.

The beta functions above can be checked against previous results in various limits. In the limit $g^2=\lambda^2=0$ the model reduces to pure QED with $2N_f$ flavors of four-component Dirac fermions, and Eq.~(\ref{be21L})-(\ref{be23L}) together with the four-loop contribution~\cite{SuppMat} reproduce the four-loop QED beta function~\cite{gorishny1991}. For $e^2=g^2=0$ the model reduces to the bosonic $O(3)$ vector model; Eq.~(\ref{bl21L})-(\ref{bl23L}) with the four-loop contribution~\cite{SuppMat} agree with the known four-loop result~\cite{vladimirov1979}. Setting $e^2=0$ only, one obtains the ungauged chiral Heisenberg GNY model; our result for $\beta_{g^2}$ and $\beta_{\lambda^2}$ agrees in that limit with the four-loop beta functions recently computed for this model~\cite{zerf2017}.

One-loop beta functions for the specific case of $N_f=1$, with all couplings $e^2$, $g^2$, $\lambda^2$ nonzero, were obtained in Ref.~\cite{ghaemi2006}. Modulo trivial rescalings of the coupling constants, our result for $\beta_{e^2}$ agrees with theirs, but $\beta_{g^2}$ and $\beta_{\lambda^2}$ disagree. In Sec.~\ref{sec:LargeN}, we show that the critical exponents obtained from our beta functions agree order by order up to $\mathcal{O}(\epsilon^4,1/N_f^2)$ with continuous-$d$ results obtained in the large-$N_f$ expansion. We are thus confident that our beta functions, and the critical exponents we derive from them in the following sections, are correct.

\subsection{Quantum critical point and stability exponent}
\label{sec:QCP}

In this section we explore the N\'eel-ASL quantum critical point (QCP), and we will find it has different critical exponents from the ASL phase, which is described by conformal QED$_{3}$. The ASL phase and the N\'eel-ASL QCP are described by different conformal field theories; these correspond to different RG fixed points, characterized by a different type of stability and different critical exponents. The ASL phase~\cite{hermele2005} is an example of a critical phase, that is, a gapless phase of matter with power-law correlations and no relevant symmetry-preserving perturbations (i.e., it is a stable phase of matter). A QCP is also characterized by power-law correlations, but is not a stable phase, i.e., it has a relevant symmetry-preserving perturbation corresponding to the tuning parameter for the transition. To summarize, the fact that the exponents are different in the ASL phase and at the QCP is a generic feature of continuous transitions out of critical phases. (As another example, in the context of the AF quantum phase transition for repulsively interacting fermions on the honeycomb lattice, critical exponents go from the free-Dirac universality class in the semimetallic phase to the chiral Heisenberg GNY universality class at the QCP~\cite{herbut2006}.) This can be understood physically from the fact that additional soft degrees of freedom---order parameter fluctuations---appear at the QCP, and change the universality class from that of the critical phase itself.

To identify a possible QCP we look for stable fixed points of the RG flow on the critical ($m^2=0$) hypersurface, i.e., common zeros of the set of beta functions (\ref{be2})-(\ref{bl2}). We denote fixed-point couplings by $(e^{2}_{*},g^{2}_{*},\lambda^{2}_{*})$. At one-loop order we find eight fixed points: the Gaussian fixed point $(0,0,0)$, the conformal QED fixed point $(\frac{3\epsilon}{16N_f},0,0)$,
the $O(3)$ Wilson-Fisher fixed point $(0,0,\frac{\epsilon}{88})$, a conformal QED $\times$ $O(3)$ Wilson-Fisher fixed point $(\frac{3\epsilon}{16N_f},0,\frac{\epsilon}{88})$, 
two $O(3)$ GNY fixed points with $e^{2}_{*}=0$ and $g^{2}_{*}\neq0,\lambda^{2}_{*}\neq0$, and two $O(3)$ QED-GNY fixed points with nonzero values for all couplings. The only stable fixed point among all eight is one of the latter two, with fixed-point couplings given by
\begin{align}
e^{2}_{*}&=\frac{3}{16N_{f}}\epsilon+\mathcal{O}(\epsilon^2),\label{e2star}\\
g^{2}_{*}&=\frac{4N_{f}+9}{2N_{f}(4N_{f}+1)}\epsilon+\mathcal{O}(\epsilon^2),\label{g2star}\\
\lambda^{2}_{*}&=\frac{Y-4N_{f}-17}{176(4N_{f}+1)}\epsilon+\mathcal{O}(\epsilon^2),\label{l2star}
\end{align}
where we define
\begin{align}
Y\equiv\sqrt{16 N_{f}^2 + 488N_{f} + 1873 + \frac{1782}{N_{f}}}.
\end{align}
These three coupling constants are positive for all values of $N_{f}$---including the case $N_f=1$, appropriate for spin-1/2 microscopic degrees of freedom---as in the chiral Ising QED$_3$-GNY model~\cite{janssen2017,ihrig2018,zerf2018}; thus the fixed point is physical and describes a valid QCP. This establishes that at one-loop order, the N\'eel-ASL transition is continuous and its critical properties can be systematically calculated in the $\epsilon$ expansion. From the four-loop order results in Eq.~(\ref{betaalpha2}), a power series expansion of the critical exponents up to fourth order in $\epsilon$ can be obtained. The first exponent one can calculate is the stability exponent $\omega$, which controls leading corrections to scaling and would appear, for instance, in the subleading dependence of the AF susceptibility on the critical tuning parameter near the QCP. Mathematically, it corresponds to the smallest eigenvalue of the stability matrix, i.e., the Jacobian matrix $\mathcal{J}_{\alpha\alpha'}=\partial\beta_{\alpha^2}/\partial\alpha'^2$, $\alpha,\alpha'=e,g,\lambda$ of derivatives of the beta functions evaluated at the fixed point (\ref{e2star})-(\ref{l2star}). We find $\omega=\epsilon+\mathcal{O}(\epsilon^2)$ at one-loop order. The full four-loop result for general $N_f$ is given in Ref.~\cite{SuppMat}; here we present it only for $N_f=1$, appropriate to the N\'eel-ASL transition:
\begin{align}\label{omegaepsilon}
\omega\approx\epsilon-0.9\epsilon^2+6.742\epsilon^3-32.22\epsilon^4+\mathcal{O}(\epsilon^5).
\end{align}
In this and later $\mathcal{O}(\epsilon^4)$ expressions for the $N_f=1$ critical exponents, coefficients are given to four significant digits.

\subsection{Order parameter anomalous dimension}
\label{sec:OPanomalous}

Microscopically, the order parameter anomalous dimension $\eta_\phi$ controls the long-distance behavior of AF correlations at the QCP,
\begin{align}
\langle\boldsymbol{N}(r)\cdot\boldsymbol{N}(r')\rangle\sim\frac{1}{|r-r'|^{d-2+\eta_\phi}},
\end{align}
where $\boldsymbol{N}(r)\sim(-1)^{x+y}\b{S}_r$ and $|r-r'|$ is much greater than the lattice constant (but smaller than the system size, for a finite-size system approaching the thermodynamic limit). In a physical spin system this exponent would be extracted from an analysis of the width of the $(\pi,\pi)$ spin structure factor, measured in neutron scattering. In the field theory it is computed by evaluating $\gamma_\phi$ in Eq.~(\ref{gammaX}) at the $O(3)$ QED-GNY fixed point (\ref{e2star})-(\ref{l2star}),
\begin{align}\label{etaphiRG}
\eta_\phi\equiv\gamma_\phi(e_*^2,g_*^2,\lambda_*^2).
\end{align}
As for the beta functions, we express the anomalous dimensions as sums of contributions at fixed loop order,
\begin{align}
\gamma_X=\gamma_X^\text{(1L)}+\gamma_X^\text{(2L)}+\gamma_X^\text{(3L)}+\gamma_X^\text{(4L)}.
\end{align}
We find
\begin{align}
\gamma_\phi^\text{(1L)}&=2N_fg^2,\\
\gamma_\phi^\text{(2L)}&=160\lambda^4-\frac{7N_f}{4}g^4+10N_fe^2g^2,\\
\gamma_\phi^\text{(3L)}&=-3520\lambda^6-600N_fg^2\lambda^4+\frac{N_f}{8}(41-144\zeta_3)e^2g^4\nn\\
&\phantom{=}+\frac{N_f}{128}(624N_f-131+240\zeta_3)g^6+50N_fg^4\lambda^2\nn\\
&\phantom{=}-\frac{N_f}{2}(119+64N_f-144\zeta_3)e^4g^2,
\end{align}
with the four-loop contribution $\gamma_\phi^\text{(4L)}$ given electronically~\cite{SuppMat}. At one-loop order, we find
\begin{align}
\eta_{\phi}&=\frac{4N_f+9}{4N_f+1}\epsilon+\mathcal{O}(\epsilon^2).
\end{align}
Evaluating the full four-loop result~\cite{SuppMat} for $N_f=1$, we obtain
\begin{align}
\eta_{\phi}\approx2.6\epsilon+1.993\epsilon^2+1.963\epsilon^3-4.169\epsilon^4+\mathcal{O}(\epsilon^5).
\end{align}

\subsection{Correlation length exponent}

In addition to the anomalous dimension $\gamma_\phi$ of the order parameter field $\boldsymbol{\phi}$, we also compute $\gamma_{m^2}\equiv\gamma_{\phi^2}-\gamma_\phi$ which appears in the beta function for the scalar mass squared,
\begin{align}
\mu\frac{dm^2}{d\mu}=-(2+\gamma_{m^2})m^2.
\end{align}
Evaluated at the QCP, the anomalous dimension $\gamma_{m^2}$ is related to the correlation length exponent $\nu$,
\begin{align}\label{OneOverNu}
1/\nu=2+\gamma_{m^2}(e_*^2,g_*^2,\lambda_*^2),
\end{align}
which controls the divergence of the zero-temperature correlation length upon approach to the QCP. The contributions to $\gamma_{m^2}$ up to three-loop order are given by
\begin{align}
\gamma_{m^2}^\text{(1L)}&=-2N_fg^2-40\lambda^2,\\
\gamma_{m^2}^\text{(2L)}&=800\lambda^4+\frac{11N_f}{4}g^4+80N_fg^2\lambda^2-10N_fe^2g^2,\\
\gamma_{m^2}^\text{(3L)}&=-99840\lambda^6+\frac{N_f}{128}(-2672N_f+1923-1392\zeta_3)g^6\nn\\
&\phantom{=}+\frac{5N_f}{2}(48N_f-109-72\zeta_3)g^4\lambda^2-1320N_fg^2\lambda^4\nn\\
&\phantom{=}+\frac{N_f}{2}(64N_f+119-144\zeta_3)e^4g^2\nn\\
&\phantom{=}+\frac{3N_f}{8}(-131+208\zeta_3)e^2g^4\nn\\
&\phantom{=}+\frac{5N_f}{2}(408-384\zeta_3)e^2g^2\lambda^2,
\end{align}
and the four-loop contribution $\gamma_{m^2}^\text{(4L)}$ is given in electronic format~\cite{SuppMat}. At one-loop order, we find
\begin{align}
1/\nu&=2 - \frac{68 N_f + 113 + 5Y}{22(4N_f+1)}\epsilon+\mathcal{O}(\epsilon^2),
\end{align}
while the full four-loop result~\cite{SuppMat} for $N_f=1$ is
\begin{align}
1/\nu\approx2-4.577\epsilon+5.089\epsilon^2-35.81\epsilon^3+298.7\epsilon^4+\mathcal{O}(\epsilon^5).
\end{align}

\subsection{CDW and VBS susceptibility exponents}
\label{sec:FermionBilinears}

Besides the N\'eel order parameter field $\boldsymbol{\phi}$, which can be identified with the bilinear $\overline{\Psi}\boldsymbol{\sigma}\Psi$, other fermion bilinears will develop universal power-law correlations at the QCP. In the original $d=3$ theory with two-component spinors $\psi_{\alpha\sigma}$, these are Lorentz-invariant bilinears of the form $\overline{\psi}\tau_i\sigma_j\psi$ where the $\tau_i$ are Pauli matrices acting on the nodal/valley index $\alpha=\pm$. Besides the spin-triplet N\'eel bilinear $\overline{\Psi}\bsigma\Psi=\overline{\psi}\tau_3\bsigma\psi$ already considered, in this section we consider the three spin-singlet bilinears of the form $\overline{\psi}\tau_i\psi$, $i=1,2,3$. The $\overline{\psi}\psi$ spin-singlet bilinear will be discussed in Sec.~\ref{sec:QAH}.

The physical meaning of these bilinears can be ascertained from an analysis of their transformation properties under space-group and time-reversal symmetries via the projective symmetry group (PSG)~\cite{wen2002} (see Appendix~\ref{app:PSG}). First, the transformation properties of the spin-triplet bilinear $\overline{\psi}\tau_3\bsigma\psi$ agree with those of the microscopic N\'eel order parameter and confirm the identification obtained in the naive continuum limit (Sec.~\ref{sec:square}). The spin-singlet bilinear $\overline{\psi}\tau_3\psi=\overline{\Psi}\Psi$ transforms in the same way as the N\'eel bilinear under space-group symmetries but oppositely under time reversal; it thus corresponds to a staggered density or CDW order parameter for the lattice $c$ fermions,
\begin{align}\label{CDWOP}
\mathcal{O}_\text{CDW}(r)=
(-1)^{x+y}\sum_{\sigma}c_{r\sigma}^\dag c_{r\sigma}^{\phantom{\dagger}}.
\end{align}
As discussed in Ref.~\cite{hermele2005}, in the language of an underlying spin model this operator is symmetry equivalent to the staggered component of the density of skyrmions in the N\'eel field. The spin-singlet bilinears $\overline{\psi}\tau_1\psi$ and $\overline{\psi}\tau_2\psi$ transform into each other under $C_4$ rotations, are both even under time reversal, and transform oppositely under reflections and unit translations. In four-component notation with the choice of $4\times 4$ gamma matrices introduced in Sec.~\ref{sec:square}, these can be written as $\overline{\psi}\tau_1\psi=i\overline{\Psi}\Gamma_3\Psi$ and $\overline{\psi}\tau_2\psi=-i\overline{\Psi}\Gamma_5\Psi$. These two bilinears can thus be identified with the two components of the VBS order parameter $\boldsymbol{V}\equiv(V_x,V_y)$, where $V_x\sim i\overline{\Psi}\Gamma_5\Psi$ and $V_y\sim i\overline{\Psi}\Gamma_3\Psi$ correspond to columnar VBS order in the $x$ and $y$ directions, respectively. In the microscopic theory (\ref{Hlattice}), these can be defined as~\cite{Meng2019}
\begin{align}
V_x(r)=(-1)^x\b{S}_r\cdot\b{S}_{r+\hat{x}},\hspace{5mm}
V_y(r)=(-1)^y\b{S}_r\cdot\b{S}_{r+\hat{y}}.
\end{align}
The microscopic CDW and VBS operators are functions of on-site bilinears of the form $c_{r\sigma}^\dag c_{r\sigma'}^{\phantom{\dagger}}$ and are thus manifestly gauge invariant.

From the operator identifications above we therefore expect that at the QCP, the CDW and VBS order parameters should develop universal power-law correlations at long distances $|r-r'|\gg 1$ (measured in units of the lattice constant),
\begin{align}
\langle\mathcal{O}_\text{CDW}(r)\mathcal{O}_\text{CDW}(r')\rangle&\sim\frac{1}{|r-r'|^{2\Delta_\text{CDW}}},\label{CDWcorrelations}\\
\langle\boldsymbol{V}(r)\cdot\boldsymbol{V}(r')\rangle
&\sim\frac{1}{|r-r'|^{2\Delta_\text{VBS}}},\label{VBScorrelations}
\end{align}
where
\begin{align}
\Delta_\text{CDW}&=\Delta_{\overline{\Psi}\Psi},\label{DeltaCDW}\\
\Delta_\text{VBS}&=\Delta_{i\overline{\Psi}\Gamma_5\Psi}=\Delta_{i\overline{\Psi}\Gamma_3\Psi}.\label{DeltaVBS}
\end{align}
In the language of an underlying spin model, the VBS and CDW correlation functions would correspond to four-spin and six-spin correlation functions, respectively, which can in principle be measured by two-magnon and three-magnon inelastic light scattering, respectively~\cite{fleury1969}. Note that these correlations are already power-law in the ASL phase, as observed numerically for the N\'eel and VBS order parameters in Ref.~\cite{Meng2019}. However, the exponents in Eq.~(\ref{DeltaCDW})-(\ref{DeltaVBS}), as well as the exponent $\Delta_\text{AF}=\Delta_\phi$ for AF spin-spin correlations (i.e., the scaling dimension of the order parameter field $\boldsymbol{\phi}$), reflect the particular conformal field theory associated with the N\'eel-ASL QCP and are thus predicted to be \emph{different} than those in the ASL phase, which corresponds to the pure conformal QED$_3$ fixed point.

We further observe that while $i\overline{\Psi}\Gamma_5\Psi$ is a Lorentz (pseudo)scalar in $d=4$ dimensions, $i\overline{\Psi}\Gamma_3\Psi$ is not, but rather corresponds to one component of a Lorentz vector $i\overline{\Psi}\Gamma_\mu\Psi$. However, both are Lorentz scalars in $d=3$ dimensions. Therefore, the $\epsilon$ expansion will predict different scaling dimensions for $i\overline{\Psi}\Gamma_5\Psi$ and $i\overline{\Psi}\Gamma_3\Psi$, although they are expected to have the same scaling dimension in the physical $d=3$ theory since they transform into each other under the action of microscopic symmetries. For the purposes of determining $\Delta_\text{VBS}$, in the $\epsilon$ expansion we will therefore only calculate the dimension of the Lorentz pseudoscalar $i\overline{\Psi}\Gamma_5\Psi$, which does not have the unphysical property of transforming as a Lorentz vector in $d=4$ dimensions. We remind the reader that, as explained in the first paragraph of Sec.~\ref{sec:epsilon}, the entirety of Sec.~\ref{sec:epsilon} (and in fact the entire paper, except for Sec.~\ref{sec:kagome}) is concerned with the N\'eel-ASL transition on the square lattice, for which the $\epsilon$-expansion is a meaningful approach. All the exponents are well defined in physical $d = 2 + 1$ dimensions. We are adopting a conservative approach, where in the $\epsilon$-expansion analysis we only calculate the scaling dimension of quantities that remain Lorentz invariant when continuing from three to four spacetime dimensions. By contrast, in the $1/N_f$ expansion in $d=3$ the scaling dimensions of $i\overline{\Psi}\Gamma_5\Psi$ and $i\overline{\Psi}\Gamma_3\Psi$ should agree order by order in $1/N_f$. On the other hand, $\Delta_\text{CDW}$ can be meaningfully identified with $\Delta_{\overline{\Psi}\Psi}$ computed in both the $\epsilon$ and $1/N_f$ expansions with four-component spinors, since $\overline{\Psi}\Psi$ is a Lorentz scalar in both $d=4$ and $d=3$ dimensions.

The presence of $\Gamma_5$ can create issues in the $\epsilon$-expansion if one insists on preserving four-dimensional identities such as $\Gamma_5=\frac{1}{4!}\epsilon_{\mu\nu\lambda\sigma}\Gamma_\mu\Gamma_\nu\Gamma_\lambda\Gamma_\sigma$, which cannot be easily generalized to noninteger $d$~\cite{thooft1972}. In our case, we are interested in describing a quantum critical point in $d=3$, for which the intrinsically four-dimensional tensor $\epsilon_{\mu\nu\lambda\sigma}$ does not exist. It is therefore sufficient to use the naive prescription according to which $\{\Gamma_5,\Gamma_\mu\}=0$ in all $d$ dimensions~\cite{chanowitz1979}. In this prescription evanescent operators~\cite{dugan1991} are not generated when renormalizing the $i\overline{\Psi}\Gamma_5\Psi$ operator. Evanescent operators can be generated when renormalizing four-fermion operators~\cite{dipietro2018}, but we will not concern ourselves with the renormalization of such operators here.

We now turn to the computation of the scaling dimensions $\Delta_{\overline{\Psi}\mathcal{M}\Psi}$ of fermion bilinears $\overline{\Psi}\mathcal{M}\Psi$ (here $\mathcal{M}$ is the identity matrix or $i\Gamma_5$). As discussed above, such scaling dimensions describe the universal power-law decay in real space of various equal-time correlation functions (i.e., thermodynamic susceptibilities) at the quantum critical point. The way those scaling dimensions are calculated here is similar to how a linear-response susceptibility is usually defined/measured in statistical mechanics, i.e., one turns on a weak perturbation and computes the response to this perturbation, in the limit that the strength of the perturbation goes to zero. Here, to compute the fermion bilinear scaling dimensions, one must add a small, perturbative fermion mass term (really, turn on a weak staggered or VBS order), and compute the ratio of the induced order to the applied perturbation. (Since the added mass is infinitesimally weak, the fermions remain Dirac fermions, albeit slightly massive.) As usual with linear response, however, the ratio itself is a property of the system in the absence of the perturbation. Here, the susceptibility exponents are universal properties of the quantum critical point itself, where there are no mass terms and the fermions are gapless. To put these ideas into practice, we follow the method discussed in Ref.~\cite{zerf2018}, whereby the bilinear is added to the bare (renormalized) Lagrangian with a coefficient $M_0$ ($M$), the two coefficients being related by $M=M_0\mu^{-1}Z_M^{-1}Z_\Psi$.  The renormalized mass $M$ obeys the RG equation
\begin{align}
\mu\frac{dM}{d\mu}=(-1-\gamma_M+\gamma_\Psi)M,
\end{align}
where $\gamma_M=\mu(d\ln Z_M/d\mu)$. Since $M$ is the coefficient of a gauge-invariant operator, the combination $\gamma_M-\gamma_\Psi$ must be gauge invariant, even though $\gamma_M$ and $\gamma_\Psi$ individually are not. The scaling dimension of the bilinear is then given by
\begin{align}\label{DimBilinear}
\Delta_{\overline{\Psi}\mathcal{M}\Psi}=d-1-\gamma_{\overline{\Psi}\mathcal{M}\Psi}(e_*^2,g_*^2,\lambda_*^2),
\end{align}
where $\gamma_{\overline{\Psi}\mathcal{M}\Psi}\equiv\gamma_M-\gamma_\Psi$. By contrast with the chiral Ising QED-GNY model~\cite{zerf2018}, in which a $\overline{\Psi}\Psi$ insertion (i.e., adding this term to the Lagrangian) breaks a $\mathbb{Z}_2$ chiral symmetry and radiatively induces a relevant $\phi^3$ interaction in the renormalized Lagrangian, here the continuous $O(3)$ symmetry ensures that no additional operators which might mix with $\overline{\Psi}\mathcal{M}\Psi$ are radiatively induced by an insertion of this operator~\footnote{A radiatively induced interaction is one which is not present in the original Lagrangian but appears at higher orders in perturbation theory due to loop diagrams.}.

The CDW susceptibility exponent is given in terms of the normal mass operator anomalous dimension $\gamma_{\overline{\Psi}\Psi}$, given up to three-loop order by
\begin{align}
\gamma_{\overline{\Psi}\Psi}^\text{(1L)}&=6e^2-\frac{9}{4}g^2,\\
\gamma_{\overline{\Psi}\Psi}^\text{(2L)}&=-18e^2g^2-\frac{1}{3}(40N_f-9)e^4+\frac{3}{64}(56N_f+59)g^4,\\
\gamma_{\overline{\Psi}\Psi}^\text{(3L)}&=-\frac{1}{27}[8N_f(140N_f+621-648\zeta_3)-3483]e^6\nn\\
&\phantom{=}+\frac{3}{1024}[8N_f(88N_f-397)-847-2256\zeta_3]g^6\nn\\
&\phantom{=}-\frac{3}{16}[N_f(-137+144\zeta_3)-6(43+22\zeta_3)]e^2g^4\nn\\
&\phantom{=}+\frac{9}{16}(160N_f-109+336\zeta_3)e^4g^2-75g^4\lambda^2\nn\\
&\phantom{=}+435g^2\lambda^4.
\end{align}
The four-loop term $\gamma_{\overline{\Psi}\Psi}^\text{(4L)}$ is given electronically~\cite{SuppMat}. Contributions are found to be independent of the gauge-fixing parameter $\xi$ at each loop order, which constitutes a strong check on the calculation. Evaluating Eq.~(\ref{DimBilinear}) at the QCP (\ref{e2star})-(\ref{l2star}), we obtain at one-loop order
\begin{align}\label{DeltaPsiPsi}
\Delta_{\overline{\Psi}\Psi}=3+\frac{9-N_f(4N_f+1)}{N_f(4N_f+1)}\epsilon+\mathcal{O}(\epsilon^2),
\end{align}
with the full four-loop expression given in Ref.~\cite{SuppMat}. For $N_f=1$, we obtain
\begin{align}
\Delta_{\overline{\Psi}\Psi} \approx 3 + 0.8 \epsilon - 2.984 \epsilon^2 - 4.352 \epsilon^3 - 10.49 \epsilon^4
+\mathcal{O}(\epsilon^5).
\end{align}

Similarly, the VBS susceptibility exponent is obtained from the axial mass operator anomalous dimension $\gamma_{i\overline{\Psi}\Gamma_5\Psi}$. Up to three-loop order we obtain
\begin{align}
\gamma_{i\overline{\Psi}\Gamma_5\Psi}^\text{(1L)}&=6e^2+\frac{3}{4}g^2,\\
\gamma_{i\overline{\Psi}\Gamma_5\Psi}^\text{(2L)}&=-\left(\frac{40N_f}{3}-3\right)e^4+6e^2g^2
-\frac{3}{64}(8N_f-11)g^4,\\
\gamma_{i\overline{\Psi}\Gamma_5\Psi}^\text{(3L)}&=\left[-\frac{8}{27}N_f(140N_f+621-648\zeta_3)+129\right]e^6\nn\\
&\phantom{=}+\frac{3}{1024}[8N_f(-40N_f+67)+509-528\zeta_3]g^6\nn\\
&\phantom{=}+\frac{3}{16}[N_f(-19+48\zeta_3)+44-108\zeta_3]e^2g^4\nn\\
&\phantom{=}+\frac{3}{16}(64N_f+13-720\zeta_3)e^4g^2+15g^4\lambda^2\nn\\
&\phantom{=}-105g^2\lambda^4,
\end{align}
with the four-loop contribution $\gamma_{i\overline{\Psi}\Gamma_5\Psi}^\text{(4L)}$ given in electronic format~\cite{SuppMat}. At one-loop order for general $N_f$, we find
\begin{align}
\Delta_{i\overline{\Psi}\Gamma_5\Psi}=3-\frac{2N_f(4N_f+7)+9}{2N_f(4N_f+1)}\epsilon+\mathcal{O}(\epsilon^2),
\end{align}
while for $N_f=1$ at four-loop order, we obtain
\begin{align}
\Delta_{i\overline{\Psi}\Gamma_5\Psi}\approx 3-3.1\epsilon-3.145\epsilon^2+10.78\epsilon^3-85.32\epsilon^4
+\mathcal{O}(\epsilon^5).
\end{align}
The full four-loop expression for general $N_f$ is given in Ref.~\cite{SuppMat}.

Finally, for $N_f>1$, in addition to the flavor-singlet bilinears $\overline{\Psi}\Psi$ and $i\overline{\Psi}\Gamma_5\Psi$ one can define an $SU(N_f)$ flavor-adjoint bilinear,
\begin{align}\label{AdjBilinear}
\overline{\Psi}T_A\Psi=\sum_{i,j=1}^{N_f}\overline{\Psi}_iT_A^{ij}\Psi_j,\hspace{5mm}
A=1,\ldots,N_f^2-1,
\end{align}
where $T_A$ are the Hermitian generators of $SU(N_f)$. This operator does not exist for $N_f=1$ and thus has no physical meaning in the context of the N\'eel-ASL transition, but the computation of its scaling dimension for $N_f>1$ is an interesting problem in quantum field theory. The corresponding anomalous dimension $\gamma_{\overline{\Psi}T_A\Psi}$ is given up to four-loop order in Ref.~\cite{SuppMat}. At one-loop order, $\Delta_{\overline{\Psi}T_A\Psi}$ is equal to the flavor-singlet, normal mass bilinear dimension (\ref{DeltaPsiPsi}). A difference in scaling dimensions between singlet and adjoint bilinears begins appearing only at four-loop order,
\begin{align}\label{DiffSingAdj}
\Delta_{\overline{\Psi}\Psi}-\Delta_{\overline{\Psi}T_A\Psi}=
\frac{9(1-3\zeta_3)(4N_f+9)^3}{32N_f^3(4N_f+1)^3}\epsilon^4+\mathcal{O}(\epsilon^5).
\end{align}
As discussed in Ref.~\cite{zerf2018}, the difference between singlet and adjoint mass dimensions comes from bilinear insertions in closed fermion loops appearing in the loop expansion of the fermion two-point function; such closed loops vanish for adjoint insertions since the $SU(N_f)$ generators are traceless, but are generally nonzero for singlet insertions. However, here the tracelessness of the spin $SU(2)$ generators appearing at each Yukawa vertex implies that nonvanishing singlet insertions in closed fermion loops occur at higher loop order than in the chiral Ising QED-GNY model, since such closed loops must contain an even number of Yukawa vertices to not vanish in the chiral Heisenberg case.

\subsection{Fermion bilinear dimensions in the chiral Heisenberg GNY model}
\label{sec:O3GNY4Lbilinears}

Turning off the gauge coupling $e^2$ in the Lagrangian (\ref{O3QEDGNY}), we obtain the pure chiral Heisenberg GNY model, which for $N_f=1$ describes the universal critical properties of the semimetal-AF insulator quantum phase transition in the half-filled Hubbard model on the honeycomb and $\pi$-flux square lattices~\cite{sorella1992,herbut2006,honerkamp2008,herbut2009,sorella2012,ulybyshev2013,
assaad2013,janssen2014,parisen2015,otsuka2016,buividovich2018}. In this limit, the Dirac fermions can be interpreted as real electronic excitations in a semimetal, rather than fractionalized spinon excitations in a spin system.
Whereas the usual exponents ($\nu$, $\eta_\phi$, and the fermion anomalous dimension $\eta_\Psi$) for the chiral Heisenberg GNY model have been computed both in the $\epsilon$-~\cite{rosenstein1993,zerf2017} and large-$N_f$~\cite{gracey2018c} expansions, to our knowledge scaling dimensions of fermion bilinears have thus far not been computed. In particular, the $\overline{\Psi}\Psi$ bilinear corresponds in the long-wavelength limit to the CDW order parameter, i.e., Eq.~(\ref{CDWOP}) on the $\pi$-flux square lattice, and the Semenoff mass~\cite{semenoff1984} on the honeycomb lattice. Correlations of this bilinear can in principle be computed in sign-problem-free QMC simulations of the half-filled repulsive Hubbard model on the aforementioned lattices, and at the semimetal-AF QCP should decay at long distances according to Eq.~(\ref{CDWcorrelations}) with an exponent $\Delta_\text{CDW}=\Delta_{\overline{\Psi}\Psi}^\text{cH-GNY}$ given by
\begin{align}
\Delta_{\overline{\Psi}\Psi}^\text{cH-GNY} &\approx 3 - 0.1 \epsilon - 0.4278 \epsilon^2 + 
 0.2186 \epsilon^3 - 1.063 \epsilon^4 \nn\\
 &\phantom{\approx} + \mathcal{O}(\epsilon^5).
\end{align}
For the chiral Heisenberg GNY model with general $N_f$, we obtain at one-loop order
\begin{align}
\Delta_{\overline{\Psi}\Psi}^\text{cH-GNY} = 3 + \frac{7-8N_f}{2(4N_f+1)}\epsilon+\mathcal{O}(\epsilon^2).
\end{align}
A four-loop expression for general $N_f$ is given electronically in Ref.~\cite{SuppMat}.

Similarly, the Hermitian $i\overline{\Psi}\Gamma_3\Psi$ and $i\overline{\Psi}\Gamma_5\Psi$ bilinears can be viewed as the two components of the columnar VBS order parameter on the $\pi$-flux square lattice, or alternatively as the real and imaginary parts of a $\mathbb{Z}_3$-symmetric complex order parameter describing Kekul\'e VBS order on the honeycomb lattice~\cite{hou2007,ryu2009,roy2010}. At the semimetal-AF QCP of the half-filled repulsive Hubbard model on those lattices the correlation function of these operators should decay as in Eq.~(\ref{VBScorrelations}), with the exponent $\Delta_\text{VBS}=\Delta_{i\overline{\Psi}\Gamma_5\Psi}^\text{cH-GNY}$ given by
\begin{align}
\Delta_{i\overline{\Psi}\Gamma_5\Psi}^\text{cH-GNY} &\approx 3 - 1.3 \epsilon - 0.1674 \epsilon^2 + 
 0.01146 \epsilon^3 - 0.1148 \epsilon^4 \nn\\
 &\phantom{\approx} + \mathcal{O}(\epsilon^5).
\end{align}
For general $N_f$, the one-loop result is
\begin{align}
\Delta_{i\overline{\Psi}\Gamma_5\Psi}^\text{cH-GNY}=3-\frac{8N_f+5}{2(4N_f+1)}\epsilon+\mathcal{O}(\epsilon^2),
\end{align}
and the full four-loop result is given in electronic format~\cite{SuppMat}. By contrast with the corresponding exponents at the N\'eel-ASL QCP, which would describe multi-spin correlations in a microscopic spin system, at the semimetal-AF insulator QCP these exponents could be accessed by measuring ordinary two-point correlations of one-body operators for electrons, e.g., density-density and current-current correlations.

Turning finally to the $SU(N_f)$ flavor-adjoint bilinear (\ref{AdjBilinear}), the four-loop diagrams responsible for the difference (\ref{DimBilinear}) in scaling dimensions between the singlet and adjoint bilinears are $\mathcal{O}(e^2g^6)$ and vanish for the pure chiral Heisenberg GNY model with $e^2=0$. Thus for the latter model this difference in scaling dimensions can at most be $\mathcal{O}(\epsilon^5)$.

\section{Large-$N_f$ expansion}
\label{sec:LargeN}

To complement the $\epsilon$-expansion studies just discussed, we now turn to the large-$N_f$ expansion (for a recent review, see Ref.~\cite{GraceyReview2018}). In contrast with Sec.~\ref{sec:epsilon}, where $\epsilon=4-d$ is treated as a formal expansion parameter but $N_f$ is arbitrary, here we work in fixed $d$ dimensions but treat $1/N_f$ as a formal expansion parameter. That is, we consider a system with $N_{f}$ flavors of $SU(2)$ doublets of four-component Dirac fermions (i.e., a total of $2N_f$ four-component Dirac fermions), where we assume $N_{f}\gg1$, and perform perturbation theory in powers of $1/N_{f}$; at the end of the calculation we then consider the physically appropriate limit $N_{f}\rightarrow1$.  (Larger values of $N_f$ may be physically accessible in systems of large-spin ultracold fermionic atoms such as $^{87}$Sr~\cite{gorshkov2010,desalvo2010}.)
The large-$\Nf$ formalism we use has been detailed in Refs.~\cite{gracey1992,gracey1993c,*gracey2018b,gracey1993b,gracey2018} for the chiral Ising QED-GN(Y) model. 
We refer the reader to those articles for more details as well as to the original papers of Vasil'ev {\it et al.}~\cite{Vasilev1981,Vasilev1981b}, 
where the large-$\Nf$ critical point approach was developed for scalar theories with $N_f$ flavors in $d$ dimensions. 
In essence, the zeroth-order ($N_{f}=\infty$) approximation corresponds to calculating the gauge-field and scalar-field propagators in the random-phase approximation (RPA), and corrections at higher orders in $1/N_{f}$ come from diagrams containing self-energy and vertex corrections computed with the RPA-resummed scalar-field and gauge-field propagators. Thus, the diagrams are grouped according to their powers in $1/N_{f}$, irrespective of their loop order, which is the classification used for grouping diagrams in the $\epsilon$ expansion.
The elegance of the large-$N_{f}$ method is such that one carries out all computations directly at the $d$-dimensional fixed point where there is scaling behavior. 
The core Lagrangian for determining large-$\Nf$ critical exponents is that which governs the universality class of the $O(3)$ QED-GNY fixed point of the four-dimensional Lagrangian (\ref{O3QEDGNY}) used for the perturbative computations. 
The essence of that Lagrangian is that it involves the kinetic terms of the fields as well as the $3$-point interactions between the matter and force fields. 
By the latter we include not only the photon but also the scalar field $\phi^a$. For our case there are only two such interactions and the Euclidean Lagrangian of the universal theory used for the large-$\Nf$ analysis is
\begin{equation}
\mathcal{L} = \overline{\Psi}_i \partialslash \Psi_i +
\tilde{\phi}^a \overline{\Psi}_i \halfs \sigma^a \Psi_i  
-i\tilde{A}_\mu \overline{\Psi}_i \Gamma_\mu \Psi_i + \ldots,
\label{laguniv}
\end{equation}
where sums over repeated indices $i=1,\ldots,N_f$ and $a=1,2,3$ are understood.
We have not included any additional terms since there would be an infinite number of operators built from the three fields. 
Only a few of these operators would be relevant in a fixed dimension; in particular, the kinetic terms for the scalar and gauge fields in Eq.~(\ref{O3QEDGNY}) are irrelevant at criticality in the large-$N_f$ limit. Also the coupling constants have been rescaled out of the interactions to produce the new fields $\tilde{\phi}^a$ and $\tilde{A}_\mu$ since the formalism applies at criticality and there the 
couplings do not run~\cite{Vasilev1981,Vasilev1981b}. 
Excluding one or the other of the two  interactions in Eq.~(\ref{laguniv}) would lead to critical exponents of a different universality class. 
At criticality the canonical dimensions of the fields are fixed by the dimensionlessness of the action in $d$ dimensions and this leads 
to the asymptotic scaling forms of the propagators~\cite{gracey1992,gracey2018},
\begin{align}
\langle \Psi_i(x) \overline{\Psi}_j(y) \rangle & \sim 
\frac{(\xslash-\yslash) A \delta_{ij}}{((x-y)^2)^{\hat{\alpha}}} ,\nn\\
\langle \phi^a(x) \phi^b(y) \rangle & \sim 
\frac{B_\phi \delta^{ab}}{((x-y)^2)^{\hat{\beta}_\phi}} , \nonumber \\
\langle A_\mu(x) A_\nu(y) \rangle & \sim 
\frac{B_A}{((x-y)^2)^{\hat{\beta}_A}} \nn\\
&\hspace{-4mm} \times \left[ \delta_{\mu\nu} +
\frac{2\hat{\beta}_A}{(2\mu-2\hat{\beta}_A-1)}
\frac{(x-y)_\mu (x-y)_\nu}{(x-y)^2} \right], 
\label{propcoord}
\end{align}
in the approach to the fixed point in coordinate space. 
Here $A$, $B_A$ and  $B_\phi$ are coordinate-independent amplitudes and the full dimension of each  field is defined by
\begin{equation}
\hat{\alpha} = \mu + \halfs \eta,~~~
\hat{\beta}_A = 1 - \eta - \chi_A,~~~
\hat{\beta}_\phi = 1 - \eta - \chi_\phi,
\end{equation}
where $\chi_A$ and $\chi_\phi$ are the anomalous dimensions of the respective vertices of (\ref{laguniv}) and we retain the definition 
$d$~$=$~$2\mu$ of Refs.~\cite{gracey2018,Vasilev1981} for consistency with other large-$\Nf$ work using this  formalism. 
To avoid confusion with the notation for the $\beta$-functions we use a hat for the full dimension of the field exponents.
The vertex dimensions as well as the other exponents such as $\eta$ are functions of $d$ and $\Nf$ and each exponent can be expanded in a series of the form
\begin{align}\label{etaseries}
\eta = \sum_{n=1}^\infty \frac{\eta_n}{\Nf^n} .
\end{align} 
Moreover these critical exponents are related to the $\epsilon$-expansion of  their respective RG functions at the $O(3)$ QED-GNY fixed point found in Sec.~\ref{sec:QCP}. 
As such this provides a non-trivial check on the four-loop  renormalization. 
Since the large-$\Nf$ formalism of Refs.~\cite{Vasilev1981,Vasilev1981b} operates purely at the critical point all our $1/\Nf$ results will be exclusively in the 
Landau gauge. The gauge parameter, $\xi$, of our linear gauge fixing can be regarded as a second coupling constant and at criticality its associated 
RG function, which is in effect a $\beta$-function, defines a critical coupling. In this case it corresponds to the Landau gauge.  

\subsection{Critical exponents}

Using this formalism we have extended the $\mathcal{O}(1/\Nf^2)$ computations of Ref.~\cite{gracey2018} to the case where the Pauli matrices appear in the scalar-fermion vertex. 
Since Refs.~\cite{gracey1993b,gracey2018} already record the values for all the large-$\Nf$ master integrals contributing to the scaling forms of the various Green's 
functions needed to find the operator dimension at $\mathcal{O}(1/\Nf^2)$, we merely record the outcome of this straightforward exercise. 
The key building block is the fermion anomalous dimension, which at leading order is 
\begin{equation}\label{etaFermion1}
\eta_1 = - \frac{(4\mu^3-6\mu^2-3\mu+4)}{8\mu(\mu-1)} 
\frac{\Gamma(2\mu-1)}{\Gamma^3(\mu)\Gamma(1-\mu)},
\end{equation}
in Landau gauge, where $\Gamma(z)$ is the gamma function. Having established this result, the two leading-order vertex dimensions are required for the solution of the $\mathcal{O}(1/\Nf^2)$ Schwinger-Dyson equations. We find
\begin{equation}
\chi_{\phi,1} = - \frac{\mu(4\mu^2-2\mu-1)\eta_1}
{(4\mu^3-6\mu^2-3\mu+4)}  ,~~~
\chi_{A,1} = - \eta_1.
\end{equation}
The simplicity of the latter equation is a reflection that in the critical large-$\Nf$ setup the photon has a full dimension of unity which correctly shows the consistency of the Ward-Takahashi identity. In the RG approach this is equivalent to the all-orders statement that $\gamma_A=\epsilon$ at any fixed point with $e_*^2\neq 0$, which follows from Eq.~(\ref{be2}).
In addition we have computed $1/\nu$ to $\mathcal{O}(1/\Nf)$ and found
\begin{equation}\label{NuInverseLargeN}
\frac{1}{\nu} = 2 \mu - 2 - 
\frac{4 ( 3 \mu^2 - 4 \mu + 2 )( 2 \mu - 1 ) \eta_1 }
{(4 \mu^3-6 \mu^2-3 \mu+4)\Nf} + \mathcal{O} \left( 1/\Nf^2 \right).
\end{equation}

With the leading order exponents for $\eta$, $\chi_A$, and $\chi_\phi$ we have
evaluated several exponents to the next order finding 
\begin{widetext}
\begin{eqnarray}
\eta_2 &=& \biggl[ 3 \mu (2 \mu-1) (\mu-1) (4\mu^3-6\mu^2-3\mu+4)
\left[ \psi^\prime(\mu-1) - \psi^\prime(1) \right]  - 12 (\mu-1)^2 (2\mu^2+\mu-2)
\hat{\Psi}(\mu) \nonumber \\
&& \hspace{2mm}  + \frac{16 \mu^6 - 96 \mu^5 + 156 \mu^4 - 18 \mu^3 - 141 \mu^2 
+ 114 \mu - 28}{\mu-1} 
\biggr] \frac{\eta_1^2}{(4\mu^3-6\mu^2-3\mu+4)^2}, \label{etaFermion2} \\
\chi_{\phi,2} &=& \biggl[ 4 \mu (\mu-1) (2\mu^2+\mu-2)
\hat{\Psi}(\mu) - \frac{( 48 \mu^7 - 160 \mu^6 + 104 \mu^5 + 134 \mu^4 - 188 \mu^3 + 31 \mu^2
+ 28 \mu - 4)}{\mu-1}  \nonumber \\
&& \hspace{2mm} - 3 ( 16 \mu^3 - 4 \mu^2 - 14 \mu + 7 ) \mu^2 (\mu-1)
\left[ \psi^\prime(\mu-1) - \psi^\prime(1) \right] 
\biggr] \frac{\eta_1^2}{(4\mu^3-6\mu^2-3\mu+4)^2},
\label{eta2chi2}
\end{eqnarray}
\end{widetext}
where $\psi(z)=\Gamma'(z)/\Gamma(z)$ is the Euler digamma function and we define
\begin{align}
\hat{\Psi}(\mu)=\psi(2 \mu-1) - \psi(1) + \psi(1-\mu) - \psi(\mu-1).
\end{align}

As these expressions are valid in $d$ dimensions, expanding them in powers of $\epsilon$ with 
$d=4-\epsilon$ and comparing with the corresponding
RG functions at criticality we find full agreement, which is
a solid check on both the perturbative and large-$\Nf$ computations. The fermion anomalous dimension $\eta$ obtained from Eqs.~(\ref{etaFermion1}) and (\ref{etaFermion2}) agrees with the Landau-gauge ($\xi=0$) fermion anomalous dimension obtained from the four-loop RG function $\gamma_\Psi$ defined in Eq.~(\ref{gammaX}). The exponent $\hat{\beta}_\phi$ corresponds to the full scaling dimension of the scalar field, which is given by $(d-2+\eta_\phi)/2$ in $d$ dimensions where the anomalous dimension $\eta_\phi$ corresponds to the RG function $\gamma_\phi$ at criticality [Eq.~(\ref{etaphiRG})]. Finally, Eq.~(\ref{NuInverseLargeN}) agrees with the inverse correlation length exponent found in Eq.~(\ref{OneOverNu}). In $d=3$ dimensions we obtain 
\begin{eqnarray}
\left. \eta \right|_{d=3} &=& - \frac{1}{3\pi^2 \Nf} - 
\frac{( 3 \pi^2 + 80 )}{9 \pi^4 \Nf^2} +
\mathcal{O} \left( 1/\Nf^3 \right), \nonumber \\
\left. \eta_\phi \right|_{d=3} &=& 1 + \frac{32}{3 \pi ^2 \Nf} 
+ \frac{95 \pi ^2-368}{3 \pi ^4 N_{f}^2} +
\mathcal{O} \left( 1/\Nf^3 \right), \nonumber \\
\left. \frac{1}{\nu} \right|_{d=3} &=& 1 - \frac{44}{3\pi^2 \Nf} + 
\mathcal{O} \left( 1/\Nf^2 \right), \label{LargeNfExponents}
\end{eqnarray}
or, numerically
\begin{eqnarray}
\left. \eta \right|_{d=3} &=& - \frac{0.03377}{\Nf} - 
\frac{0.1250}{\Nf^2} +
\mathcal{O} \left( 1/\Nf^3 \right), \nonumber \\
\left. \eta_{\phi}\right|_{d=3} &=& 1 + \frac{1.081}{\Nf} + 
\frac{1.949}{N_{f}^2} +
\mathcal{O} \left( 1/\Nf^3 \right), \nonumber \\ 
\left. \frac{1}{\nu} \right|_{d=3} &=& 1 - \frac{1.486}{\Nf} + 
\mathcal{O} \left( 1/\Nf^2 \right) ,
\end{eqnarray}
using $\eta_{\phi}=2\left[2-\mu-\left(\eta+\chi_\phi\right)\right]$.

\subsection{CDW and VBS susceptibility exponents}

The final exercise is to determine the scaling dimension of fermion bilinears in the large-$N_f$ expansion. We begin with the flavor-singlet bilinear $\overline{\Psi}\Psi$, which determines the CDW susceptibility exponent [see Eqs.~(\ref{CDWcorrelations}) and (\ref{DeltaCDW})], as well as the adjoint bilinear $\overline{\Psi}T_A\Psi$. We shall compute these at
$\mathcal{O}(1/\Nf^2)$, which requires the expression for $\eta_2$. In this case there is 
a subtle difference between the setup for Eq.~(\ref{laguniv}) and that of the 
theory where the Pauli matrix of the scalar-fermion vertex is absent. The 
latter was considered in Ref.~\cite{gracey2018} and in that case the $\mathcal{O}(1/\Nf^2)$ 
expressions for the adjoint and singlet operator dimensions were different.
This was because of the graphs of Figure $7$ of Ref.~\cite{gracey2018}. These correspond 
to the insertion of the operator in a closed fermion loop. For the adjoint 
case these graphs are absent due to the trace over the matrix of the operator. 
The operator of the singlet case in Ref.~\cite{gracey2018} is the fermion mass operator 
and the corresponding graphs contribute leading to different $\mathcal{O}(1/\Nf^2)$ 
exponents. However, in the case of Eq.~(\ref{laguniv}) the presence of the Pauli 
matrix in the scalar-fermion vertex means that there is no contribution from 
the corresponding graphs of Figure $7$ of Ref.~\cite{gracey2018}. This does not mean that the 
singlet and adjoint operator dimensions are equivalent in the perturbative 
case. In fact they are different, with the discrepancy apparent at 
$\mathcal{O}(1/\Nf^3)$, as can be proved explicitly by expanding Eq.~(\ref{DiffSingAdj}) in inverse powers of $N_f$. Computations at $\mathcal{O}(1/\Nf^3)$ are however beyond the reach of the current large-$\Nf$ formalism. 
The outcome is that we have the leading-order exponent
\begin{equation}
\eta_{{\cal O},1} = - \frac{\mu(4\mu^2-10\mu+7)\eta_1}
{(4\mu^3-6\mu^2-3\mu+4)},  
\end{equation}
and 
\begin{widetext}
\begin{eqnarray}
\eta_{{\cal O},2} &=& - \biggl[ 3 \mu^2 (2\mu-1) (\mu-1) (4\mu^2-10\mu+7)
\left[ \psi^\prime(\mu-1) - \psi^\prime(1) \right] + 12 \mu (\mu-1) (2\mu^2+\mu-2)\hat{\Psi}(\mu) \nonumber \\
&& \hspace{5mm} + \frac{16 \mu^6 - 128 \mu^5 + 236 \mu^4 - 186 \mu^3 
+ 67 \mu^2 - 4 \mu - 4}{\mu-1} \biggr] \frac{\eta_1^2}{(4\mu^3-6\mu^2-3\mu+4)^2},
\label{etaop2}
\end{eqnarray}
\end{widetext}
at next-to-leading order, for both singlet and adjoint bilinears.
We have checked that the $\epsilon$-expansion of both are in full agreement
with the four-loop order results obtained from Eq.~(\ref{DimBilinear}). The next stage is to deduce the three-dimensional values 
\begin{equation}\label{DeltaPsiPsiLargeNf}
\left. \Delta_{\overline{\Psi}\Psi} \right|_{d=3} =
2 - \frac{4}{3\pi^2 \Nf} + \frac{4 ( 12 - \pi^2  )}{3 \pi^4 \Nf^2} +
\mathcal{O} \left( 1/\Nf^3 \right),
\end{equation}
or
\begin{equation}
\left. \Delta_{\overline{\Psi}\Psi} \right|_{d=3} = 2 -
\frac{0.1351}{\Nf} + \frac{0.02916}{\Nf^2} + 
\mathcal{O} \left( 1/\Nf^3 \right) ,
\label{expqedichgnd3}
\end{equation}
where
\begin{equation}\label{SingletBilinearDimLargeN}
\Delta_{\overline{\Psi}\Psi} = 2 \mu - 1 + \eta + \eta_{{\cal O}} .
\end{equation}
In contrast to the corresponding expression in Ref.~\cite{gracey2018} the coefficient of 
the $\mathcal{O}(1/\Nf^2)$ term is significantly smaller than that at $\mathcal{O}(1/\Nf)$. This 
may mean that exponent estimates for relatively low values of $\Nf$ could be 
reasonably reliable. However, this should not be interpreted as a general statement regarding the relative reliability of the large-$N_{f}$ expansion at small $N_{f}$ for the Heisenberg versus Ising theories, since the coefficients of $\mathcal{O}(1/N_f^3)$ and higher-order terms remain unknown in these theories and thus cannot be compared.

We next turn to the axial mass operator $i\overline{\Psi}\Gamma_5\Psi$, which controls the decay of VBS correlations [see Eqs.~(\ref{VBScorrelations}) and (\ref{DeltaVBS})] at the QCP. Expressing its scaling dimension as $\Delta_{i\overline{\Psi}\Gamma_5\Psi}=2\mu-1+\eta+\eta_{\mathcal{O}_5}$, the leading $\mathcal{O}(1/N_f)$ correction is 
\begin{equation}
\eta_{\mathcal{O}_5,1}= -\frac{\mu(4\mu^2+2\mu-5)\eta_1}{(4\mu^3-6\mu^2-3\mu+4)},
\end{equation}
with the next-to-leading order term given by
\begin{widetext}
\begin{eqnarray}
\eta_{\mathcal{O}_5,2}&=&
-\biggl[3(4\mu^2+2\mu-5)\mu^2(2\mu-1)(\mu-1)[\psi^{\prime}(\mu-1)-\psi^{\prime}(1)]-12(2\mu^2+\mu-2)\mu(\mu-1)\hat{\Psi}(\mu)\nonumber\\
&& \hspace{5mm} +\frac{16\mu^6-32\mu^5+20\mu^4+18\mu^3-65\mu^2+44\mu-4}{\mu-1}
\biggr]\frac{\eta_{1}^2}{(4\mu^{3}-6\mu^{2}-3\mu+4)^2}.
\end{eqnarray}
\end{widetext}
In $d=3$ dimensions, this evaluates to
\begin{align}\label{DeltaPsiGamma5PsiLargeNf}
\Delta_{i\overline{\Psi}\Gamma_5\Psi}=2-\frac{22}{3\pi^2N_f}+\frac{2(36-11\pi^{2})}{3\pi^{4}N_{f}^2}+\mathcal{O}(1/N_f^3).
\end{align}
If the result for arbitrary $d$ is expanded in powers of $\epsilon$, where $d=4-\epsilon$, then the series that ensues agrees with the large-$N_{f}$ expansion of the four-loop order result in Eq.~(\ref{DimBilinear}).

\subsection{QAH susceptibility exponent and Aslamazov-Larkin diagrams}
\label{sec:QAH}

We now turn to the spin-singlet bilinear $\overline{\psi}\psi=i\overline{\Psi}\Gamma_3\Gamma_5\Psi$, which is odd under time reversal $\mathcal{T}$ and reflection $R_x$ (i.e., parity in 2+1 dimensions) as shown in Appendix~\ref{app:PSG}. Because of the presence of the $\Gamma_3$ matrix, this bilinear is not Lorentz invariant in $d=4-\epsilon$ dimensions but is Lorentz invariant in strict $d=3$ dimensions, where the spacetime index $\mu=0,1,2$ only. As a result, we do not calculate its dimension in the $\epsilon$ expansion, but can compute it in the large-$N_f$ expansion in fixed $d=3$, using the relation $\Gamma_5=\Gamma_0\Gamma_1\Gamma_2\Gamma_3$. Physically, this operator corresponds to the continuum limit of a QAH or Haldane mass term~\cite{haldane1988}, first discussed in the context of the $\pi$-flux phase on the square lattice by Wen, Wilczek, and Zee~\cite{wen1989}:
\begin{align}
\mathcal{O}_\text{QAH}(r)\sim i(-1)^{x+y}\sum_\sigma&\bigl(c_{r\sigma}^\dag W_{r,r+\b{a}_1} c_{r+\b{a}_1,\sigma}\nn\\
&\phantom{\bigl(}-c_{r\sigma}^\dag W_{r,r+\b{a}_2} c_{r+\b{a}_2,\sigma}\bigr)+\mathrm{h.c.}
\end{align}
The pure imaginary, diagonal next-nearest-neighbor hopping terms break time-reversal symmetry and give a mass to the Dirac fermions. We define two Wilson line operators as
\begin{align}\label{OQAH}
W_{r,r+\b{a}_1}&=e^{i\theta_{r,r+\hat{x}}}e^{i\theta_{r+\hat{x},r+\hat{x}-\hat{y}}}
+e^{i\theta_{r,r-\hat{y}}}e^{i\theta_{r-\hat{y},r+\hat{x}-\hat{y}}},\nn\\
W_{r,r+\b{a}_2}&=e^{i\theta_{r,r+\hat{x}}}e^{i\theta_{r+\hat{x},r+\hat{x}+\hat{y}}}
+e^{i\theta_{r,r+\hat{y}}}e^{i\theta_{r+\hat{y},r+\hat{x}+\hat{y}}},
\end{align}
which obey $W_{r,r+\b{a}_1}^\dag=W_{r+\b{a}_1,r}$ and $W_{r,r+\b{a}_2}^\dag=W_{r+\b{a}_2,r}$. Such Wilson line insertions~\cite{hermele2005} are required to ensure that $\mathcal{O}_\text{QAH}(r)$ is invariant under lattice $U(1)$ gauge transformations $c_{r\sigma}\rightarrow e^{i\varphi_r}c_{r\sigma}$, $\theta_{r,r+\hat{\mu}}\rightarrow\theta_{r,r+\hat{\mu}}+\varphi_r-\varphi_{r+\hat{\mu}}$, and their particular form has been chosen to preserve all lattice symmetries: one can check that Eq.~(\ref{OQAH}) transforms precisely as the continuum $\overline{\psi}\psi$ bilinear under the microscopic PSG transformations in Eq.~(\ref{PSG}). As result we expect that at the N\'eel-ASL QCP, correlations of the QAH operator (\ref{OQAH}) as computed in QMC would decay in power-law fashion at long distances,
\begin{align}
\langle\mathcal{O}_\text{QAH}(r)\mathcal{O}_\text{QAH}(r')\rangle\sim\frac{1}{|r-r'|^{2\Delta_\text{QAH}}},
\end{align}
with
\begin{align}\label{DeltaQAH}
\Delta_\text{QAH}=\Delta_{i\overline{\Psi}\Gamma_3\Gamma_5\Psi},
\end{align}
where $\Delta_{i\overline{\Psi}\Gamma_3\Gamma_5\Psi}$ is the dimension of the continuum bilinear.

In the preceding subsections critical exponents have been computed in arbitrary $d$-dimensional spacetime utilizing the fact that the trace of the product $\Gamma_{\mu_1}\cdots\Gamma_{\mu_{2n+1}}$ of an odd number of $4\times 4$ gamma matrices is zero. In fixed $d=3$ dimensions with two-component spinors, this assumption is invalid since $\tr\gamma_\mu\gamma_\nu\gamma_\lambda\propto\epsilon_{\mu\nu\lambda}$, which was shown to give additional contributions to the large-$N_f$ critical exponents of the chiral Ising QED$_3$-GNY model from Aslamazov-Larkin diagrams~\cite{boyack2019,benvenuti2019}. In the four-component formulation with $d=3$, we expect Aslamazov-Larkin diagrams to again contribute in the presence of a $\Gamma_3\Gamma_5$ insertion, since $\tr\Gamma_3\Gamma_5\Gamma_\mu\Gamma_\nu\Gamma_\lambda\propto\epsilon_{\mu\nu\lambda}$. The calculation of the Aslamazov-Larkin diagrams for the chiral-Ising case was performed in Ref.~\cite{boyack2019}, and so those results can be adapted to the present case by incorporating the additional $SU(2)$ spin structure present in the Feynman rules. As a result of the modified Feynman rules, the large-$N_f$ scalar propagator is twice the value of that appearing in Ref.~\cite{boyack2019}, while the photon propagator is now one half of its value in the chiral-Ising case. In reference to the Feynman diagrams in Ref.~\cite{boyack2019}, we note that since there are two spin components, the diagram in Fig.~4 is multiplied by two, whereas the spin factor and the modified photon propagator factor cancel one another to 
render Figs.~5(a-e) equal to their original value. For the diagrams in Figs.~6(a-c), the Pauli matrices in the Yukawa vertex now lead to an additional factor of $\frac{1}{4}\mathrm{tr}\sigma^{a}\sigma^{b}\delta^{ab}=3/2$, and thus the diagrams as a whole are $2\times3/2$ larger than their original values.  The diagrams in Figs.~7(a,d) still give zero total contribution. As a result, the QAH susceptibility exponent in Eq.~(\ref{DeltaQAH}) is given to $\mathcal{O}(1/N_f)$ by
\begin{equation}
\Delta_{i\overline{\Psi}\Gamma_3\Gamma_5\Psi}=2+\frac{44}{3\pi^2N_{f}} +\mathcal{O} \left(1/\Nf^2\right) \nonumber \\.
\end{equation}

\subsection{Fermion bilinear dimensions in the chiral Heisenberg GN(Y) model}
\label{sec:O3GNYlargeNf}

While the fermion ($\eta$) and boson ($\eta_\phi$) anomalous dimensions as well as the inverse correlation length exponent $1/\nu$ have been computed in the large-$N_f$ expansion for the pure chiral Heisenberg GN(Y) model~\cite{gracey2018c}, the scaling dimensions of fermion bilinears have thus far not been computed using this method. Using the same formalism as in previous subsections, we have computed the dimension of the flavor-singlet $\overline{\Psi}\Psi$ and flavor-adjoint $\overline{\Psi}T_A\Psi$ bilinears at $\mathcal{O}(1/N_f^{2})$, which are the same at this order, as well as the dimension of the axial bilinear $i\overline{\Psi}\Gamma_5\Psi$ at $\mathcal{O}(1/N_f^{2})$. The general expressions are again Eq.~(\ref{SingletBilinearDimLargeN}) and its axial counterpart, together with Eq.~(\ref{etaseries}), but where the exponents $\eta$, $\eta_\mathcal{O}$, and $\eta_{\mathcal{O}_5}$ must be computed at the critical fixed point of the pure chiral Heisenberg GN(Y) theory. The fermion exponents $\eta_1$ and $\eta_2$ were computed previously~\cite{gracey2018c}, but are reproduced here for convenience and accounting for the change in convention regarding the definition of flavor number ($N$ vs $N_f$). We obtain
\begin{align}
\eta_1^\text{cH-GNY}&=-\frac{3\Gamma(2\mu-1)}{4\mu\Gamma(1-\mu)\Gamma(\mu-1)\Gamma^2(\mu)}, \nn \\
\eta_{\mathcal{O},1}^\text{cH-GNY}=-\eta_{\mathcal{O}_5,1}^\text{cH-GNY}&=\frac{\mu}{\mu-1}\eta_1^\text{cH-GNY}, 
\end{align}
at $\mathcal{O}(1/N_f)$, and
\begin{align}
\eta_2^\text{cH-GNY}&=\biggl[\frac{2\mu-3}{3(\mu-1)}\Psi(\mu)+\frac{4\mu^2-6\mu+1}{2\mu(\mu-1)^2}\biggr]\left(\eta_1^\text{cH-GNY}\right)^2, \nn \\
\eta_{\mathcal{O},2}^\text{cH-GNY}&=\biggl[\frac{\mu(2\mu-3)}{3(\mu-1)^2}\hat{\Psi}(\mu)-\frac{\mu(4\mu-3)}{3(\mu-1)^3}\biggr]\left(\eta_1^\text{cH-GNY}\right)^2, \nn \\
\eta_{\mathcal{O}_5,2}^\text{cH-GNY}&=-\biggl[\frac{\mu(2\mu-3)}{3(\mu-1)^2}\hat{\Psi}(\mu)+\frac{\mu}{(\mu-1)^3}\biggr]\left(\eta_1^\text{cH-GNY}\right)^2,
\end{align}
at $\mathcal{O}(1/N_f^2)$, defining
\begin{align}
\Psi(\mu)=\psi(2\mu-1)-\psi(1)+\psi(2-\mu)-\psi(\mu).
\end{align}
In $d=4-\epsilon$ dimensions the results for $\Delta_{\overline{\Psi}\Psi}^\text{cH-GNY}$ and $\Delta_{i\overline{\Psi}\Gamma_5\Psi}^\text{cH-GNY}$ thus found agree with the corresponding four-loop results in Sec.~\ref{sec:O3GNY4Lbilinears}. The latter are obtained from Eq.~(\ref{DimBilinear}), but evaluated at the pure GNY fixed point. For $d=3$, we find
\begin{align}
\Delta_{\overline{\Psi}\Psi}^\text{cH-GNY}&=2+\frac{4}{\pi^2 N_f}-\frac{32}{3\pi^4 N_f^2}+\mathcal{O}(1/N_f^3),\label{SingletLargeNcHGNY} \\
\Delta_{i\overline{\Psi}\Gamma_5\Psi}^\text{cH-GNY}&=2-\frac{2}{\pi^2 N_f}-\frac{32}{3\pi^4 N_f^2}+\mathcal{O}(1/N_f^3).\label{AxialLargeNcHGNY}
\end{align}
As discussed in Sec.~\ref{sec:O3GNY4Lbilinears}, for $N_f=1$ those exponents control the decay of CDW and VBS correlations at the semimetal-AF QCP in the repulsive Hubbard model on the honeycomb and $\pi$-flux square lattices, and are in principle accessible to sign-problem-free QMC simulations.

\subsection{Monopole operators}
\label{sec:monopole}

The microscopic theory studied in Ref.~\cite{Meng2019} is defined on a lattice and thus has a compact $U(1)$ gauge group, as a result of the magnetic flux being defined only modulo $2\pi$ on the lattice. In such theories total magnetic flux is not conserved, unlike in noncompact $U(1)$ gauge theories, and monopole operators---gauge-invariant local operators which insert $2\pi q$ magnetic flux at a point in space, with $q\in\mathbb{Z}$---generically appear in the effective low-energy Lagrangian~\cite{polyakov1975,polyakov1977,polyakov1987}, provided they are invariant under all microscopic symmetries. If allowed monopole operators are relevant under the RG, they will proliferate, gap out the $U(1)$ gauge field, and confine excitations with nonzero gauge charge (here, the fermionic spinons); if irrelevant, compactness effects disappear at long distances. In a phase where the fermions are gapped by a symmetry-breaking order parameter, e.g., the N\'eel phase in our case, monopoles are known to be relevant and do proliferate~\cite{polyakov1975,polyakov1977,polyakov1987}, leading to a conventional AF phase with no propagating fractionalized excitations (spinons or low-energy gauge fields). When the fermions are gapless, such as in the ASL phase or at the N\'eel-ASL QCP, the monopoles are ``dressed'' by these gapless degrees of freedom and their scaling dimensions change. Those scaling dimensions can be computed in the large-$N_f$ expansion using the state-operator correspondence of conformal field theory~\cite{borokhov2002}; a $2\pi q$ monopole operator is irrelevant if its scaling dimension $\Delta_{2\pi |q|}$ is greater than 3.

In the ASL phase, described by conformal QED$_{3}$, in the large-$N_f$ limit the scaling dimension of the smallest ($q=\pm 1$) monopole operator allowed by the symmetries of the square lattice is $\Delta_{2\pi}^\text{ASL}=1.06N_f-0.0383+\mathcal{O}(1/N_f)$~\cite{Pufu2014}, where $N_{f}$ is, according to our convention, the number of $SU(2)$ doublets of four-component Dirac spinors. (Monopoles with larger $|q|$ have larger scaling dimensions, at least at leading order in the $1/N_f$ expansion.) Considering only the first two terms in the large-$N_f$ expansion would suggest that monopoles are relevant for $N_f=1$, which would destabilize the ASL phase. However, the QMC results~\cite{Meng2019,Meng2019b} find a stable ASL phase with power-law correlations in accordance with the predictions of the noncompact theory. There are thus two possibilities: (1) the $\mathcal{O}(1/N_f)$ and higher-order terms in $\Delta_{2\pi}^\text{ASL}$ become important at small $N_f$ and cause the monopoles to be irrelevant for $N_f=1$; (2) the monopoles are formally relevant, but the bare monopole fugacity $g_0$ is such that the length scale $L^*\sim a g_0^{-1/(3-\Delta_{2\pi}^\text{ASL})}$ beyond which monopole proliferation is observable (with $a$ the lattice constant) is much greater than the system sizes currently accessible in QMC simulations. (Equivalently, in the thermodynamic limit $L\rightarrow\infty$ monopole proliferation would occur only below a very small crossover temperature $T^*\sim c/L^*$ with $c$ some characteristic velocity.) In the latter scenario, critical scaling characteristic of the ASL phase would appear in an extended crossover regime $a\ll L\ll L^*$.

Turning to the N\'eel-ASL QCP itself, scaling dimensions of monopole operators at this critical point can again in principle be computed in the large-$N_f$ expansion. Such a calculation was recently performed~\cite{dupuis2019} for the theory in Eq.~(\ref{LGWnoncollinear}), describing the transition from the ASL to the $\b{q}=0$ noncollinear state on the kagom\'e lattice. However, as already discussed, this theory is distinct from the theory (\ref{L2+1square}) of the N\'eel-ASL transition on the square lattice. Generalizing to $N_f$ numbers of $SU(2)$ doublets of four-component Dirac fermions, which is required for a large-$N_f$ computation of monopole scaling dimensions, the theory (\ref{L2+1square}) has an $SU(2)_\text{spin}\times SU(N_f)_\text{flavor}\times U(1)_\text{nodal}$ symmetry, while the theory (\ref{LGWnoncollinear}) has a larger $SU(2)_\text{spin}\times SU(2N_f)_\text{flavor-nodal}$ symmetry. (On both lattices, and with our definition of $N_f$, the ASL phase has an enhanced $SU(4N_f)$ symmetry.) That is because, as discussed in Sec.~\ref{sec:kagome}, the bilinear that condenses on the kagom\'e lattice is proportional to the identity in nodal space, while the bilinear that condenses on the square lattice contains a $\tau_3$ nodal Pauli matrix and is only invariant under $U(1)$ nodal rotations about that direction. As a result, the large-$N_f$ monopole scaling dimensions computed for theory (\ref{LGWnoncollinear}) in Ref.~\cite{dupuis2019} cannot be directly transposed to the N\'eel-ASL transition, since the two critical theories have different flavor symmetries. On the other hand, the QMC results~\cite{Meng2019} suggest that a continuous N\'eel-ASL transition occurs at $N_f=1$. Indeed, Ref.~\cite{Meng2019} calculated the flux energy per plaquette, for various values of the coupling strength $J$. This was observed to be a continuous quantity, which moreover was reproducible with increasing system size, and the absence of any singularity at the critical $J_{c}$ suggests a continuous phase transition. While one expects from the state-operator correspondence that $\Delta_{2\pi}^\text{N\'eel-ASL}\propto N_f$ in the large-$N_f$ limit, plus subleading corrections in $1/N_f$, two scenarios consistent with the QMC results can be envisaged as above, with the added uncertainty resulting from the unknown proportionality coefficient. (1) $\Delta_{2\pi}^\text{N\'eel-ASL}<3$ at $N_f=1$, and the transition becomes first order in the thermodynamic limit (assuming the ASL phase is itself stable) but the length scale $L^{**}\sim ag_0^{-1/(3-\Delta_{2\pi}^\text{N\'eel-ASL})}$ beyond which this happens is much greater than the system size $L$; critical behavior with the exponents computed here is observed for $a\ll L\ll L^{**}$. (2) $\Delta_{2\pi}^\text{N\'eel-ASL}>3$ at $N_f=1$ and true critical behavior persists in the thermodynamic limit.

In summary, the ultimate fate of monopoles in the deep infrared limit for $N_f=1$ remains unknown, both in the ASL phase and at the N\'eel-ASL QCP. The critical exponents computed in this work describe either true critical behavior in the thermodynamic limit, or quasicritical behavior below crossover lengths $L^*$ and $L^{**}$ which are larger than system sizes currently accessible in numerics.

\section{Resummed critical exponents}
\label{sec:Pade}

Mathematically, the $\epsilon$-expansion and the large-$N_{f}$ expansion both produce asymptotic series, with a radius of convergence of zero. 
Resummation procedures, which are consistent ways of assigning finite values to formally divergent series, are thus necessary to extract physical results from the finite-order expressions derived analytically in Sections~\ref{sec:epsilon} and~\ref{sec:LargeN}.
We now apply two particular approximate resummation methods to obtain estimates of critical exponents in fixed $d=3$ for general $N_f\geq 1$. A standard method to extrapolate finite-order $\epsilon$-expansion results to the physical dimension $d=3$ is the use of Pad\'e approximants~\cite{KleinertPhi4}. For a given loop order $L$, the (one-sided) Pad\'e approximants are defined by  
\begin{equation}
[m/n](\epsilon)=\frac{\sum_{i=0}^m a_i\epsilon^i}{1+\sum_{j=1}^nb_{j}\epsilon^{j}},
\end{equation}
where $m$ and $n$ are two positive integers obeying $m+n=L$. The coefficients $a_i$ and $b_j$ are determined such that expanding the above function in powers of $\epsilon$ to $\mathcal{O}(\epsilon^{L})$ reproduces the $\epsilon$-expansion results. We also use the Pad\'e-Borel method~\cite{KleinertPhi4}, which provides an alternative method to resummation of the $\epsilon$-expansion results. For an expansion given by $\Delta(\epsilon)=\sum_{k=0}^{\infty}\Delta_{k}\epsilon^k$, the Borel sum is defined as $B_{\Delta}(\epsilon)=\sum_{k=0}^{\infty}\frac{1}{k!}\Delta_{k}\epsilon^{k}$. 
The Borel transform is the following exact result:
\begin{equation}
\Delta(\epsilon)=\int_{0}^{\infty}dt\ e^{-t}B_{\Delta}(\epsilon t).
\end{equation}
Since the $\epsilon$-expansion coefficients are known only to fourth order, in the above expression $B_{\Delta}$ is replaced by a Pad\'e approximant at a given order. In the following we present the results of Pad\'e extrapolation of the $\epsilon$-expansion for general $N_f$, as well as the corresponding Pad\'e-Borel results for $N_f=1$ only. We also apply those two methods to the resummation of the $1/N_f$-expansion series, as done for instance in Ref.~\cite{gracey2018}, for the case of $N_f=1$ only.
In regard to the appropriateness of applying resummation techniques, we note that the large-$N_{f}$ expansion already at $\mathcal{O}(1/N_f)$ agrees quite well with numerical QMC results for the ASL phase with $N_{f} = 1$ in our notation (see Fig.~7 in Ref.~\cite{Meng2019}, which is $N_{f} = 2$ in their notation). Since the ASL phase with $N_{f} = 1$ is already well captured by the large-$N_{f}$ expansion, it is physically reasonable to investigate applying this method to phase transitions out of the ASL phase.

\begin{figure}[t]
\centering\includegraphics[width=8.5cm,height=4.5cm,clip]{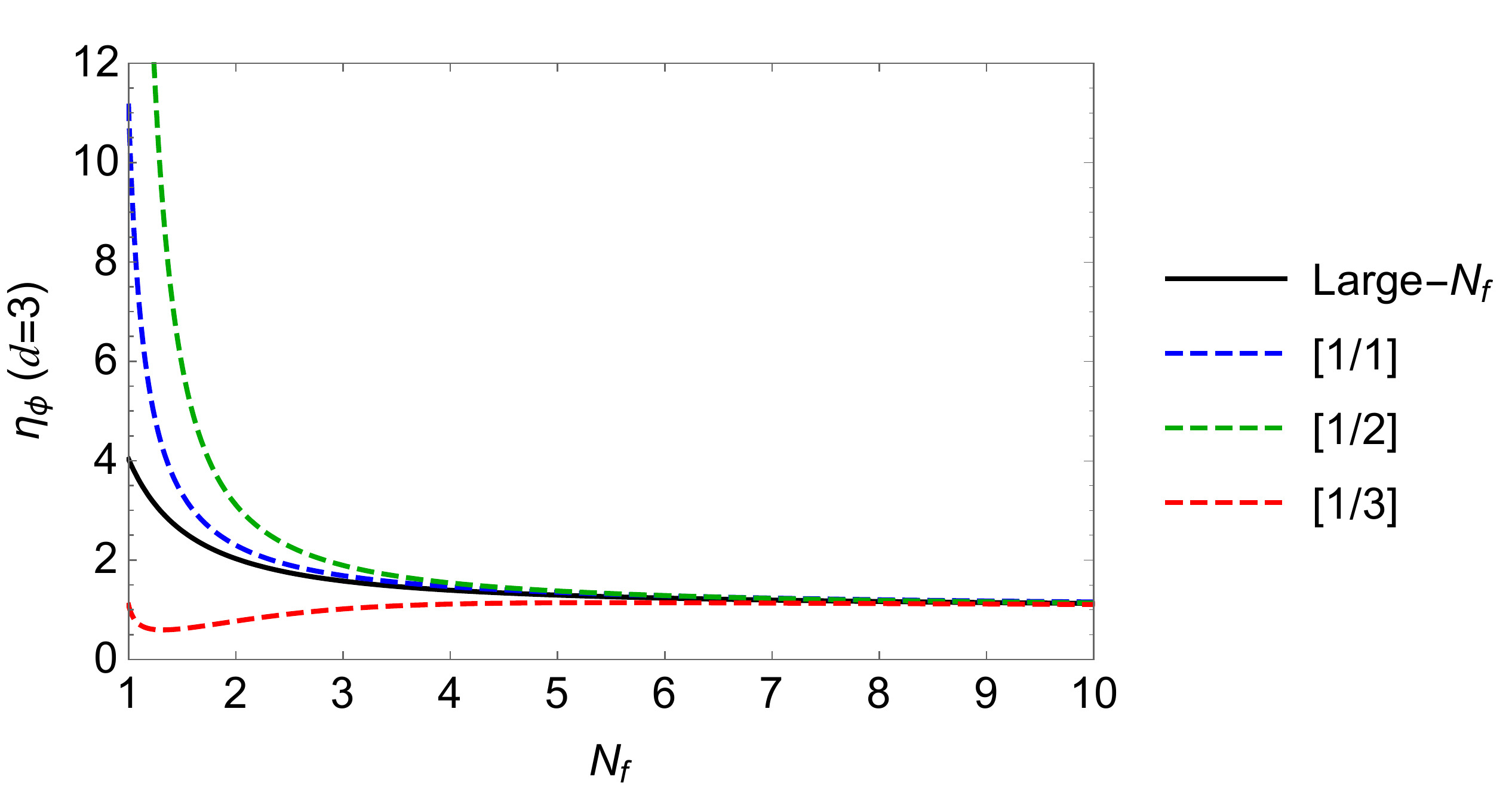}
\caption{Colored lines: Pad\'e approximants for $\eta_{\phi}$ in $d=3$ as a function of $N_{f}$ at two (blue), three (green), and four-loop (red) orders; black line: large-$N_f$ result from Eq.~(\ref{LargeNfExponents}).}
\label{fig:etaphi_3d}
\end{figure}

\begin{figure}[t]
\centering\includegraphics[width=8.5cm,height=4.5cm,clip]{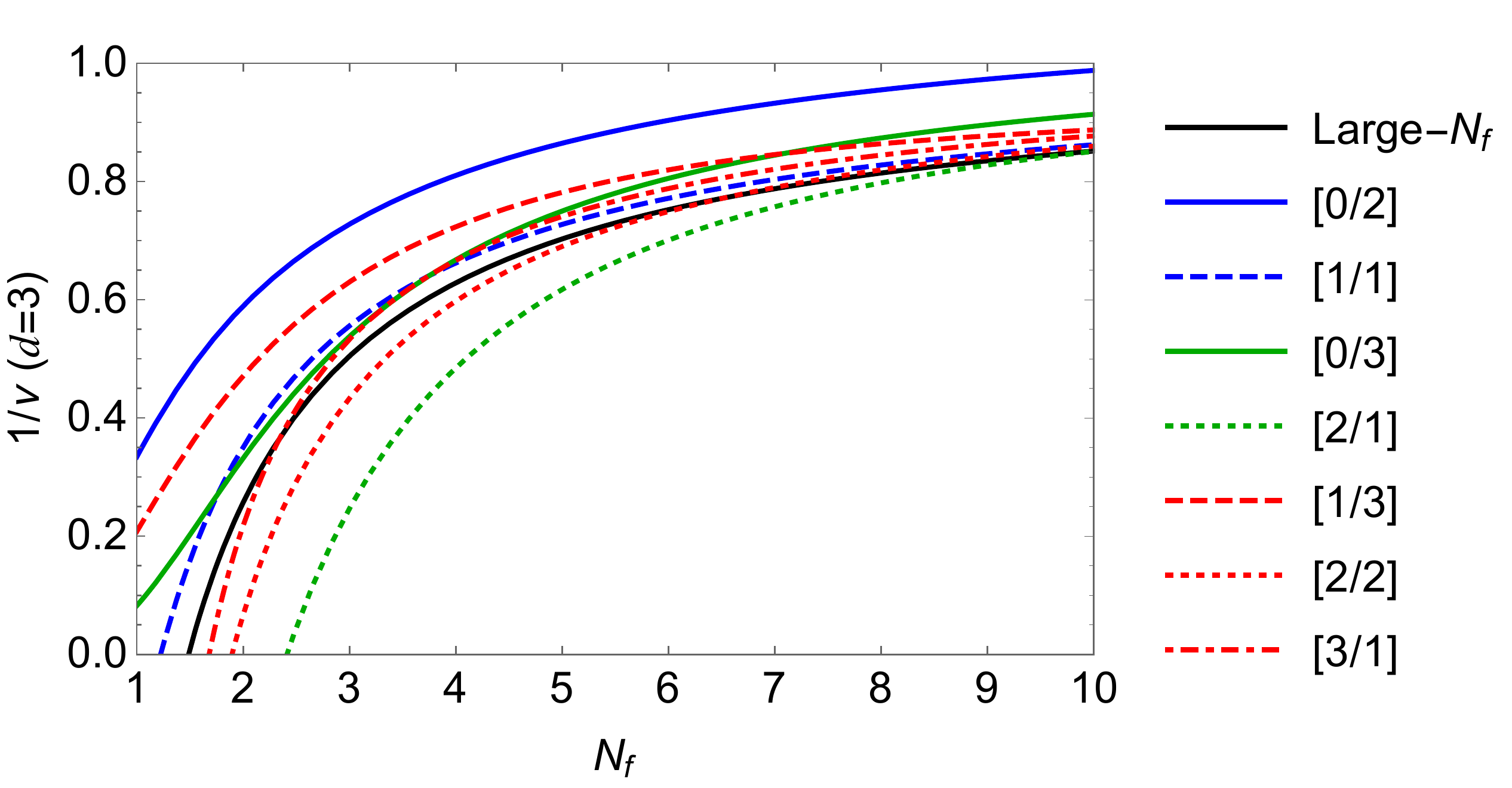}
\caption{Colored lines: Pad\'e approximants for $1/\nu$ in $d=3$ as a function of $N_{f}$ at two (blue), three (green), and four-loop (red) orders; black line: large-$N_f$ result from Eq.~(\ref{LargeNfExponents}).}
\label{fig:nuInv_3d}
\end{figure}

\begin{figure}[t]
\centering\includegraphics[width=8.5cm,height=4.5cm,clip]{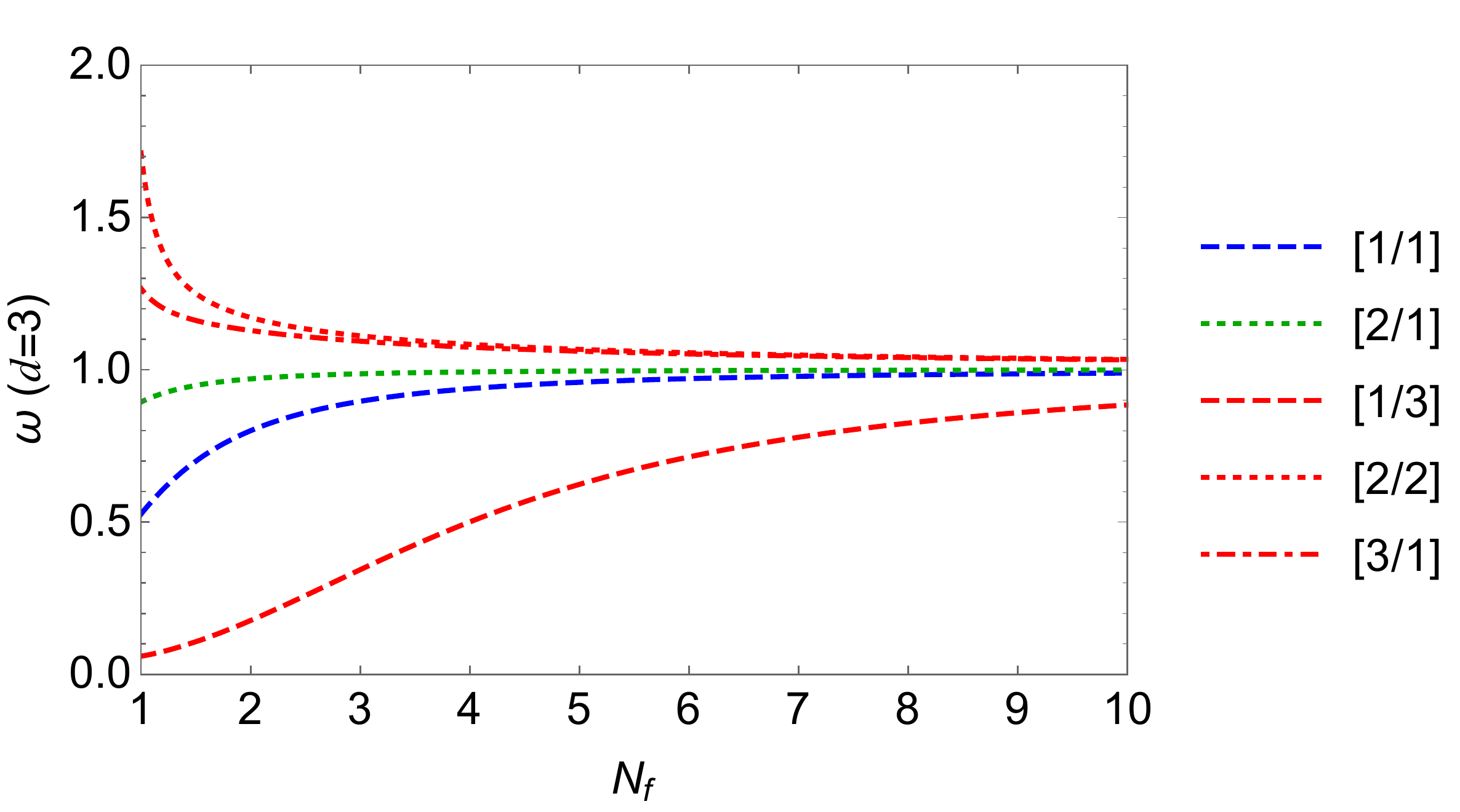}
\caption{Pad\'e approximants for $\omega$ in $d=3$ as a function of $N_{f}$ at two (blue), three (green), and four-loop (red) orders.}
\label{fig:omega_3d}
\end{figure}

\subsection{Chiral Heisenberg QED$_3$-GNY model}

The Pad\'e approximants for $\eta_{\phi}$, $1/\nu$, and $\omega$ at two-loop (blue), three-loop (green), and four-loop (red) orders are shown in Figs.~\ref{fig:etaphi_3d}, \ref{fig:nuInv_3d}, and \ref{fig:omega_3d} respectively. Only those Pad\'e approximants that are nonsingular in the extrapolation region $0<\epsilon<1$ are shown in the figures. With increasing $N_f$ the various approximants generally converge towards each other as well as towards the $d=3$ large-$N_f$ result (black line). Furthermore, in Fig.~\ref{fig:nuInv_3d} the spread of values predicted by the approximants decreases with increasing loop order for a fixed $N_f$. However, at smaller values of $N_f$ there is a discernible deviation amongst the various Pad\'e approximants. A similar phenomenon was apparent for the chiral Ising QED-GNY theory~\cite{ihrig2018,zerf2018} as well as for pure QED~\cite{dipietro2017}, whereas Pad\'e approximants for pure GNY models are comparatively better behaved~\cite{mihaila2017,zerf2017,ihrig2018b}. Physically this can be understood from the fact that the disordered phase of pure GNY models (a Dirac semimetal) is adiabatically connected to a system of noninteracting Dirac fermions regardless of the value of $N_f$, whereas in QED-GNY models the disordered phase (the ASL) consists of a system of mutually interacting Dirac fermions and gauge fields which becomes increasingly strongly coupled in the infrared for small $N_f$ (at least in the sense of the $1/N_f$ expansion). In particular, the large variation of the [1/3] Pad\'e approximant for both $\eta_{\phi}$ and $\omega$ in relation to the other approximants was a feature also noticed for the chiral Ising QED-GNY model~\cite{zerf2018}; if this approximant is ignored for those exponents a better overall consistency is achieved. Unitarity bounds~\cite{ferrara1974,mack1977} in conformal field theory require the scaling dimension $\Delta$ of a Lorentz scalar to obey $\Delta\geq d/2-1$. Since $\Delta_{\phi}=(d-2+\eta_{\phi})/2$ and $\Delta_{\phi^2}=d-1/\nu$, this condition imposes $\eta_{\phi}\geq 0$, which is satisfied by all Pad\'e approximants, and $1/\nu\leq d/2+1$. While all the approximants for $1/\nu$ satisfy the latter criterion in $d=3$, some of them give unphysical negative extrapolation values for small values of $N_f$.

\begin{figure}[t]
\centering\includegraphics[width=8.5cm,height=4.5cm,clip]{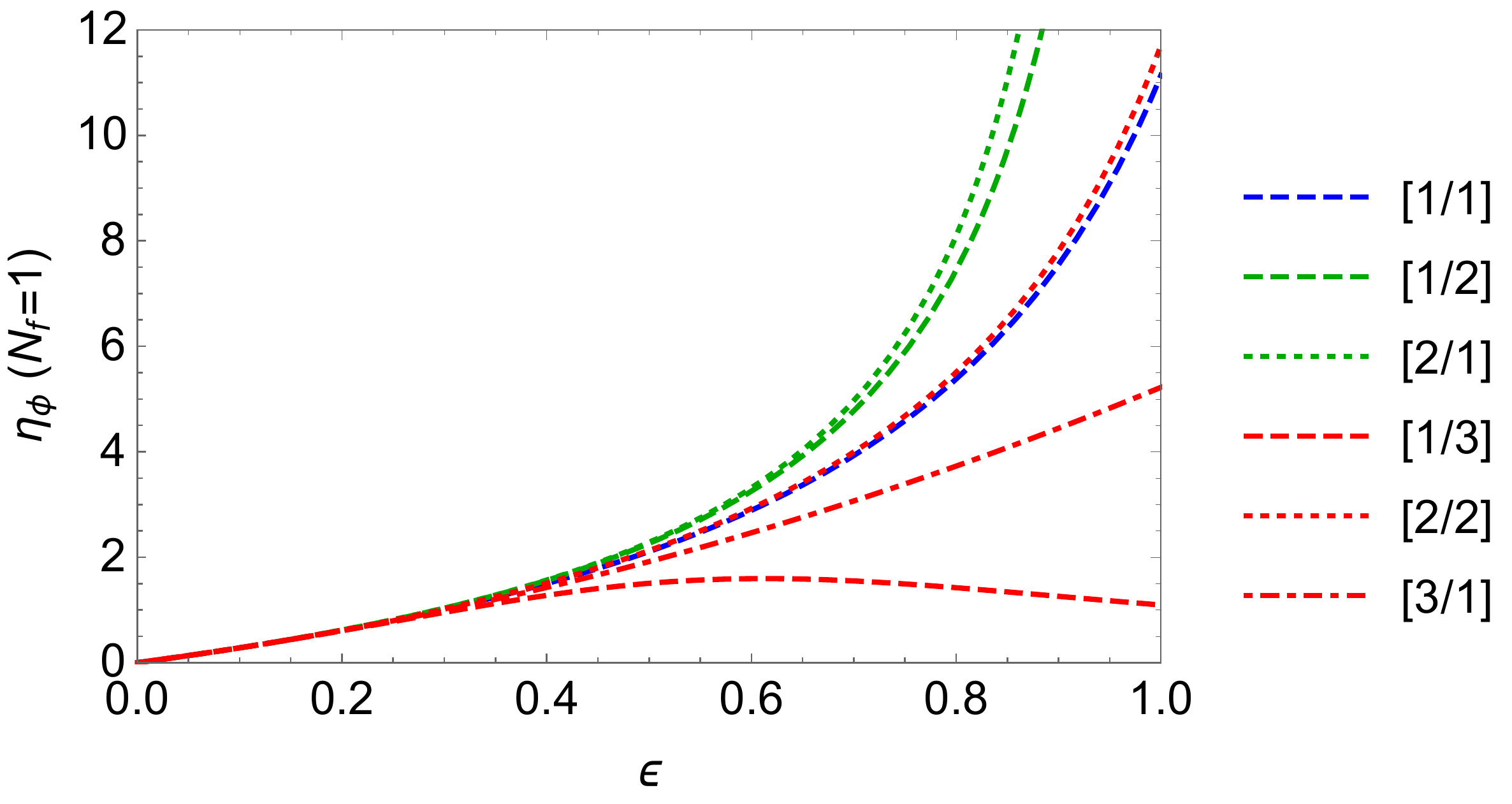}
\caption{Pad\'e approximants for $\eta_{\phi}$ as a function of $\epsilon$ for $N_{f}=1$ at two (blue), three (green), and four-loop (red) orders.}
\label{fig:etaphi_N1}
\end{figure}

\begin{figure}[t]
\centering\includegraphics[width=8.5cm,height=4.5cm,clip]{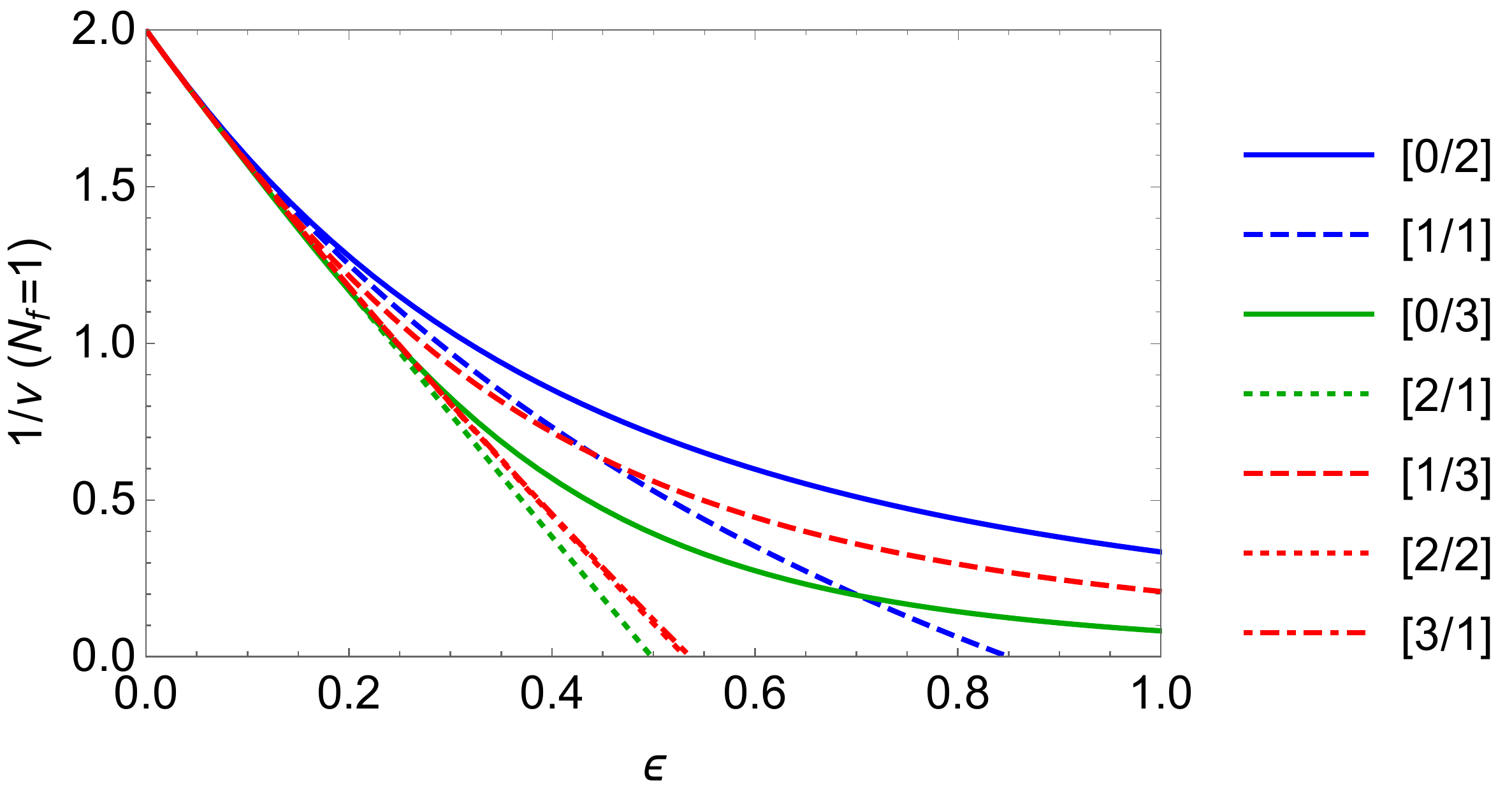}
\caption{Pad\'e approximants for $1/\nu$ as a function of $\epsilon$ for $N_{f}=1$ at two (blue), three (green), and four-loop (red) orders.}
\label{fig:nuInv_N1}
\end{figure}

\begin{figure}[t]
\centering\includegraphics[width=8.5cm,height=4.5cm,clip]{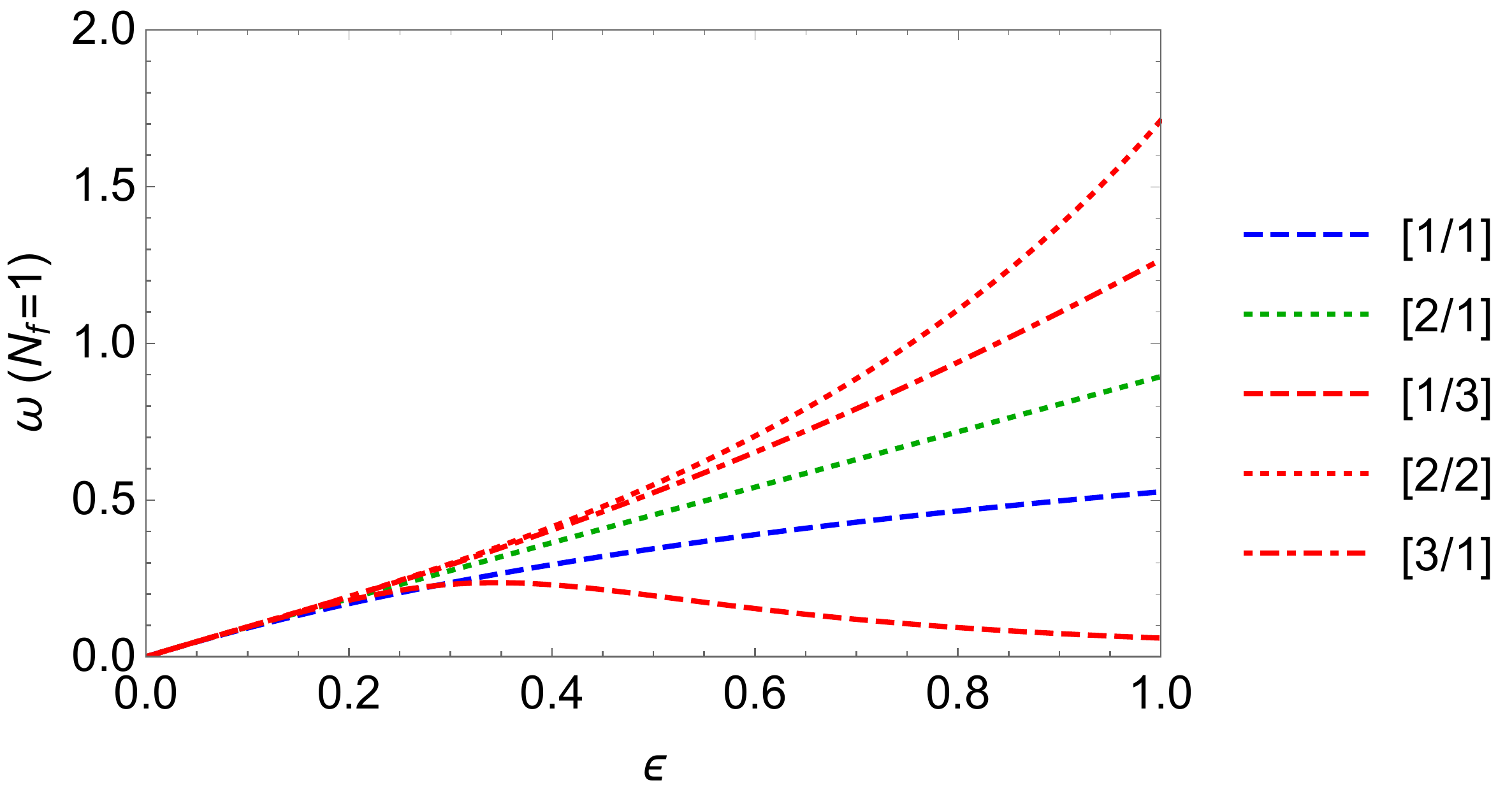}
\caption{Pad\'e approximants for $\omega$ as a function of $\epsilon$ for $N_{f}=1$ at two (blue), three (green), and four-loop (red) orders.}
\label{fig:omega_N1}
\end{figure}

For fixed fermion flavor number $N_{f}=1$, the behavior of the Pad\'e approximants as functions of $\epsilon$ is shown in Figs.~\ref{fig:etaphi_N1}-\ref{fig:omega_N1}. 
As expected, one can observe that all approximants converge to the same result in the $\epsilon\rightarrow0$ limit. The $\epsilon$ dependence is beneficial to illustrate (1) that there are no singularities in the extrapolation region $0 < \epsilon < 1$, otherwise the extrapolation would not be well defined, and (2) to show that certain extrapolations violate unitarity bounds before the $\epsilon\rightarrow1$ limit is reached.
In Fig.~\ref{fig:etaphi_N1} there is significant variation---as much as two orders of magnitude---among the various approximants at $\epsilon=1$ for $\eta_{\phi}$, although the [1/1] and [2/2] approximants produce similar estimates. (The [2/2] approximant was not plotted in Fig.~\ref{fig:etaphi_3d} as it contained singularities in the extrapolation region $0<\epsilon<1$ for certain values of $N_f$ in the range considered.)
In contrast to the present results, for the case of the pure chiral Heisenberg GNY model, where the gauge field is absent, the Pad\'e approximants for $\eta_{\phi}$ shown in Ref.~\cite{zerf2017} for small $N_f$
were in good agreement with each other as well as with QMC and functional RG results. This suggests that gauge fluctuations are a decisive feature in the critical behavior of the scalar field. 
The stability exponent $\omega$ exhibits less variability than $\eta_{\phi}$, as seen in Fig.~\ref{fig:omega_N1}, but better conformity between approximants was again exhibited in the pure chiral Heisenberg GNY case~\cite{zerf2017}.
This reiterates the point that for small $N_{f}$ gauge fluctuations tend to destabilize the system. The distinctive behavior of the [1/3] approximant at small $N_f$ for $\eta_\phi$ and $\omega$, already discussed above, is again illustrated by its unique nonmonotonic dependence on $\epsilon$ at $N_f=1$ in Figs.~\ref{fig:etaphi_N1} and \ref{fig:omega_N1}, which was also observed in the chiral Ising QED-GNY case (see Ref.~\cite{zerf2018}, as well as unpublished results for $\omega$ vs $\epsilon$). Dropping this approximant reduces the range of extrapolated values for those exponents, but not sufficiently so as to allow us to produce a meaningful quantitative estimate. By contrast, the three approximants [0/2], [0/3], [1/3], while at different loop orders, produce a reasonably consistent estimate of $1/\nu$ for $N_f=1$ in the range $\sim$~$0.1$ - $0.3$ (Fig.~\ref{fig:nuInv_N1}). Interestingly, in the chiral Ising QED-GNY model the same set of three approximants also produced positive values of $1/\nu$ at $N_f=1$, and in the same order ($(1/\nu)_{[0/2]}>(1/\nu)_{[1/3]}>(1/\nu)_{[0/3]}$), but spread over a wider range ($\sim$~$0.05$ - $0.7$). Similarly, the [2/1], [2/2], and [3/1] approximants produce unphysical negative values of $1/\nu$ at $\epsilon=1$ in both models.

\begin{figure}[t]
\centering\includegraphics[width=8.5cm,height=4.5cm,clip]{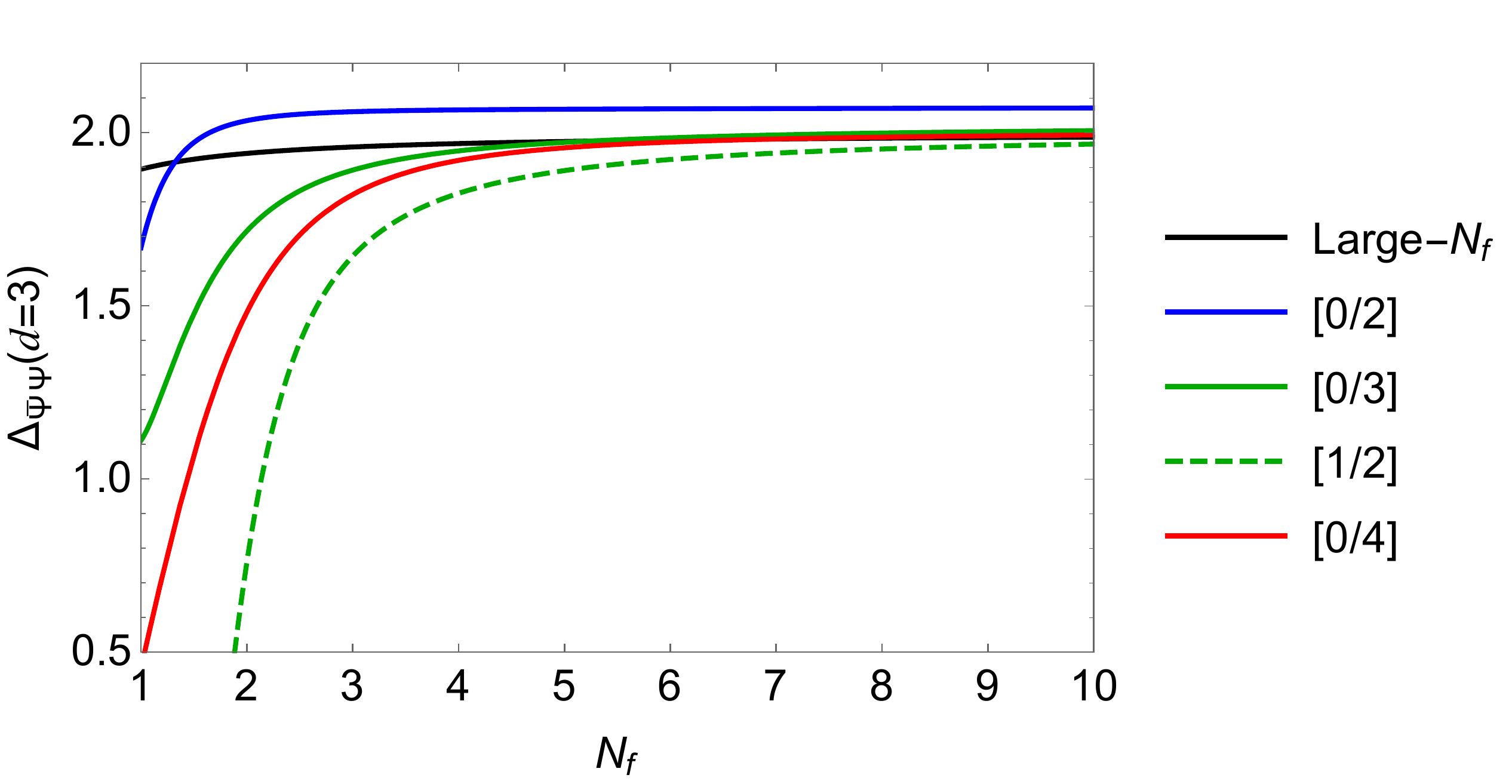}
\caption{Colored lines: Pad\'e approximants for $\Delta_{\overline{\Psi}\Psi}$ in $d=3$ as a function of $N_{f}$ at two (blue), three (green), and four-loop (red) orders; black line: large-$N_f$ result from Eq.~(\ref{DeltaPsiPsiLargeNf}).}
\label{fig:deltaSinglet_3d}
\end{figure}

\begin{figure}[t]
\centering\includegraphics[width=8.5cm,height=4.5cm,clip]{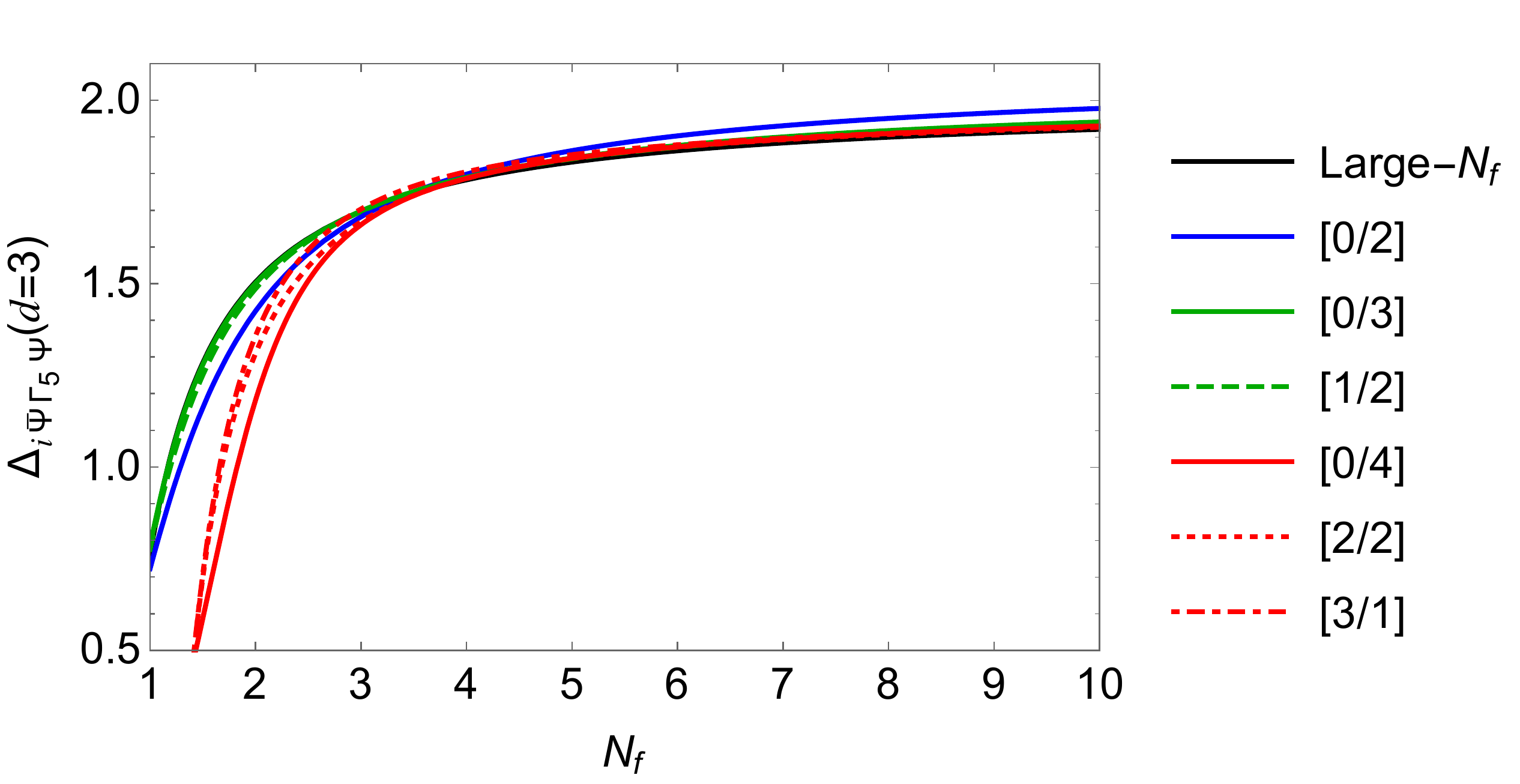}
\caption{Colored lines: Pad\'e approximants for $\Delta_{i\overline{\Psi}\Gamma_{5}\Psi}$ in $d=3$ as a function of $N_{f}$ at two (blue), three (green), and four-loop (red) orders; black line: large-$N_f$ result from Eq.~(\ref{DeltaPsiGamma5PsiLargeNf}).}
\label{fig:deltaAxial_3d}
\end{figure}

\begin{figure}[t]
\centering\includegraphics[width=8.5cm,height=4.5cm,clip]{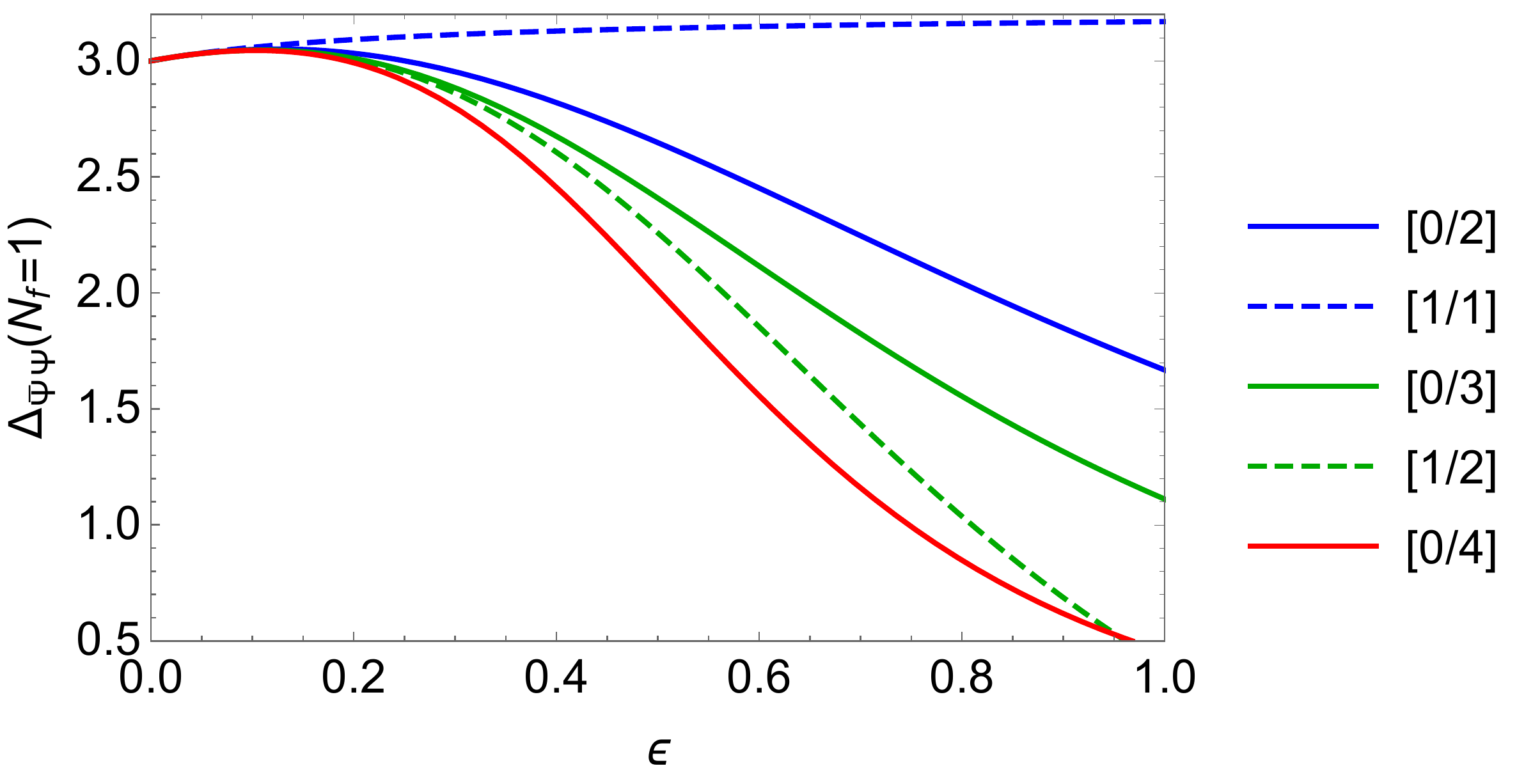}
\caption{Pad\'e approximants for $\Delta_{\overline{\Psi}\Psi}$ as a function of $\epsilon$ for $N_{f}=1$ at two (blue), three (green), and four-loop (red) orders.}
\label{fig:deltaSinglet_N1}
\end{figure}

\begin{figure}[t]
\centering\includegraphics[width=8.5cm,height=4.5cm,clip]{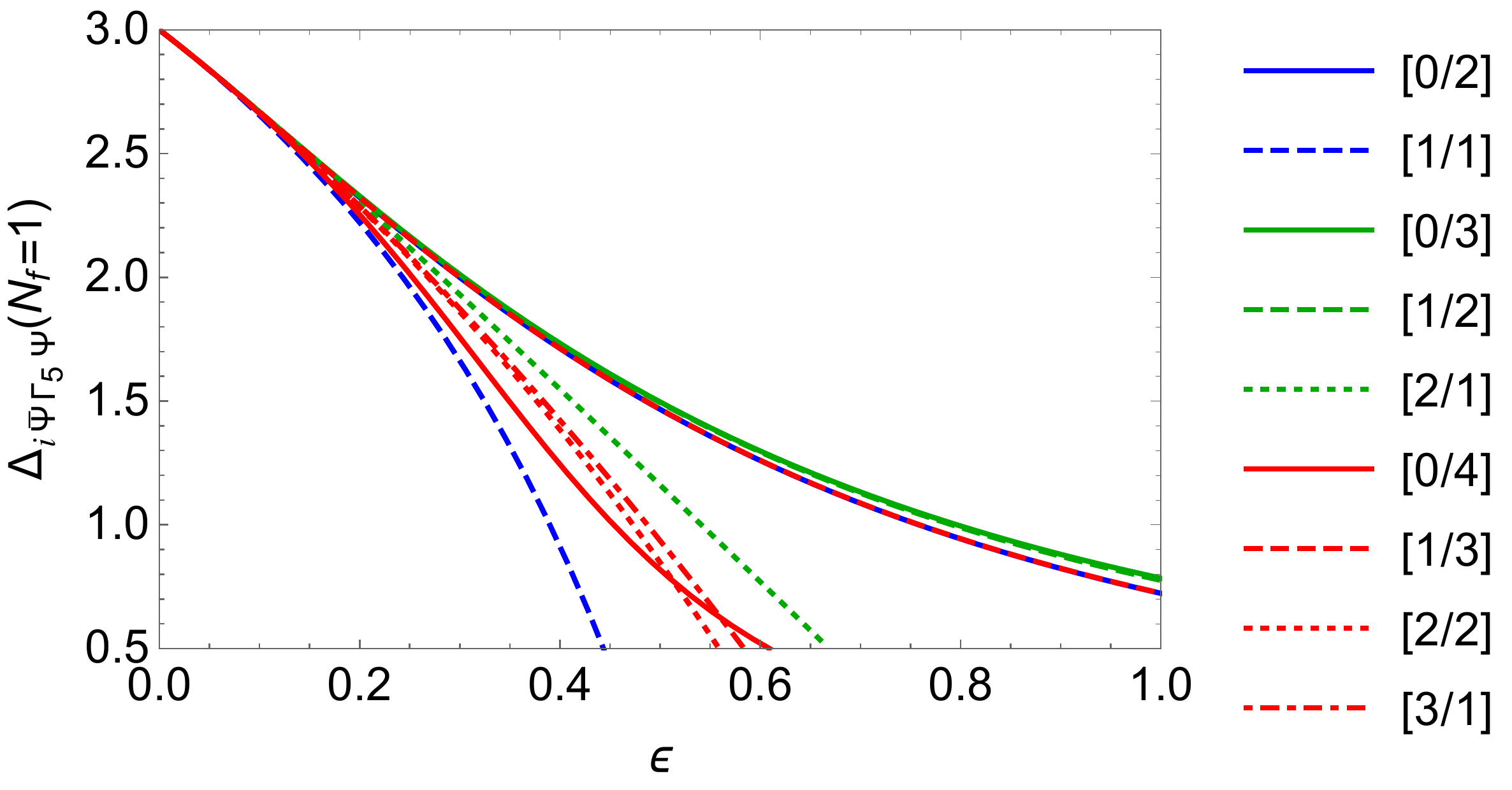}
\caption{Pad\'e approximants for $\Delta_{i\overline{\Psi}\Gamma_{5}\Psi}$ as a function of $\epsilon$ for $N_{f}=1$ at two (blue), three (green), and four-loop (red) orders.}
\label{fig:deltaAxial_N1}
\end{figure}

Turning now to the scaling dimensions $\Delta_{\overline{\Psi}\Psi}$  and $\Delta_{i\overline{\Psi}\Gamma_{5}\Psi}$ of fermion bilinears, corresponding for $N_f=1$ to CDW and VBS susceptibility critical exponents, respectively, the Pad\'e approximants are shown for fixed $d=3$ and arbitrary $N_{f}$ in Figs.~\ref{fig:deltaSinglet_3d}-\ref{fig:deltaAxial_3d}, and for $N_{f}=1$ and arbitrary $\epsilon$ in Figs.~\ref{fig:deltaSinglet_N1}-\ref{fig:deltaAxial_N1}. In the large-$\Nf$ limit both scaling dimensions asymptote to $2$, however, for small values of $\Nf$ there is a discernible variation in the approximants as for the exponents considered previously. As we only consider scalar or pseudoscalar bilinears, unitarity bounds require $\Delta_{\overline{\Psi}\mathcal{M}\Psi}\geq1/2$, which is satisfied by all approximants for $N_f\geq 2$, but violated for $N_f=1$ by the [1/2] and [0/4] approximants in Fig.~\ref{fig:deltaSinglet_3d} (see also Fig.~\ref{fig:deltaSinglet_N1} for the [0/4] approximant) and all four-loop approximants in Fig.~\ref{fig:deltaAxial_3d}. However, some of those approximants do satisfy the unitarity requirement for $N_f=1$ after Borel resummation (see Table~\ref{tab:Pade_EpsilonExp}). There is a higher degree of convergence between approximants at a given loop order for the axial mass bilinear as compared to the normal mass bilinear, as can be seen by comparing Fig.~\ref{fig:deltaSinglet_3d} and Fig.~\ref{fig:deltaAxial_3d}. Furthermore, the [0/2] (two-loop), [0/3] and [1/2] (three-loop), and [1/3] (four-loop) approximants all give very similar values for the axial mass bilinear dimension at $N_f=1$ (Fig.~\ref{fig:deltaAxial_N1}), in the range $\sim$~$0.7$ - $0.8$. (The [1/3] approximant was excluded from Fig.~\ref{fig:deltaAxial_3d} for the same reason as that mentioned for $\eta_\phi$.)

In Table~\ref{tab:Pade_EpsilonExp} we collect the numerical extrapolated values for the critical exponents $\eta_{\phi}$, $1/\nu$, and $\omega$, and for the fermion bilinear dimensions $\Delta_{\overline{\Psi}\Psi}$ and $\Delta_{i\overline{\Psi}\Gamma_5\Psi}$, as obtained from the $\epsilon$-expansion results using both Pad\'e and Pad\'e-Borel resummation, for the particular case $N_{f}=1$.  A high variability in the extrapolation values for $\eta_{\phi}$ is manifest, as already discussed, whereas the variability is significantly less for the other exponents. Supplementing the nonnegative Pad\'e estimates for $1/\nu$ with Pad\'e-Borel resummation extends the range of obtained values somewhat, predicting $1/\nu$ in the range $\sim$~$0.1$ - $0.6$. Similarly, the Pad\'e-Borel values for the fermion bilinear dimensions are systematically higher than the corresponding Pad\'e estimates. Overall, the range of extrapolated values for the exponents, apart from $\eta_{\phi}$, is comparable to that found in the chiral-Ising QED-GNY case. The use of more sophisticated resummation methods~\cite{mihaila2017} may potentially improve those estimates, but typically requires the introduction of a number of free parameters which effectively adds to the uncertainty. Another potentially beneficial approach to improve exponent estimates in three dimensions is the use of two- and four-dimensional perturbation theory~\cite{ihrig2018b}; however, at present there is a lack of two-dimensional studies on the chiral-Heisenberg QED$_{3}$ model. 

\begin{table}[t]
\centering
\begin{tabular}{|l||c|c|c|c|c|}
      \hline
      $N_{f} = 1$      		    & $\eta_\phi$        & $1/\nu$     & $\omega$       & $\Delta_{\overline{\Psi}\Psi}$  & $\Delta_{i\overline{\Psi}\Gamma_{5}\Psi}$ \\
          \hline\hline
      $[0/2]$        			    & $\times$           & 0.334         & $\times$          & 1.67					 & 0.723 \\
      $[0/2]_{\textrm{PB}}$ 
                       			   & $\times$            & 0.636         & $\times$          & 2.34 					& 1.34 \\
      \hline
      $[1/1]$       			   & 11.1                   & $\times$    &  0.526             & 3.17 					& $\times$ \\
      $[1/1]_{\textrm{PB}}$   
                        			   & $\times$           & $\times$     & 0.578              & 3.23 					& $\times$ \\
    
      \hline
      $[0/3]$         		   & $\times$            & 0.0826      & $\times$        & 1.11 					& 0.784 \\
      $[0/3]_{\textrm{PB}}$  
                         			   & $\times$            & 0.534       &  $\times$      & $\times$ 					& 1.20 \\
      \hline
      $[1/2]$          		   & 39.4                 & $\times$    & $\times$       & 0.393					& 0.776 \\
      $[1/2]_{\textrm{PB}}$          
                         			   & $\times$          & $\times$    & $\times$       & 2.39					& 0.787\\
       \hline
      $[2/1]$       			  & 135                  & $\times$    & 0.894            & $\times$				        & $\times$ \\
      $[2/1]_{\textrm{PB}}$                 
                        			   & $\times$         & $\times$    &  0.862          & $\times$				        & $\times$ \\
   
       \hline
      $[0/4]$      			  & $\times$          &  $\times$   & $\times$       & 0.452					& 0.0986\\
      $[0/4]_{\textrm{PB}}$   
                      			   & $\times$        & 0.492          &  $\times$     & 2.05 					& 1.10\\
       \hline
      $[1/3]$       			  & 1.09               &  0.208         & 0.0596       & $\times$ 				 & 0.724\\
      $[1/3]_{\textrm{PB}}$     
                        			  & 2.26               & 0.236          & 0.664         & 2.34					& $\times$ \\
      \hline 
      $[2/2]$       			  & 11.7               & $\times$     & 1.71           & $\times$				& $\times$ \\
      $[2/2]_{\textrm{PB}}$     
                       			   & $\times$       & $\times$     & $\times$     & $\times$				& $\times$ \\
      \hline
      $[3/1]$      			   & 5.22            & $\times$      & 1.27           & $\times$				& $\times$ \\
      $[3/1]_{\textrm{PB}}$    
                       			  & 5.29            & $\times$       & 1.48           & $\times$				& $\times$ \\
      \hline
\end{tabular}
      \caption{Pad\'e and Pad\'e-Borel resummations of the $\epsilon$-expansion expressions for $d=3$ and $N_{f}=1$. Approximants which are either singular in the domain $0<\epsilon<1$, undefined, or negative, are denoted by $\times$.}
      \label{tab:Pade_EpsilonExp}
\end{table}

In Table~\ref{tab:Pade_LargeN} we present extrapolated values of critical exponents as obtained from the large-$\Nf$-expansion results, again for the particular case $N_{f}=1$. In addition to the critical exponents $\eta_{\phi}$ and $1/\nu$ and the fermion bilinear dimensions considered in Table~\ref{tab:Pade_EpsilonExp}, we also present extrapolated results for the scaling dimension $\Delta_{i\overline{\Psi}\Gamma_3\Gamma_5\Psi}$ which corresponds to the QAH susceptibility exponent. The leading-order approximants for $1/\nu$ fall within the range predicted by resummation of the $\epsilon$-expansion result. The fermion bilinear results are in good agreement with one another and their uniformity is slightly better than that observed in the chiral-Ising QED$_3$-GNY case~\cite{gracey2018}. The extrapolated values for $\Delta_{\overline{\Psi}\Psi}$ are highly uniform and fall within the broad range of values predicted by resummation of the $\epsilon$ expansion; by contrast, apart from the [0/2] Pad\'e approximant, the large-$N_{f}$ based extrapolated values for $\Delta_{i\overline{\Psi}\Gamma_5\Psi}$ are higher than the values predicted by the $\epsilon$-expansion resummation of this quantity. None of the possible approximants for $\eta_{\phi}$ at this order exist, and thus no quantitative statements about this exponent can be made.

\begin{table}[t]
\centering
\begin{tabular}{|l||c|c|l|c|c|}
      \hline
           $N_{f} = 1$                                     & $\eta_\phi$  & $1/\nu$        & $\Delta_{\overline{\Psi}\Psi}$   & $\Delta_{i\overline{\Psi}\Gamma_{5}\Psi}$       & $\Delta_{i\overline{\Psi}\Gamma_{3}\Gamma_{5}\Psi}$\\
          \hline
      $[0/1]$                                            & $\times$      & 0.4022         & 1.873                                        & 1.458   								  & 7.783\\
      $[0/1]_{\textrm{PB}}$ 
                                                             & $\times$      & 0.5192         & 1.880                                        & 1.540 								  & $\times$\\
      \hline
      $[0/2]$                                           & $\times$      & --                   & 1.891                                       & 1.138  					                  	  & -- \\
      $[0/2]_{\textrm{PB}}$    
                                                            & $\times$       & --                 & $\times$                                   & 1.351					          		  & -- \\
                                                            
      \hline 
      $[1/1]$                                           & $\times$     & --                  & 1.889                                         & $\times$								  & -- \\
      $[1/1]_{\textrm{PB}}$  
                                                           & $\times$      & --                  & 1.887                                         & $\times$ 							 & -- \\
     \hline
\end{tabular}
      \caption{Pad\'e and Pad\'e-Borel resummations of the large-$N_{f}$ expressions for $d=3$ and $N_{f}=1$. Approximants which are either singular in the domain $N_f\geq 1$, undefined, or negative, are denoted by $\times$. The exponents that are unknown beyond $\mathcal{O}(1/N_f)$ are denoted by $-$, for these quantities only one approximant can be used.}
      \label{tab:Pade_LargeN}
\end{table}

\subsection{Pure chiral Heisenberg GNY model}

\begin{figure}[t]
\centering\includegraphics[width=8.5cm,height=4.5cm,clip]{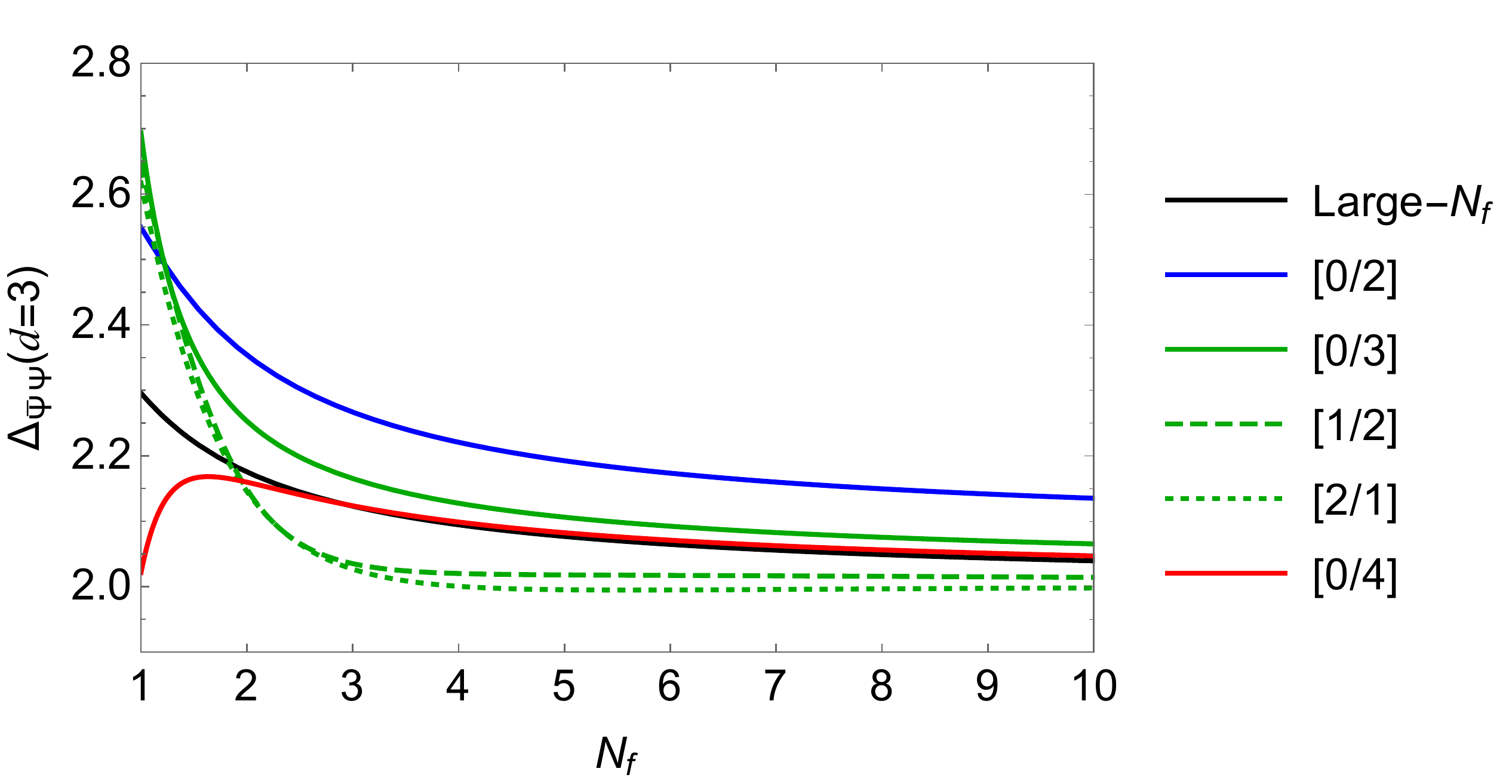}
\caption{Colored lines: Pad\'e approximants for $\Delta_{\overline{\Psi}\Psi}$ in $d=3$ as a function of $N_{f}$  in the chiral Heisenberg GNY model at two (blue), three (green), and four-loop (red) orders; black line: large-$N_f$ result from Eq.~(\ref{SingletLargeNcHGNY}).}
\label{fig:cHGNYdeltaSinglet_3d}
\end{figure}

\begin{figure}[t]
\centering\includegraphics[width=8.5cm,height=4.5cm,clip]{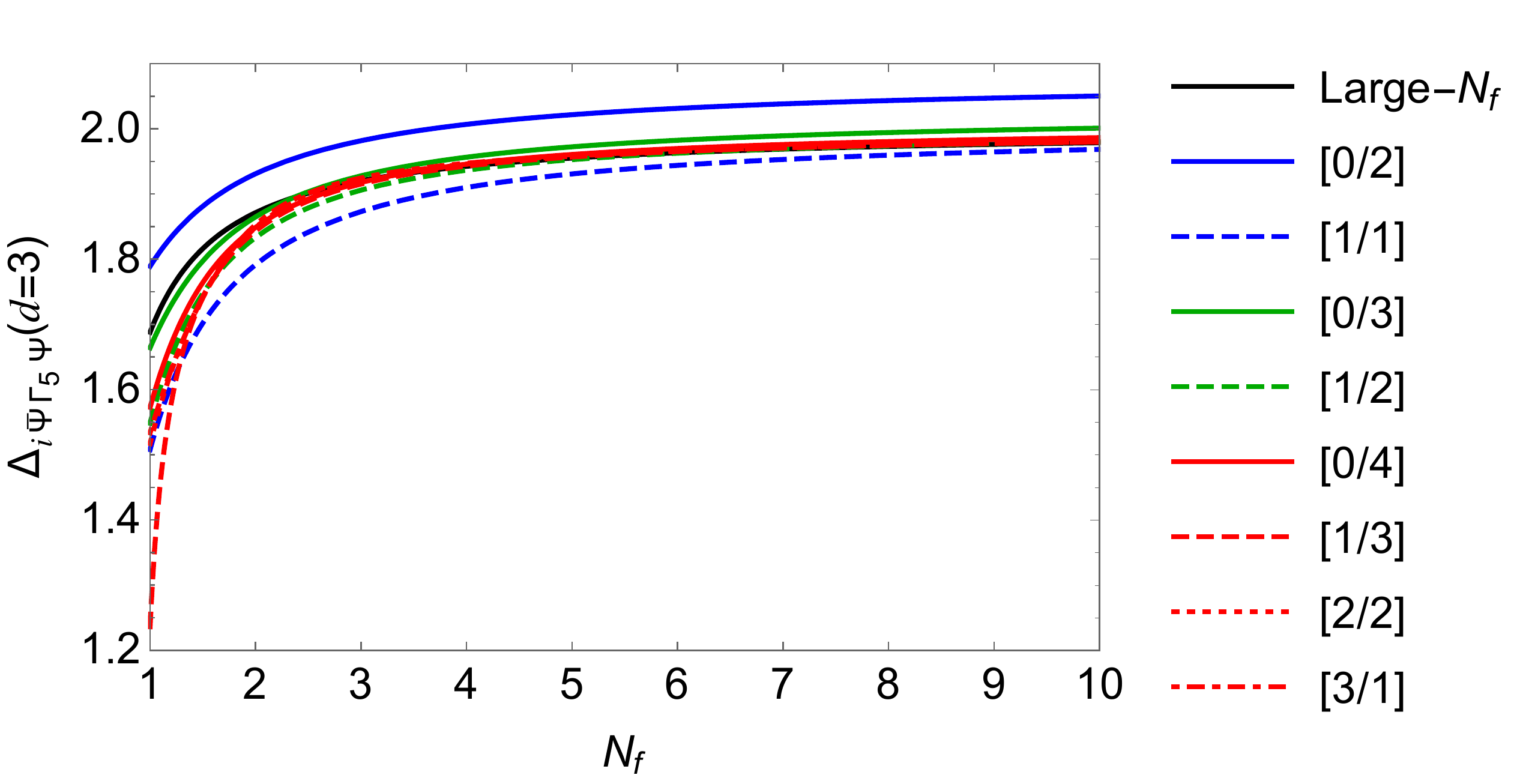}
\caption{Colored lines: Pad\'e approximants for $\Delta_{i\overline{\Psi}\Gamma_{5}\Psi}$ in $d=3$ as a function of $N_{f}$  in the chiral Heisenberg GNY model at two (blue), three (green), and four-loop (red) orders; black line: large-$N_f$ result from Eq.~(\ref{AxialLargeNcHGNY}).}
\label{fig:cHGNYdeltaAxial_3d}
\end{figure}

We finally turn to the resummation of $\epsilon$-expansion and large-$N_f$ expressions for the fermion bilinear scaling dimensions in the pure chiral Heisenberg GNY model. Figures~\ref{fig:cHGNYdeltaSinglet_3d} and \ref{fig:cHGNYdeltaAxial_3d} show the results of Pad\'e extrapolation of the $\epsilon$-expansion expressions for the normal and axial mass operator dimensions, respectively, as a function of $N_f$. The corresponding large-$N_f$ expressions, Eqs.~(\ref{SingletLargeNcHGNY}) and (\ref{AxialLargeNcHGNY}), are plotted on the same graphs for comparison purposes. The spread of values for $\Delta_{\overline{\Psi}\Psi}$ is appreciable, but comparatively less than for the same quantity in the presence of the gauge field (Fig.~\ref{fig:deltaSinglet_3d}). This state of affairs was also observed in the chiral Ising QED-GNY model~\cite{zerf2018}. Remarkably, the large-$N_f$ result and the [0/4] Pad\'e approximant give very similar results for $N_f\geq 2$. The spread of values is even smaller, and the convergence of the approximants with increasing loop order better, for $\Delta_{i\overline{\Psi}\Gamma_5\Psi}$.

\begin{figure}[t]
\centering\includegraphics[width=8.5cm,height=4.5cm,clip]{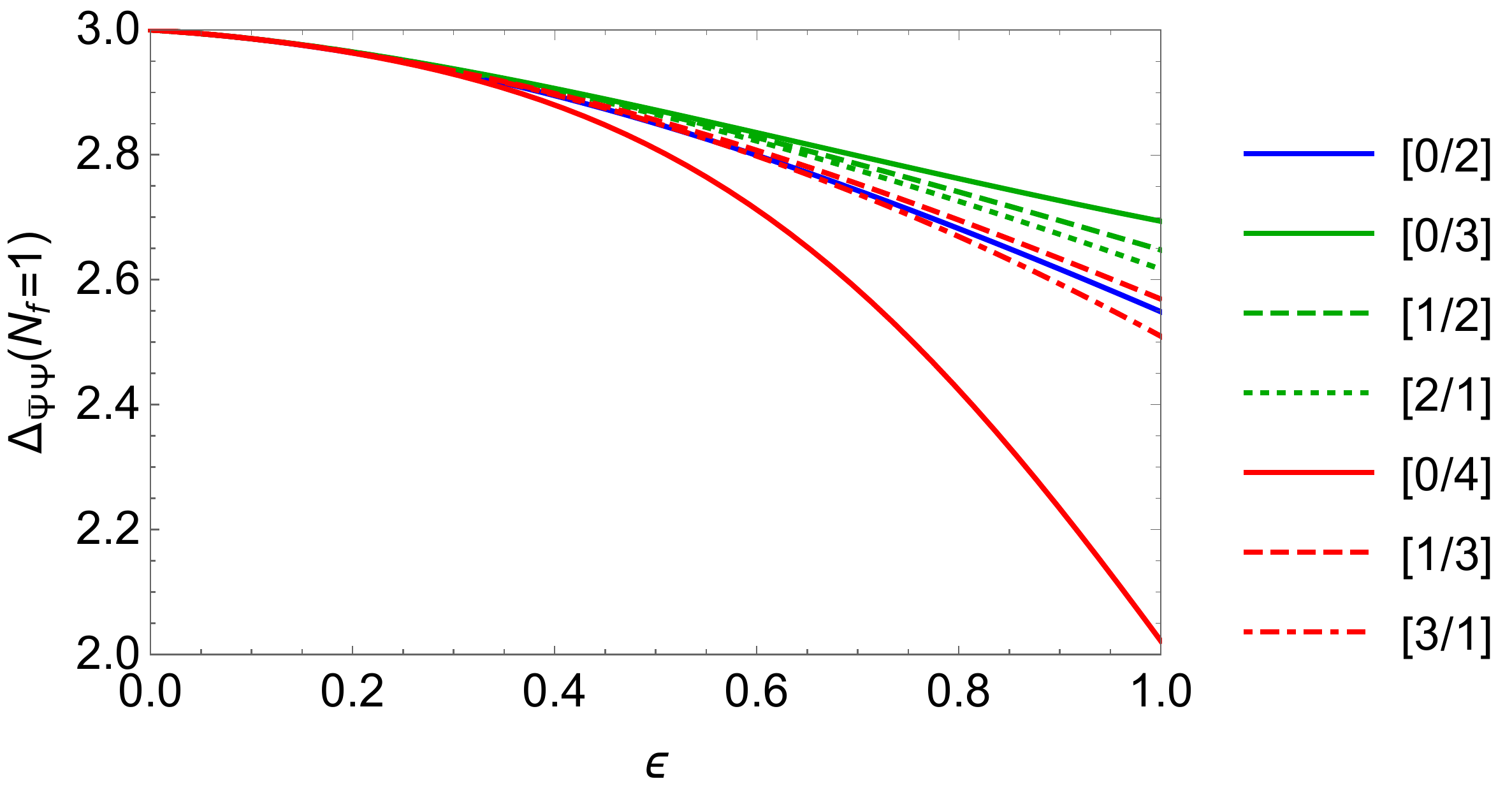}
\caption{Pad\'e approximants for $\Delta_{\overline{\Psi}\Psi}$ as a function of $\epsilon$ for $N_{f}=1$ in the chiral Heisenberg GNY model at two (blue), three (green), and four-loop (red) orders.}
\label{fig:cHGNYdeltaSinglet_N1}
\end{figure}

\begin{figure}[t]
\centering\includegraphics[width=8.5cm,height=4.5cm,clip]{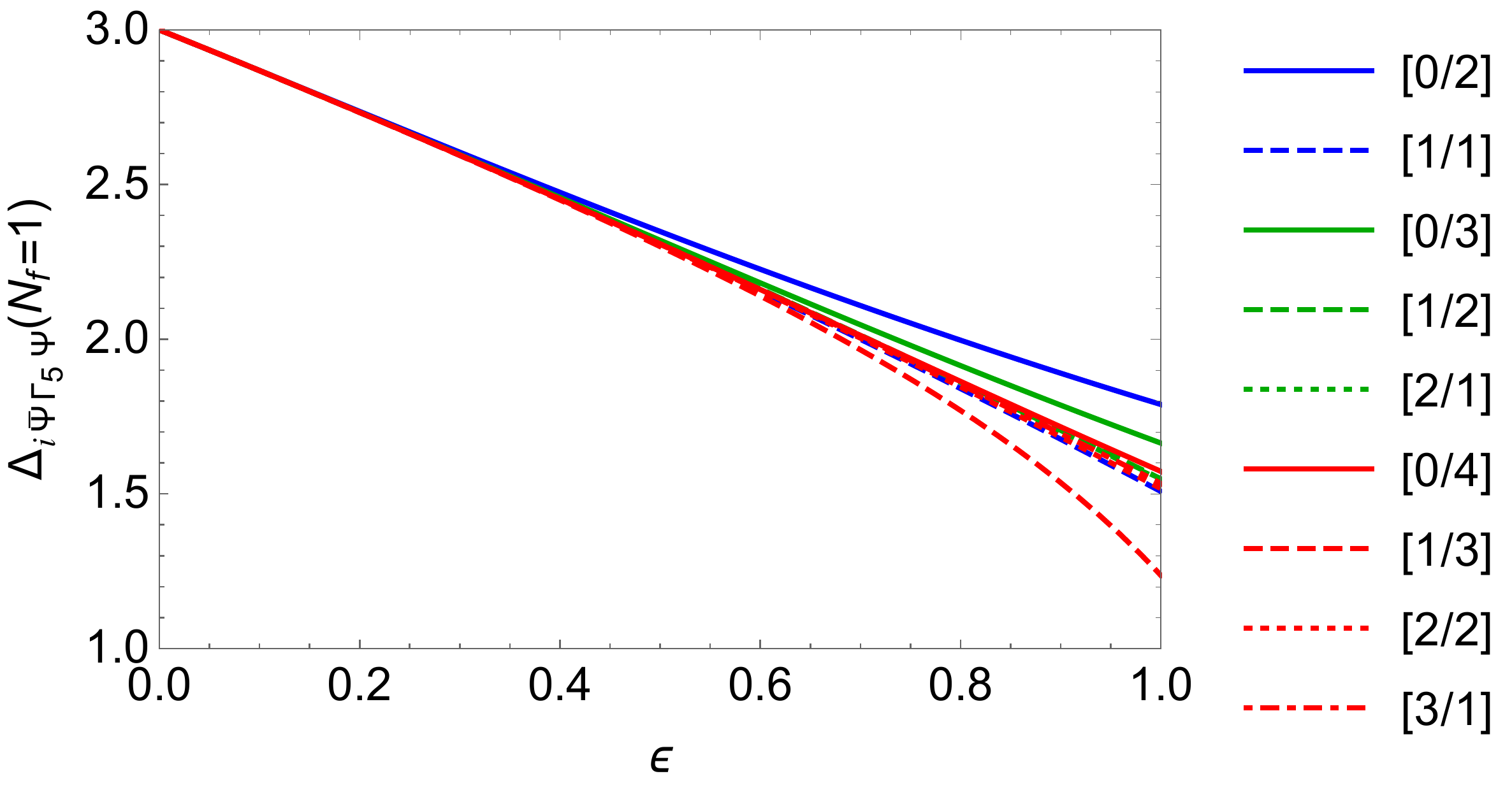}
\caption{Pad\'e approximants for $\Delta_{i\overline{\Psi}\Gamma_{5}\Psi}$ as a function of $\epsilon$ for $N_{f}=1$ in the chiral Heisenberg GNY model at two (blue), three (green), and four-loop (red) orders.}
\label{fig:cHGNYdeltaAxial_N1}
\end{figure}

In Figs.~\ref{fig:cHGNYdeltaSinglet_N1} and \ref{fig:cHGNYdeltaAxial_N1} we plot the Pad\'e approximants for $N_f=1$ as a function of $\epsilon$, and in Table~\ref{tab:Pade_EpsilonExp_cHGNY} we present numerical results for the $N_f=1$ Pad\'e and Pad\'e-Borel approximants in $d=3$. With the exception of the [0/4] approximant without Borel resummation, the results give a reasonably consistent estimate for $\Delta_{\overline{\Psi}\Psi}$ in the range $\sim$~$2.5$ - $2.7$, and for $\Delta_{i\overline{\Psi}\Gamma_5\Psi}$ in the range $\sim$~$1.2$ - $2.0$. For comparison, in Table~\ref{tab:Pade_LargeN_cHGNY} we display the Pad\'e and Pad\'e-Borel approximants obtained from the large-$N_f$ expansion results. While the approximants for the axial mass dimension fall within the range predicted by the $\epsilon$-expansion results, those for the normal mass dimension fall either within the same range (the [0/1] approximant) or give somewhat lower values ($\sim$~$2.1$ - $2.3$).

\begin{table}[t]
\centering
\begin{tabular}{|l||c|c|}
      \hline
      $N_{f} = 1$                             & $\Delta_{\overline{\Psi}\Psi}$       &$\Delta_{i\overline{\Psi}\Gamma_{5}\Psi}$    \\
          \hline\hline
      $[0/2]$                                    & 2.55                                             & 1.79 \\
      $[0/2]_{\textrm{PB}}$ 
                                                     & 2.66                                              & 2.01 \\
      \hline
      $[1/1]$       			           & $\times$                               	& 1.51 \\
      $[1/1]_{\textrm{PB}}$   
                           			  & $\times$           			        & $\times$ \\
    
      \hline
      $[0/3]$                                    & 2.69            					& 1.66 \\
      $[0/3]_{\textrm{PB}}$  
                             			  & $\times$	 				& 1.92 \\
      \hline
      $[1/2]$                                   &  2.65             				& 1.55 \\
      $[1/2]_{\textrm{PB}}$         
                           		          &  2.69			 		        & 1.56 \\
       \hline
      $[2/1]$            		          & 2.62	                  			& 1.54 \\
      $[2/1]_{\textrm{PB}}$                 
                             			  & 2.61                				& 1.54 \\
   
       \hline
      $[0/4]$            		          & 2.02	         				& 1.57 \\
      $[0/4]_{\textrm{PB}}$   
                            			  & 2.54	                 			& 1.87  \\
       \hline
      $[1/3]$          		          & 2.57	   	                                 & 1.24  \\
      $[1/3]_{\textrm{PB}}$     
                            	             	  & 2.67 	    				       & $\times$ \\
      \hline 
      $[2/2]$            			  & $\times$ 				       & 1.52 \\
      $[2/2]_{\textrm{PB}}$     
                             			  &  $\times$ 		                       & $\times$ \\
      \hline
      $[3/1]$            		          & 2.51		     			      & 1.53 \\
      $[3/1]_{\textrm{PB}}$    
                            	                   & 2.52  				              & 1.53 \\
      \hline
\end{tabular}
    \caption{Pad\'e and Pad\'e-Borel resummations of the $\epsilon$-expansion expressions for $d=3$ and $N_f=1$ in the chiral Heisenberg GNY model. Approximants which are either singular in the domain $0<\epsilon<1$, undefined, or negative, are denoted by $\times$.}
      \label{tab:Pade_EpsilonExp_cHGNY}
\end{table}

\begin{table}[t]
\centering
\begin{tabular}{|l||c|c|}
      \hline
      $N_{f} = 1$                                          & $\Delta_{\overline{\Psi}\Psi}$   & $\Delta_{i\overline{\Psi}\Gamma_{5}\Psi}$  \\
          \hline
      $[0/1]$                                                 & 2.508                                       & 1.816 \\								
      $[0/1]_{\textrm{PB}}$ 
                                                                  &  $\times$                                 & 1.829 \\						
      \hline
      $[0/2]$                                                 &  2.239                                      & 1.715 \\									
      $[0/2]_{\textrm{PB}}$    
                                                                  & 2.168                                      & 1.752 \\									  
                                                            
      \hline
      $[1/1]$                                                 & 2.319                                       & 1.559 \\								 
      $[1/1]_{\textrm{PB}}$  
                                                                  & 2.325                                       & $\times$ \\									
      \hline
\end{tabular}
      \caption{Pad\'e and Pad\'e-Borel resummations of the large-$N_{f}$ expressions for $d=3$ and $N_{f}=1$ in the chiral Heisenberg GNY model. Approximants which are either singular in the domain $N_f\geq 1$, undefined, or negative, are denoted by $\times$.}
      \label{tab:Pade_LargeN_cHGNY}
\end{table}

\section{Conclusion} 
\label{sec:Conclusion}

In this paper we provided a comprehensive analysis of the critical properties of the chiral Heisenberg QED$_3$-GNY model, motivated by a recent QMC study~\cite{Meng2019} of a $U(1)$ lattice gauge theory with spinful fermions on the 2D square lattice that observed a direct transition between a $U(1)$ deconfined phase, adiabatically connected to the ASL, and an antiferromagnetic N\'eel-ordered phase. Using the $\epsilon$ expansion below four spacetime dimensions to four-loop order, we showed the existence of a {\it bona fide} critical fixed point in the chiral Heisenberg QED$_3$-GNY model for an arbitrary number $N_f$ of $SU(2)$ doublets of four-component Dirac fermions. The existence of a critical fixed point for $N_f=1$ establishes that the N\'eel-ASL transition should be continuous, in agreement with the numerical results in Ref.~\cite{Meng2019}. This fixed point was also shown to exist in the large-$N_f$ expansion in fixed $2<d<4$ spacetime dimensions. The $\epsilon$-expansion was used to compute several critical exponents to $\mathcal{O}(\epsilon^4)$, including the order parameter anomalous dimension $\eta_\phi$, the inverse correlation length exponent $1/\nu$, and the stability critical exponent $\omega$. We additionally computed the scaling dimensions of two Dirac fermion bilinears, the normal mass operator $\overline{\Psi}\Psi$ and the axial mass operator $i\overline{\Psi}\Gamma_5\Psi$, which for $N_f=1$ had the interpretation of CDW and VBS susceptibility exponents at the N\'eel-ASL transition. These $\epsilon$-expansion results were supplemented by computations in the large-$N_f$ expansion. The exponents $\eta_\phi$, $\Delta_{\overline{\Psi}\Psi}$, and $\Delta_{i\overline{\Psi}\Gamma_5\Psi}$ were computed to $\mathcal{O}(1/N_f^2)$, while $1/\nu$ was computed to $\mathcal{O}(1/N_f)$. Additionally, the QAH susceptibility exponent, controlled by the scaling dimension of the time-reversal- and parity-odd bilinear $i\overline{\Psi}\Gamma_3\Gamma_5\Psi$, was computed to $\mathcal{O}(1/N_f)$. The CDW and VBS exponents were also computed at the critical point of the pure chiral Heisenberg GNY model, which for $N_f=1$ describes the semimetal-AF insulator transition in the honeycomb lattice and the $\pi$-flux square lattice. Pad\'e and Pad\'e-Borel resummation techniques were subsequently applied to obtain numerical estimates of all critical exponents in $d=3$ for general $N_f$, with a special emphasis on $N_f=1$. The critical exponents computed here are in principle accessible to sign-problem-free QMC simulations~\cite{Meng2019}, and we hope to compare the estimates obtained here to numerical studies of critical properties at the N\'eel-ASL QCP in the near future. Regarding possible experimental realizations, besides solid-state frustrated magnets we take note of the recent progress in the quantum simulation of lattice gauge theories~\cite{zohar2015}. (1+1)D QED was recently simulated experimentally in an ion-trap setup~\cite{martinez2016}, and concrete proposals exist to engineer lattice QED in (2+1)D~\cite{zohar2013}. The latter would represent a direct physical realization of Hamiltonian (\ref{Hlattice}).

Besides its applications to condensed matter physics, the critical fixed point of the chiral Heisenberg QED$_3$-GNY model is of interest as an example of (2+1)D conformal field theory. Recent years have witnessed a resurgence of interest in such theories, due in large part to highly successful numerical implementations of the conformal bootstrap program (for a recent review, see, e.g., Ref.~\cite{poland2019}). These have led to the determination of $d=3$ critical exponents with unprecedented accuracy in various models of interest to both statistical mechanics/condensed matter physics and high-energy physics, such as the 3D Ising~\cite{el-showk2012} and $O(N)$ vector~\cite{kos2014} models, the chiral Ising GNY model~\cite{iliesiu2016}, and QED$_3$~\cite{chester2016b,li2018}. We hope that this work will stimulate the study of chiral Ising/XY/Heisenberg QED$_3$-GNY models with the conformal bootstrap technique, which may lead to precise determinations of critical exponents that could be compared with $\epsilon$- and large-$N_f$ expansion results presented in this and previous work.

\acknowledgements

We gratefully acknowledge I. Affleck, \'E. Dupuis, S. Giombi, L. Janssen, I. R. Klebanov, X.-Y. Song, K. Wamer, and W. Witczak-Krempa for helpful discussions. RB was supported by the Theoretical Physics Institute at the University of Alberta. The work of JAG was supported by a DFG Mercator Fellowship and he thanks the
Mathematical Physics Group at Humboldt University, Berlin for its hospitality where part of this work was carried out. JM was supported by NSERC grant \#RGPIN-2014-4608, the CRC Program, CIFAR, and the University of Alberta. This project has received funding from the European Union's Horizon 2020 research and innovation programme under the Marie Sk\l{}odowska-Curie grant agreement No.~764850 (SAGEX).

\appendix
\numberwithin{equation}{section}
\numberwithin{figure}{section}

\section{Symmetry transformation properties of fermion bilinears}
\label{app:PSG}

In this Appendix we determine how Dirac fermion bilinears of the form $\overline{\psi}\tau_i\sigma_j\psi$, that are invariant under $d=3$ Lorentz transformations in the continuum theory, transform under the symmetries of the microscopic Hamiltonian (\ref{Hlattice}). (The $\tau_i$ matrices act in nodal ($\pm$) space, while the $\sigma_j$ matrices act in spin space.) This in turn allows us to associate to each bilinear a microscopic operator with the same symmetry properties, but defined in terms of the original lattice fermions $c_{r\sigma},c_{r\sigma}^\dag$, whose long-distance correlations at the N\'eel-ASL QCP will be governed by the scaling dimension of the bilinear at the chiral Heisenberg QED$_3$-GNY fixed point.

The microscopic symmetries of interest are the symmetries of the $p4m$ space group of the square lattice, generated by the four-fold rotation $C_4$ about a site and the reflection $R_x$ about the $yz$ plane and through a site, the unit lattice translations $T_x$ and $T_y$, and time-reversal symmetry $\mathcal{T}$. The model also has a particle-hole symmetry at half filling, but we will not discuss it further. Choosing the two Bravais lattice vectors for the $\pi$-flux phase in Fig.~\ref{fig:square} as $\b{a}_1=\hat{\b{x}}-\hat{\b{y}}$ and $\b{a}_2=\hat{\b{x}}+\hat{\b{y}}$, and denoting the annihilation operator $c_{\b{R}A\sigma}$ ($c_{\b{R}B\sigma}$) for a fermion of spin $\sigma$ on Bravais lattice site $\b{R}=n_1\b{a}_1+n_2\b{a}_2$, $n_1,n_2\in\mathbb{Z}$ and sublattice A (sublattice B) by $c_{(n_1,n_2)A\sigma}$ ($c_{(n_1,n_2)B\sigma}$), the fermion annihilation operator transforms as
\begin{align}
C_4\colon&c_{(n_1,n_2)A\sigma}\rightarrow c_{(-n_2,n_1)A\sigma},\label{symm1}\\
&c_{(n_1,n_2)B\sigma}\rightarrow c_{(-n_2,n_1-1)B\sigma},\\
R_x\colon&c_{(n_1,n_2)A\sigma}\rightarrow c_{(-n_2,-n_1)A\sigma},\\
&c_{(n_1,n_2)B\sigma}\rightarrow c_{(-n_2,-n_1)B\sigma},\\
T_x\colon&c_{(n_1,n_2)A\sigma}\rightarrow c_{(n_1+1,n_2)B\sigma},\\
&c_{(n_1,n_2)B\sigma}\rightarrow c_{(n_1,n_2+1)A\sigma},\\
T_y\colon&c_{(n_1,n_2)A\sigma}\rightarrow c_{(n_1,n_2)B\sigma},\\
&c_{(n_1,n_2)B\sigma}\rightarrow c_{(n_1-1,n_2+1)A\sigma},\label{symm8}
\end{align}
as well as $c_{(n_1,n_2)A\sigma}\rightarrow(i\sigma_2)_{\sigma\sigma'}c_{(n_1,n_2)A\sigma'}$ and $i\rightarrow -i$ under $\mathcal{T}$, and likewise on the B sublattice.

\begin{table}[t]
\begin{tabular}{|c||c|c|c|c|}
\hline
 & $\overline{\psi}\bsigma\psi$ & $\overline{\psi}\tau_1\bsigma\psi$ & $\overline{\psi}\tau_2\bsigma\psi$ & $\overline{\psi}\tau_3\bsigma\psi$  \\
\hline\hline
$C_4$ & $+$ & $-\overline{\psi}\tau_2\bsigma\psi$ & $\overline{\psi}\tau_1\bsigma\psi$ & $+$ \\ \hline
$R_x$ & $-$ & $+$ & $-$ & $+$ \\ \hline
$T_x$ & $+$ & $+$ & $-$ & $-$ \\ \hline
$T_y$ & $+$ & $-$ & $+$ & $-$ \\ \hline
$\mathcal{T}$ & $+$ & $-$ & $-$ & $-$ \\ \hline
\end{tabular}
\caption{PSG transformation properties of spin-triplet Dirac fermion bilinears.}
\label{triplet}
\end{table}

To determine how the Dirac fermion bilinears $\overline{\psi}\tau_i\sigma_j\psi$ transform under the microscopic symmetries, one must first determine the PSG~\cite{wen2002} associated with the Hamiltonian (\ref{Hpiflux}), i.e., the combinations of transformations (\ref{symm1})-(\ref{symm8}) and $U(1)$ gauge transformations 
\begin{align}
c_{(n_1,n_2)A\sigma}&\rightarrow e^{i\theta_{(n_1,n_2)A}}c_{(n_1,n_2)A\sigma},\\
c_{(n_1,n_2)B\sigma}&\rightarrow e^{i\theta_{(n_1,n_2)B}}c_{(n_1,n_2)B\sigma},
\end{align}
that leave the $\pi$-flux Hamiltonian (\ref{Hpiflux}) invariant. Writing any of the previously listed symmetry transformations as $\mathcal{S} c_{r\sigma} \mathcal{S}^{-1}=S_{\sigma\sigma'}c_{S(r)\sigma'}$, where $S(r)$ denotes the transformed coordinate, the associated PSG transformation is
\begin{align}\label{PSG}
\tilde{\mathcal{S}} c_{r\sigma}\tilde{\mathcal{S}}^{-1}=G^S_rS_{\sigma\sigma'}c_{S(r)\sigma'},
\end{align}
where $G^S_r\in U(1)$. By explicit calculation one finds that the gauge transformations in the PSG can be chosen as
\begin{align}
G^{C_4}_{(n_1,n_2)A}=G^{C_4}_{(n_1,n_2)B}&=(-1)^{n_1},\\
G^{R_x}_{(n_1,n_2)A}=G^{R_x}_{(n_1,n_2)B}&=(-1)^{n_1+n_2},\\
G^{T_x}_{(n_1,n_2)A}=G^{T_x}_{(n_1,n_2)B}&=(-1)^{n_1+n_2},\\
G^{T_y}_{(n_1,n_2)A}=G^{T_y}_{(n_1,n_2)B}&=(-1)^{n_1+n_2},\\
G^\mathcal{T}_{(n_1,n_2)A}=G^\mathcal{T}_{(n_1,n_2)B}&=1.
\end{align}
Next one can determine how the PSG acts on the Dirac fermions by following the procedure detailed in Ref.~\cite{hermele2008}, i.e., by expanding the Dirac fermion fields in the $\b{q}\rightarrow 0$ limit in terms of the microscopic fermion operators, and using Eq.~(\ref{PSG}). We find
\begin{align}
C_4\colon&\psi\rightarrow e^{i\pi\mu_3/4}e^{-i\pi\tau_3/4}\psi,\label{PSGDirac1}\\
R_x\colon&\psi\rightarrow \mu_2\tau_2\psi,\\
T_x\colon&\psi\rightarrow i\tau_1\psi,\\
T_y\colon&\psi\rightarrow i\tau_2\psi,\\
\mathcal{T}\colon&\psi\rightarrow\mu_2\tau_2i\sigma_2\psi K,\label{PSGDirac5}
\end{align}
where the $\mu_i$ act on the two-component spinor (Dirac) indices and $K$ denotes complex conjugation. One can check that these are indeed symmetries of the Dirac Hamiltonian (\ref{HDirac}), provided the momentum $\b{p}$ is appropriately transformed. From Eq.~(\ref{PSGDirac1})-(\ref{PSGDirac5}) we can in turn determine the PSG transformation properties of spin-triplet (Table~\ref{triplet}) and spin-singlet (Table~\ref{singlet}) Dirac fermion bilinears.

\begin{table}[t]
\begin{tabular}{|c||c|c|c|c|}
\hline
 & $\overline{\psi}\psi$ & $\overline{\psi}\tau_1\psi$ & $\overline{\psi}\tau_2\psi$ & $\overline{\psi}\tau_3\psi$  \\
\hline\hline
$C_4$ & $+$ & $-\overline{\psi}\tau_2\psi$ & $\overline{\psi}\tau_1\psi$ & $+$ \\ \hline
$R_x$ & $-$ & $+$ & $-$ & $+$ \\ \hline
$T_x$ & $+$ & $+$ & $-$ & $-$ \\ \hline
$T_y$ & $+$ & $-$ & $+$ & $-$ \\ \hline
$\mathcal{T}$ & $-$ & $+$ & $+$ & $+$ \\ \hline
\end{tabular}
\caption{PSG transformation properties of spin-singlet Dirac fermion bilinears.}
\label{singlet}
\end{table}

\bibliography{O3QED3GNY}

\begin{thebibliography}{105}
\bibitem [{Sup()}]{SuppMat}%
  \BibitemOpen
  \href@noop {} {}\bibinfo {note} {We provide the full analytical four-loop
  expressions at general $N_f$ for the $\beta$ and $\gamma$ functions in the
  files \texttt{cHQED3GNYbetagamma.m}, \texttt{cHQED3GNYbetagamma.inc}, and for
  the critical exponents in the files \texttt{cHQED3GNYexponents.m},
  \texttt{cHQED3GNYexponents.inc}.}\BibitemShut {Stop}%
\end{thebibliography}

\end{document}